\documentclass[trackchanges,twocolumn]{aastex7}
\DeclareSymbolFont{matha}{OML}{txmi}{m}{it}% txfonts
\DeclareMathSymbol{\varv}{\mathord}{matha}{118}

\usepackage{tabularx}
\usepackage{natbib}
\usepackage{amsmath}
\usepackage{amssymb}
\usepackage{bm}
\usepackage{comment}
\usepackage{hyperref}
\usepackage[normalem]{ulem}
\usepackage{soul}
\usepackage{tablefootnote}
\usepackage{graphicx}
\usepackage{color, colortbl}
\usepackage{dcolumn}    % Align decimal points in tables
\newcolumntype{z}{D{.}{.}{4}}
\newcolumntype{d}{D{.}{.}{1}}
\newcommand{\Npar}{N_{\rm{par}}}

\definecolor{darkgreen}{rgb}{0.0,0.6,0.00}

\usepackage{CJKutf8}

\defcitealias{Baronett24}{B24}

\begin{document}
\begin{CJK}{UTF8}{mj}
\title{Bridging Unstratified and Stratified Simulations of the Streaming Instability for $\tau_s=0.1$ Grains}

%% Author List 
\correspondingauthor{Jeonghoon Lim}
\author[0000-0003-2719-6640]{Jeonghoon Lim (임정훈)
}
\affiliation{Department of Physics and Astronomy, Iowa State University, Ames, IA 50010, USA}
\email[show]{jhlim@iastate.edu}

\author[orcid=0000-0003-0412-760X,gname=Stanley,sname=Baronett]{Stanley A. Baronett}
\affiliation{Nevada Center for Astrophysics, University of Nevada, Las Vegas, Box 454002, Las Vegas, NV 89154, USA}
\affiliation{Department of Physics and Astronomy, University of Nevada, Las Vegas, Box 454002, Las Vegas, NV 89154, USA}
\email{barons2@unlv.nevada.edu}

\author[0000-0002-3771-8054]{Jacob B. Simon}
\affiliation{Department of Physics and Astronomy, Iowa State University, Ames, IA 50010, USA}
\email{jbsimon@iastate.edu}

\author[orcid=0000-0003-2589-5034,gname=Chao-Chin,sname=Yang]{Chao-Chin Yang (楊朝欽)}
\affiliation{Department of Physics and Astronomy, The University of Alabama, Box 870324, Tuscaloosa, AL 35487-0324, USA}
\email{ccyang@ua.edu}

\author[0000-0003-0801-3159]{Debanjan Sengupta}
\affiliation{Department of Astronomy, New Mexico State University, PO Box 30001, MSC 4500, Las Cruces, NM 88003-8001, USA}
\email{debanjan@nmsu.edu}

\author[0000-0001-5372-4254]{Orkan M. Umurhan}
\affiliation{SETI Institute, 389 Bernardo Way, Mountain View, CA 94043, U.S.A.}
\affiliation{Space Sciences Division, Planetary Systems Branch, NASA Ames Research Center, Mail Stop 245-3, Moffett Field, CA 94035, USA}
\affiliation{Cornell Center for Astrophysics and Planetary Sciences, Cornell University, Ithaca, NY 14853, USA}
\email{oumurhan@seti.org}

\author[0000-0002-3768-7542]{Wladimir Lyra}
\affiliation{Department of Astronomy, New Mexico State University, PO Box 30001, MSC 4500, Las Cruces, NM 88003-8001, USA}
\email{wlyra@nmsu.edu}
%% Author List 

\begin{abstract}
The streaming instability (SI), driven by aerodynamic coupling between solids and the gas under a global radial pressure gradient, concentrates solids and facilitates planetesimal formation.  Unstratified simulations are commonly used to study the SI, based on the assumption that they approximate conditions near the disk midplane. However, it remains unclear how accurately these unstratified simulations capture the midplane dust-gas dynamics in stratified disks. To address this, we examine the saturated state of the SI in stratified simulations and compare dust-gas dynamics to those in unstratified simulations across various radial pressure gradients. To this end, we consider a dimensionless dust stopping time  ($\tau_s$) of 0.1 and perform 2D axisymmetric, stratified simulations. We find that the formation of dust filaments during dust settling exhibits morphological similarities to those in unstratified simulations. Vertical gravity acts to redistribute momentum vertically in response to momentum flux, resulting in midplane velocities in the center-of-mass frame that are consistent with those from unstratified models at any given pressure gradient. Furthermore, the velocity dispersions and density distributions of the gas and dust near the midplane of our stratified simulations closely match those in unstratified simulations. While further exploration across the parameter space is needed, our results suggest that, for $\tau_s=0.1$, unstratified simulations represents well the midplane dust--gas dynamics in stratified disks before any strong clumping occurs. Consequently, our results confirm that in the saturated state, the streaming turbulence in stratified simulations behaves similarly to that in unstratified simulations for the parameter values explored here.

\end{abstract}


\section{Introduction} \label{sec:intro}
The core-accretion planet formation scenario involves a range of processes, including the growth of micron-sized dust grains into millimeter- to centimeter-sized pebbles through collisional coagulation \citep{Blum08,Guttler10,Zsom10}, the formation of kilometer-sized planetesimals (e.g., \citealt{Johansen15,simon_mass_2016,Schafer17,Li_demographics_2019}), the formation of Earth-sized planets or planetary cores through planetesimal \citep{Greenberg1978,KokuboIda1998,Raymond2009} and/or pebble accretion \citep{OrmelKlahr2010,LambrechtsJohansen2012,Bitsch+15}, and the formation of gas giants through runaway gas accretion \citep{Pollack+96,AyliffeBate2012}. Among these processes, the formation of planetesimals entails substantial increases in both size and mass and plays a key role in connecting dust grain coagulation and the formation of planetary cores.

One promising route towards planetesimal formation is through the gravitational collapse of a pebble cloud. For this to occur, pebbles need to be concentrated to sufficiently high volume density such that the mutual gravitational attraction between pebbles overcomes tidal shear and turbulent diffusion \citep{klahr_turbulence_2020,Gerbig20,gerbig_planetesimal_2023}. Several mechanisms for pebble concentration have been reported, including local pressure maxima (\citealt{johansen_rapid_2007,Carrera2021, Carrera2022, schafer_coexistence_2022, xu_turbulent_2022}), vortices (e.g., \citealt{Lyra24}), and the streaming instability (SI; e.g., \citealt{YG05,YJ07,JY07}). 

Currently, the leading candidate for planetesimal formation is the SI, a linear instability in rotating disks, arising from the mutual aerodynamic drag between dust and the gas \citep{YG05}. The instability is driven by radial pressure gradients causing the gas to deviate from Keplerian rotation, whereas the pebbles continue on Keplerian orbits.  When the pebble-to-gas column density ratio exceeds a critical value, which is dependent on pebble size, the SI rapidly concentrates pebbles into filamentary structures where the local volume density of pebbles is high enough to trigger gravitational collapse 
\citep{Johansen09b,carrera_how_2015,Yang2017,LiYoudin21,Lim24b}. Numerical simulations of the SI that implement the self-gravity of pebbles have shown that planetesimals form in such regions with masses roughly comparable to minor bodies in the Solar System \citep{johansen_rapid_2007,Johansen15,simon_mass_2016,Schafer17,abod_mass_2019,Li_demographics_2019,Gole20,Schafer2024}. Furthermore, the SI has been shown to produce binary planetesimals with angular momentum and inclination distributions in agreement with Cold Classical Kuiper Belt Objects \citep{Nesvorny2019,Nesvorny2021}.

Before triggering strong dust clumping and planetesimal formation, the SI plays a significant role in regulating the dust-gas dynamics near the disk midplane. To date, detailed investigations of these dynamics have been conducted almost exclusively in vertically unstratified simulations. For example, \citet{JY07} were the first to examine the dynamics of dust and the gas driven by the SI in such a setup. More recently, \citet{Baronett24} (hereafter, \citetalias{Baronett24}) conducted a comprehensive study of the unstratified SI in its saturated state, utilizing high grid resolution and exploring various global pressure gradients. They demonstrated that kinematic and morphological properties of dust and the gas sensitively depend on the pressure gradient. For example, they found that velocity dispersions of the gas and dust, as well as the length scale of dust filaments, increase with increasing pressure gradients for both tightly and marginally coupled solids. 

By contrast, vertically stratified simulations have primarily focused on how the pressure gradient affects SI-driven clumping, rather than the broader dust-gas dynamics. For example, \citet{BaiStone10c_pressure_gradient} found that as the pressure gradient increases, the SI requires a higher dust-to-gas column density ratio to drive strong concentration of dust. Similarly, \citet{abod_mass_2019} reported a decrease in planetesimal formation efficiency with increasing pressure gradient. While \citet{Sengupta_Umurhan23} conducted detailed investigations of turbulence in the dust layer arising from various instabilities, their study did not vary the radial pressure gradient. Therefore, despite these previous efforts, how the radial pressure gradient influences the nonlinear evolution of the SI and the resulting dust-gas dynamics remains to be investigated in stratified disks. In this study, we aim to address this gap. This question is particularly relevant given that pressure gradients vary significantly across protoplanetary disks, especially in the presence of substructures such as pressure bumps and gaps (e.g.,\citealt{ALMA2015,andrews_disk_2018,MAPSLaw21,exoALMAStadler25}).

Furthermore, since unstratified simulations are often regarded as an approximation for the disk midplane, previous studies have identified notable similarities between unstratified simulations and the midplane of stratified simulations. For example, \citet{Yang2017} found that random dust voids driven by the SI develop within the dust layer in their stratified simulations (see their Figure 1), a phenomenon also observed in the saturated state of the SI in unstratified simulations with tightly coupled dust (\citealt{JY07}; \citetalias{Baronett24}). Additionally, both unstratified (\citealt{JY07}; \citetalias{Baronett24}) and stratified \citep{LiYoudin21} simulations show that the dust density field exhibits filamentary structures with fine-scale turbulent features when the dust is tightly coupled. By contrast, when dust is marginally coupled, the dust morphology appears more undulated, with less prominent small-scale filamentary feature. 

Despite this progress, more quantitative approaches are needed to rigorously examine the connection between unstratified simulations and the midplane of stratified simulations. Motivated by these considerations, we conduct a detailed investigation into the nonlinear saturated state of the SI across various radial pressure gradients in 2D axisymmetric numerical simulations of stratified disks. Specifically, we examine dust--gas dynamics driven by the SI by analyzing the kinematic properties and density distributions of the gas and dust during the saturated state of the SI, and compare them to those in unstratified simulations. For this comparison, we adopt results from the AB models in \citetalias{Baronett24} and carry out our stratified SI simulations using physical parameters consistent with their models.  

This paper is organized as follows. In Section \ref{sec:method}, we describe our stratified models within the local shearing box approximation. Section \ref{sec:results} presents our simulation results and the comparison to unstratified models from \citetalias{Baronett24}. We discuss our findings and future considerations in Section \ref{sec:Discussion} and provide a summary in Section \ref{sec:Summary}.

%%%%%%%  Table 1%%%%%%%%%%%%%%%%%
\begin{deluxetable}{zdcdccc}
\tablecaption{Model Parameters}\label{tab:simlist}
\tablehead{
\colhead{$Z$} & \colhead{$\Pi$} & \colhead{$L_x\times L_z/(\Pi H)^2$} & \colhead{$N_{\Pi H}$} & \colhead{$n_p$} \\
%2nd row
\colhead{(1)} &
\colhead{(2)} &
\colhead{(3)} &
\colhead{(4)} &
\colhead{(5)} &
}
\startdata
0.0024 & 0.01 & $4 \times 8$  & 256 & 1 \\
0.0048 & 0.02 & $4 \times 8$  & 256 & 1 \\
0.012 & 0.05  & $4 \times 8$  & 256 & 1 \\
0.024 & 0.1   & $4 \times 8$  & 256 & 1  \\
\hline 
\multicolumn{6}{c}{\textbf{AB runs from B24}} \\
... & 0.01 & $10 \times 10$  & 204.8 & 4 \\
... & 0.02 & $5 \times 5$    & 409.6 & 4 \\
... & 0.05 & $2 \times 2$   & 1024   & 4 \\
... & 0.1  & $1 \times 1$   & 2048   & 4  \\
\enddata
\tablecomments{ Columns: (1) surface density ratio of dust particles to the gas (see Equation \ref{eq:Z}); (2) dimensionless parameter for the radial pressure gradient (see Equation \ref{eq:Pi}); (3) domain size in units of $\Pi H$; (4) the number of cells per $\Pi H$; (5) the average number of particles per cell. All of the runs listed above have a dimensionless stopping time ($\tau_s$, Equation \ref{eq:taus}) of 0.1. The simulations from \citetalias{Baronett24} are unstratified such that they used a total dust-to-gas density ratio of 1 instead of $Z$. 
}    
\end{deluxetable}
%%%%%%%  Table 1%%%%%%%%%%%%%%%%%

\section{Methods} \label{sec:method}

We simulate the coupled dynamics of the gas and dust in a protoplanetary disk using numerical methods similar to those of \citet{Lim24b}, summarized as follows. 

We use the {\sc Athena} code \citep{Stone08} to model isothermal, unmagnetized gas with the particle module developed by \citet{BaiStone10a}. We adopt the local shearing-box approximation \citep{GoldreichLynden-Bell1965} to simulate a vertically stratified small patch of a protoplanetary disk (\citealt{Hawley95}; see \citealt{stone_implementation_2010} for {\sc Athena} implementation). The shearing box is centered at a fiducial disk radius ($r$), rotates at the corresponding Keplerian frequency ($\Omega$), and uses local Cartesian coordinates ($x$, $y$, $z$) to represent the radial, azimuthal, and vertical directions, respectively. Our models are axisymmetric ($x$-$z$), meaning that all three components of velocity are present but depend only on $x$ and $z$. We apply the standard shearing-periodic boundary conditions in $x$ and outflow boundary conditions in $z$ \citep{Li18}.

The gas is initialized in vertical hydrostatic balance and has a Gaussian density profile with vertical scale height $H$. Similarly, dust particles are initially distributed with a Gaussian density profile in the vertical direction.  The initial midplane densities of gas $(\rho_{g0})$ and dust particles $(\rho_{p0})$ are given by 
\begin{equation}\label{eq:rhog0}
    \rho_{g0} = \frac{\Sigma_{g0}}{\sqrt{2\pi} H},
\end{equation}
and 
\begin{equation}\label{eq:rhop0}
    \rho_{p0} = \frac{\Sigma_{p0}}{\sqrt{2\pi} H_{p0}},
\end{equation}
respectively. Here, $\Sigma_{p0}$ ($\Sigma_{g0}$) is the initial surface density of dust particles (the gas), and $H_{p0}$ is the initial scale height of dust particles. As the vertical profile of the particle density remains nearly Gaussian throughout our simulations, the particle scale height ($H_p$) at a given time is approximated by the standard deviation of vertical positions of the particles $(z_p)$:
\begin{equation}\label{eq:Hp}   
     H_p = \sqrt{\overline{z_p^2}-\overline{z_p}^2},
\end{equation}
where the overbar denotes an average over all particles. The vertical velocities of the gas and dust particles are initially zero. In the radial direction, the positions of the particles are randomly chosen from a uniform distribution. We apply the Nakagawa--Sekiya--Hayashi (NSH) equilibrium solutions \citep{Nakagawa1986} to the horizontal (i.e., radial and azimuthal) velocities of the gas and the particles.  

We consider three physical parameters across our simulations, which are listed in Table \ref{tab:simlist} along with the domain size, the number of grid cells, and particle resolution (see below). The stopping time ($t_s$) measures the degree of coupling between the gas and dust, i.e. the timescale over which the relative velocity between the two decays due to drag. We use the dimensionless stopping time in our simulations:
\begin{equation}\label{eq:taus}
    \tau_s=t_s\Omega.
\end{equation}
We use $\tau_s=0.1$ and compare our results to unstratified AB simulations from \citetalias{Baronett24} that have the same $\tau_s$.

Second, the global pressure gradient is parameterized by 
\begin{equation}\label{eq:Pi}
   \Pi=\frac{\eta u_K}{c_s}=\frac{\eta r}{H}. 
\end{equation}
Here, $c_s$ is isotermal sound speed, $u_K$ is the Keplerian velocity, and $\eta$ is proportional to the radial pressure gradient, quantifying the deviation of the gas rotational velocity from $u_K$ due to pressure gradient support. Furthermore, the characteristic length scale of the SI is approximately $\eta r$ (or, $\Pi H$; \citealt{YG05}). We select four values of $\Pi$—0.01, 0.02, 0.05, and 0.1— to examine the effect of the pressure gradient and to compare our stratified simulations with the unstratified AB models from \citetalias{Baronett24}. We use $\Pi H$ as the unit of length in our simulations throughout this paper.

Third, we use the surface density ratio $(Z)$ in our stratified models to parameterize the abundance of dust relative to the gas, which is defined by 
\begin{equation}\label{eq:Z}
    Z=\frac{\Sigma_{p0}}{\Sigma_{g0}}.
\end{equation}
We use $Z$ and the scale height of dust particles ($H_p$) to find the average midplane dust-to-gas density ratio\footnote{We adopt Equation (\ref{eq:eps}) to compute $\epsilon$ because {\sc Athena} directly outputs $H_p$, making the calculation straightforward. This expression is equivalent to the more direct definition, $\rho_p(z=0)/\rho_g(z=0)$, if the vertical profile of $\rho_p$ is Gaussian.}
\begin{equation}\label{eq:eps}
    \epsilon=\frac{Z}{H_p/H}. 
\end{equation}

We explain our choice of $Z$ values as follows.
To ensure $\epsilon \approx 1$ in our simulations in the nonlinear saturated state---consistent with the total density ratio in the unstratified AB runs of \citetalias{Baronett24}, which represents the midplane---we first set  $\Pi = 0.05$ and determined the $Z$ value that results in $\epsilon \approx 1$. We then selected the remaining $Z$ values such that all simulations maintain the same $Z/\Pi$ ratio. This ratio, along with $\tau_s$, may be a key parameter in determining the nature of SI clumping (see \citealt{sekiya_two_2018}). Consequently, we adopt $Z$ values of 0.0024, 0.0048, 0.012, and 0.024 for $\Pi = 0.01$, 0.02, 0.05, and 0.1, respectively (see Table \ref{tab:simlist}), with $Z/\Pi=0.24$. In these four runs, we set $H_{p0}=\Pi H/2$, and the initial $\epsilon$ (i.e., at $t=0$) is then $2Z/\Pi$. We confirm that our simulations maintain $\epsilon \approx 1$ during the nonlinear saturated state of the SI (see Section \ref{sec:results:saturation}), allowing for a direct comparison of our results to those in \citetalias{Baronett24}.

%%%%%%%%% Figure %%%%%%%%%%%%
\begin{figure*}
    \includegraphics[width=\textwidth]{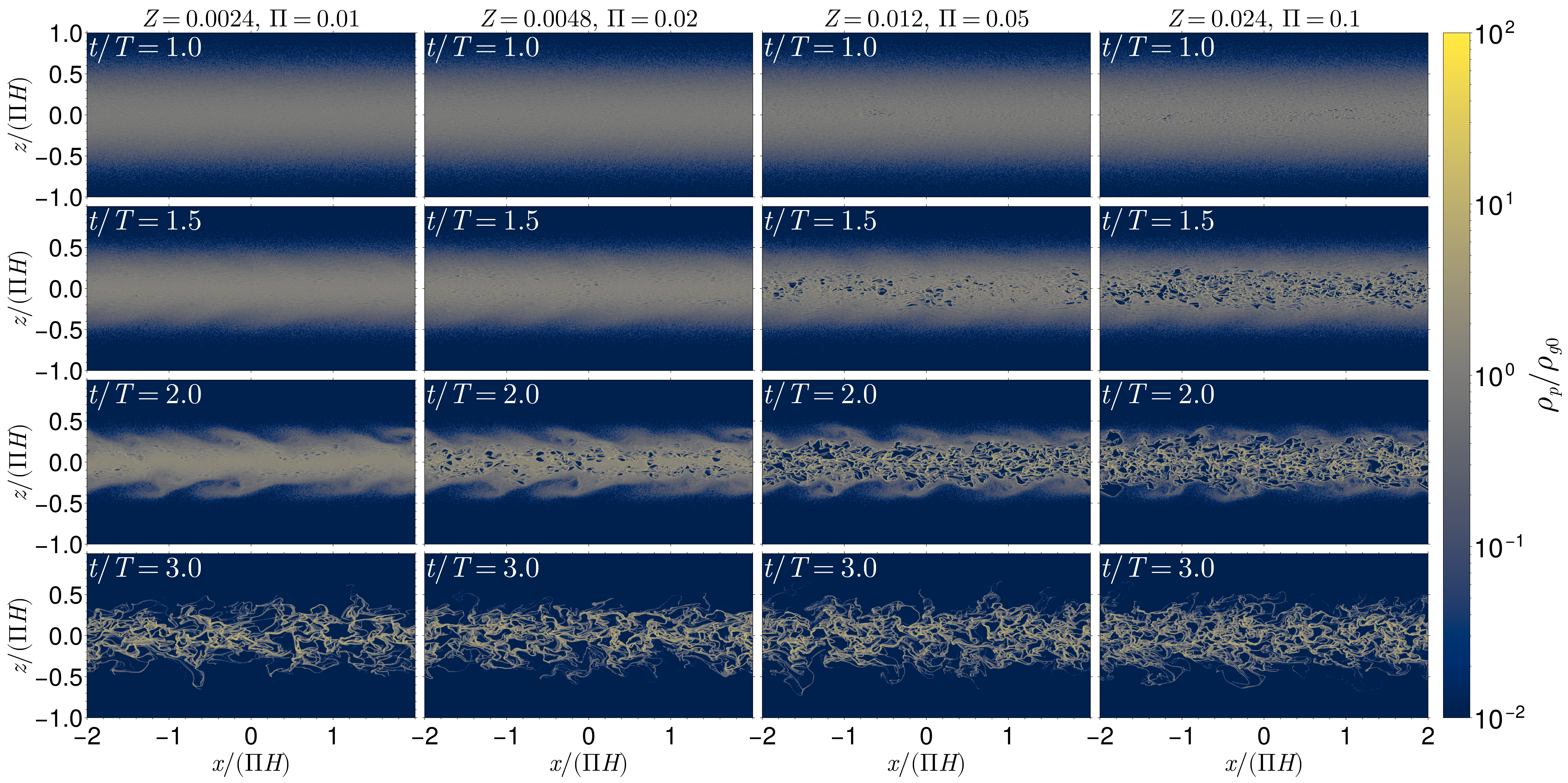}
    \caption{Evolution of dust density $(\rho_p)$ during the sedimentation phase. From left to right, columns correspond to $\Pi=0.01,~0.02,~0.05$, and $0.1$, respectively. Each row corresponds to snapshots at different times: $t/T=1,~1.5,~2.0$, and $3.0$ from top to bottom.  The vertical domain is zoomed in to $-1.0 \leq z/(\Pi H) \leq 1.0$, while the full vertical extent is $8\Pi H$. The simulations presented here have the same value of $Z/\Pi$. Dust voids emerge earlier at higher $\Pi$, appearing at the midplane around $t/T \approx 1$ for $\Pi = 0.1$ (top right) and away from the midplane around $t/T \approx 2$ for $\Pi = 0.01$ (third row, first column). These voids develop into filaments as time goes on.}
    \label{fig:AB_sed_snapshots}
\end{figure*}
%%%%%%%%% Figure %%%%%%%%%%%%

All simulations share the same domain size and grid resolution in terms of $\Pi H$. Specifically, the radial $(L_x)$ and vertical $(L_z)$ extents of the computational domain are $4\Pi H$ and $8\Pi H$, respectively. Every simulation contains $(N_x, N_z) = (1024, 2048)$ grid cells, providing 256 cells per $\Pi H$ ($N_{\Pi H} = \Pi H / \Delta x = 256$, where $\Delta x$ is the grid cell size). Since numerical simulations of the SI have been shown to be sensitive to domain size and grid resolution (\citealt{YangJohansen14, Yang2017, Li18};  \citetalias{Baronett24}; \citealt{Lim24b}), we performed a convergence test of both for $\Pi=0.05$. We found that the dispersions of density and velocities, as well as particle density distributions, are well converged for $L_x\gtrsim 2\Pi H$ and for $N_{\Pi H} \gtrsim 128$ (see Appendix~\ref{sec:appendixA} for more details).  For comparison, \citetalias{Baronett24} used $L_x = L_z = 0.1H$ and $N_x = N_z = 2048$ and did not scale their domain size with $\Pi$, resulting in a range of $1 \leq L_{x,z}/(\Pi H) \leq 10$ and of $205 \lesssim N_{\Pi H} \leq 2048$, depending on the $\Pi$ values (see Table \ref{tab:simlist} for comparison between our simulations and theirs). They performed a resolution study for $\Pi=0.01$ and 0.05 and found that simulations are well converged when $N_{\Pi H} \gtrsim 100$ and $\gtrsim 128$ for $\Pi=0.01$ and 0.05, respectively. Therefore, while there are differences in domain sizes and grid resolutions between their simulations and ours, the convergence tests in this work and \citetalias{Baronett24} suggest that these differences are unlikely to significantly impact the comparison of our results with those of \citetalias{Baronett24}.

We set the average number of dust particles per cell ($n_p$) to 1, and thus the total number of dust particles ($\Npar=n_pN_xN_z$) $\approx 2\times 10^{6}$, whereas $n_p=4$ and $\Npar \approx 1.7\times 10^7$ in AB runs of \citetalias{Baronett24}. Due to the vertical settling of dust that creates dust-free grid cells away from the midplane, $n_p$ is not a good measurement of particle resolution in stratified simulations. Instead, the number of dust particles can be measured relative to $H_p$. Given that $L_z=8\Pi H$ and $H_p \approx 0.24\Pi H$ during the saturated state (see Section \ref{sec:results:saturation}), the average number of dust particles per cell within $H_p$ is $\approx 33$ in our simulations. While we did not perform a convergence test on $n_p$, \citet{BaiStone10a} found in their unstratified AB models that dust density distribution and properties of the turbulence (e.g., gas velocities and diffusion) are only weakly dependent on $n_p$  within the range they considered $( 1 \leq n_p \leq 25 )$. This suggests that the difference in the effective (i.e., number of particles per cell in the region of interest such as the mid-plane in our runs) particle resolution between our work and that of \citetalias{Baronett24} likely has a negligible influence on our comparison.

\section{Results}\label{sec:results}
We present the results of our simulations and compare them to the unstratified AB models of \citetalias{Baronett24} in this section. We begin by examining the dust sedimentation phase in Section \ref{sec:results:sedimentation}. Section \ref{sec:results:saturation} describes how we determine the saturated state of the SI in our simulations. In Section~\ref{sec:results:actual_fields}, we present vertical profiles of density and velocity for both the gas and dust. Section \ref{sec:results:COM} focuses on the radial and azimuthal velocities of the center of mass (COM) and on the gas and dust velocities relative to the local COM frame. Section \ref{sec:results:dispersion} focuses on SI-induced turbulence by examining density and velocity dispersions near the midplane. Section \ref{sec:results:densityfield} focuses on density fields of the gas and dust particles at the midplane. Finally, Section \ref{sec:results:diffusion} presents the radial and vertical diffusion of dust particles.

%%%%%%%%% Figure %%%%%%%%%%%%
\begin{figure*}
    \includegraphics[width=\textwidth]{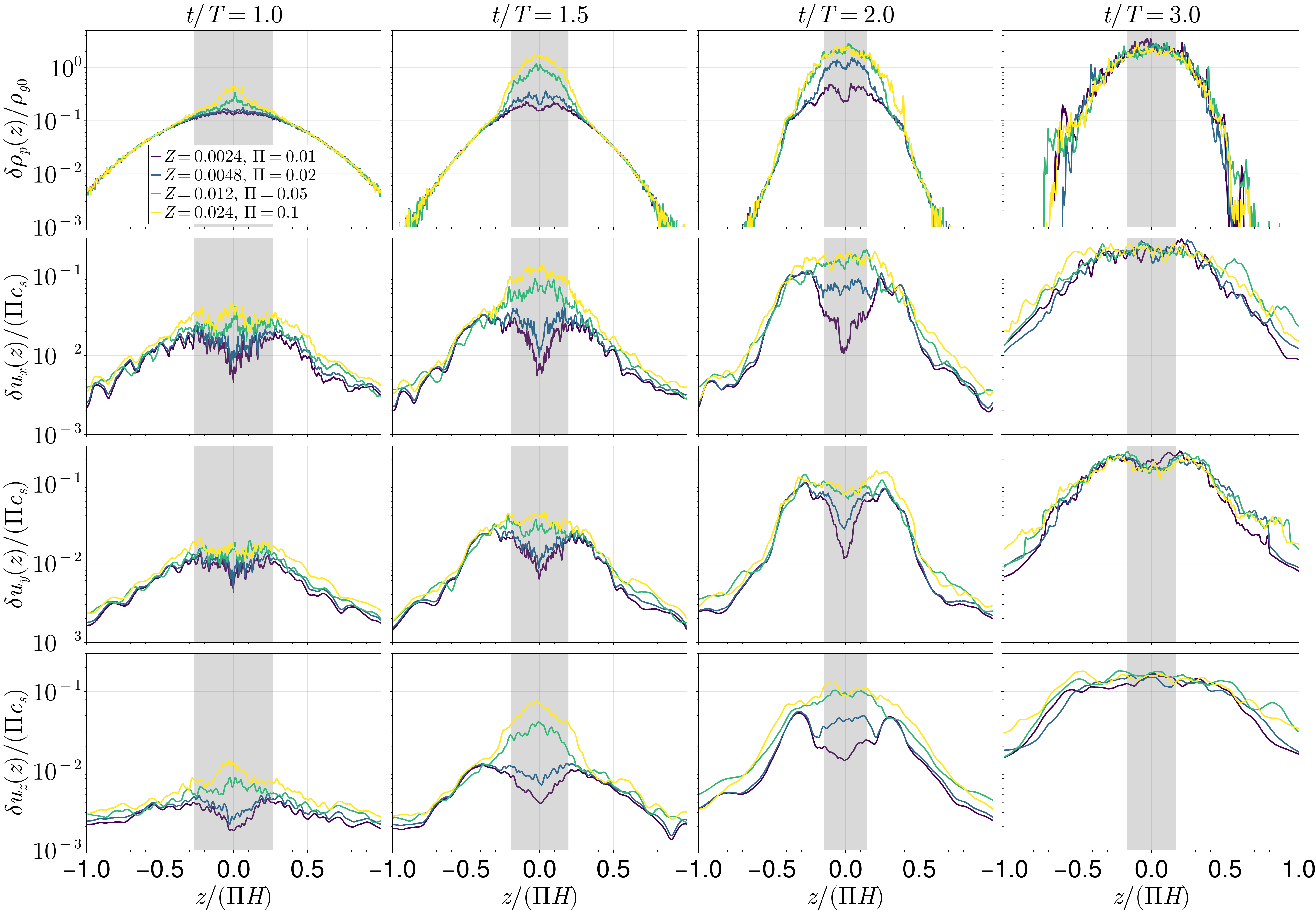}
    \caption{Dispersions of dust density normalized by the midplane gas density (top row, Equation \ref{eq:deltarhop}) and gas velocities  (normalized by $\Pi c_s$) (bottom three rows, Equation \ref{eq:deltau}) as a function of $z$ for the snapshots shown in Figure \ref{fig:AB_sed_snapshots}. Radial, azimuthal, and vertical components of the gas velocity dispersions are shown in order from the second to fourth rows. In order from left to right, panels correspond to $t/T = 1,~1.5,~2.0$, and 3.0. In each panel, darker to brighter colors represent lower to higher $\Pi$ values: 0.01, 0.02, 0.05, and 0.1. As all runs have nearly identical $H_p/(\Pi H)$ at a given time, we use the shaded region to mark $\pm H_p$ at a given time. While the density and the velocity dispersions for $\Pi$ = 0.05 and 0.1 peak near the midplane as early as $t/T = 1$, those for $\Pi$ = 0.01 and 0.02 peak near $\pm H_p$ and only become prominent by $t/T = 1.5$. After sedimentation balances turbulent diffusion ($t/T=3.0$), the density and velocity profiles look similar across different $\Pi$ values.
    }
    \label{fig:AB_sed_dispersions}
\end{figure*}
%%%%%%%%% Figure %%%%%%%%%%%%
\subsection{Sedimentation Phase}\label{sec:results:sedimentation}
Unlike in unstratified cases, dust in stratified simulations undergoes sedimentation due to vertical gravity until turbulence-driven vertical diffusion counterbalances the gravity. To illustrate how $\Pi$ influences dust sedimentation, we present snapshots of the dust density $(\rho_p)$ in Figure \ref{fig:AB_sed_snapshots}. Each column corresponds to a different $\Pi$ value: from left to right, $\Pi = 0.01, ~0.02, ~0.05$, and 0.1. From top to bottom, snapshots are taken at $t/T = 1, ~1.5, ~2.0$, and 3.0, respectively, where $T$ is the local orbital period $(2\pi \Omega^{-1})$.  We note that the $e$-folding timescale of the sedimentation is $\sim (2\pi \tau_s)^{-1}T \approx 1.6T$ for our $\tau_s$ value \citep{youdin_particle_2007}. 

Dust voids form near the midplane and develop into filamentary structures during the sedimentation period across all $\Pi$ values, resembling the evolution of dust filaments observed in unstratified simulations (\citealt{JY07}; \citetalias{Baronett24}). Although the dust filament formation occurs for all $\Pi$, the timing of their emergence varies slightly. Specifically, the voids appear as early as $t/T \approx 1.0$ for $\Pi$ = 0.05 and 0.1, whereas they emerge later at $t/T \approx 1.5$ and $\approx 2.0$ for $\Pi$ = 0.02 and 0.01, respectively. However, we caution that the onset timing of the voids may be influenced by numerics. Lower $\Pi$ values correspond to weaker initial velocity perturbations in our simulations (see Section \ref{sec:results:dispersion:temporal_evolution}), requiring slightly more time to reach the nonlinear amplitude necessary to trigger the formation of the voids compared to higher $\Pi$ runs. 

More interestingly, the location where dust voids first form appears to be influenced by $\Pi$. For $\Pi$ = 0.05 and 0.1, the voids emerge near the midplane (the two rightmost panels in the first row of Figure \ref{fig:AB_sed_snapshots}). By contrast, for $\Pi$ = 0.01 (third row, first column) and $\Pi$ = 0.02 (second row, second column), they appear slightly off the midplane. To illustrate this more clearly, we compute dispersions of dust density and gas velocities as a function of $z$. The vertical profile of the density dispersion is calculated by 
\begin{equation}\label{eq:deltarhop}
    \delta \rho_p(z) = \sqrt{\langle \rho_p^2 \rangle_x - \langle \rho_p \rangle_x^2},
\end{equation}
where $\langle \cdots \rangle_x$ denotes the spatial average along the $x$ direction. The velocity dispersions are calculated in a similar way but are weighted by gas density:
\begin{equation}\label{eq:deltau}
    \delta u_{x,y,z}(z) = \sqrt{\frac{\langle \rho_g u_{x,y,z}^2 \rangle_x}{\langle \rho_g \rangle_x} - \left(\frac{\langle \rho_g u_{x,y,z} \rangle_x}{\langle \rho_g \rangle_x}\right)^2}.
\end{equation}
Here, $u_{x,y,z}$ denotes the radial, azimuthal, and vertical velocities of the gas, respectively. Note that the azimuthal velocity (both of the gas and dust) is Keplerian-subtracted.

In Figure \ref{fig:AB_sed_dispersions}, we present the density and velocity  dispersions for the snapshots shown in Figure \ref{fig:AB_sed_snapshots}. Top panels show the density dispersion, while the bottom three panels present radial, azimuthal, and vertical velocity dispersions. From left to right, each panel corresponds to $t/T = 1,~1.5,~2.0,$ and $3.0$. In each panel, darker to brighter colors represent $\Pi = 0.01,~0.02,~0.05,$ and $0.1$, respectively. At a given time, each simulation has an approximately constant $H_p/(\Pi H)$ for the different $\Pi$ values. Thus, the shaded regions show the region $|z| < H_p$ representative of all $\Pi$ values at a given time. 

As shown in the top panels of Figure \ref{fig:AB_sed_dispersions}, the density dispersions for $\Pi = 0.05$ and $0.1$ peak at $z \approx 0$ as early as $t/T=1$. By contrast, for $\Pi = 0.01$ and $0.02$, the peak of the density dispersions emerges slightly later ($t/T \approx 1.5$ to $\approx 2.0$, respectively) and is located slightly away from $z=0$, resulting in a bimodal shape. By $t/T \approx 3.0$, the dispersion profiles become similar across different $\Pi$ values.  

The dispersion of each velocity component is closely correlated with that of the dust density. For $\Pi = 0.05$ and $0.1$, the velocity dispersions peak at the midplane between $t/T = 1.0$ and 1.5, after which they become relatively flat within $\pm H_p$. By contrast, for $\Pi = 0.01$ and $0.02$, the dispersions exhibit a bimodal shape, with peaks occurring beyond $\pm H_p$ until $t/T =3$. The smallest $\Pi$ case has the most pronounced peaks, with $\delta u_{x,y}$ being nearly an order of magnitude larger at around $\pm 2H_p$ than at the midplane at $t/T=2$.  

After dust filaments fully develop across $\Pi$ values (at  $t/T = 3$), the velocity dispersion profiles become nearly identical across $\Pi$ values. However, while the $\delta u_{x,z}$ profiles remain almost flat within the dust layer (within the shaded region), the $\delta u_y$ profile retains a slightly bimodal shape regardless of $\Pi$, with peaks located at $\pm H_p$. We come back to this point in Section \ref{sec:results:dispersion}.

%%%%%%%%% Figure %%%%%%%%%%%%
\begin{figure}
    \centering
    \includegraphics[width=\columnwidth]{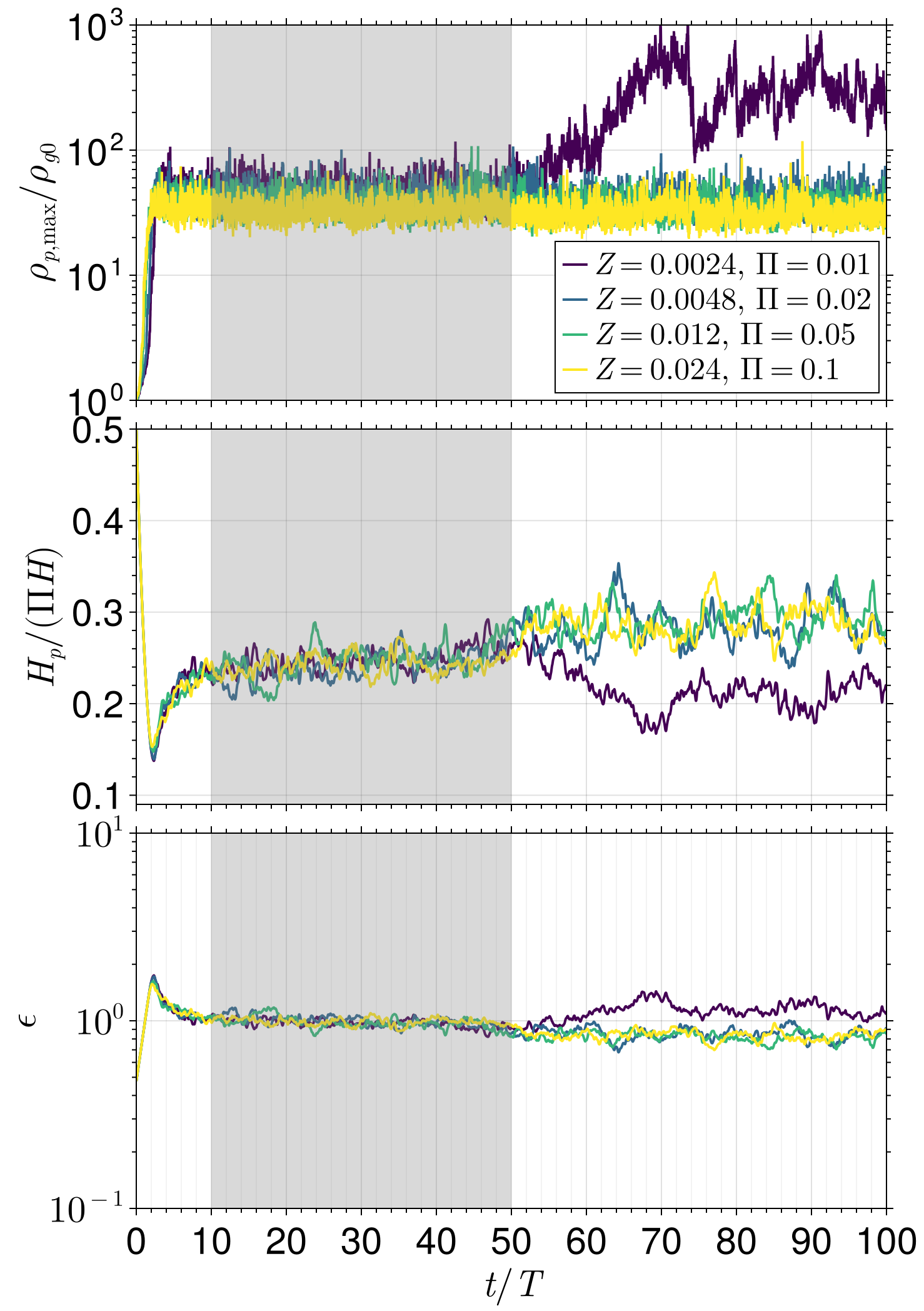}
    \caption{Time evolution of the maximum dust density normalized to the initial midplane gas density (top), dust scale~height normalized by $\Pi H$ (middle), and midplane dust-to-gas density ratio (bottom) in simulations with $\tau_s=0.1$ and $Z/\Pi=0.24$. Given that the maximum density of $\Pi=0.01$ case increases rapidly after $t/T \approx 50$, we restrict our analysis of the saturated state to the range $t/T=10$ to 50, marked by the gray shading. During the state, the time evolution remains similar across these runs, with $\epsilon \approx 1$, consistent with the unstratified AB models.}
    \label{fig:AB_DmaxHpEps}
\end{figure}
%%%%%%%%% Figure %%%%%%%%%%%%

\subsection{Saturated State of the SI}\label{sec:results:saturation}
After dust settles toward the midplane, dust--gas interactions drive turbulence. The turbulence then vertically diffuses dust away from the midplane balanced by sedimentation, which leads to a saturated state. We determine the saturated state of the SI using the diagnostics presented in Figure \ref{fig:AB_DmaxHpEps}. The figure shows the maximum density of dust particles ($\rho_{p,\textrm{max}}$), $H_p$ (in units of $\Pi H$), and $\epsilon$ as functions of time for simulations with different $\Pi$. Time is measured in units of $T$. 

In the middle panel, dust continues settling toward the midplane during the first $3T$ from the initial scale height of $0.5\Pi H$ down to $\approx 0.14\Pi H$. During the sedimentation phase, the maximum density (top panel) and the midplane density ratio (bottom panel) both increase. Afterward, $H_p$ increases until $t/T \approx 10$, after which it levels off at $\approx 0.24\Pi H$ and $\epsilon \approx 1$ is established.

%%%%%%%%% Figure %%%%%%%%%%%%
\begin{figure*}[ht!]
    \includegraphics[width=\textwidth]{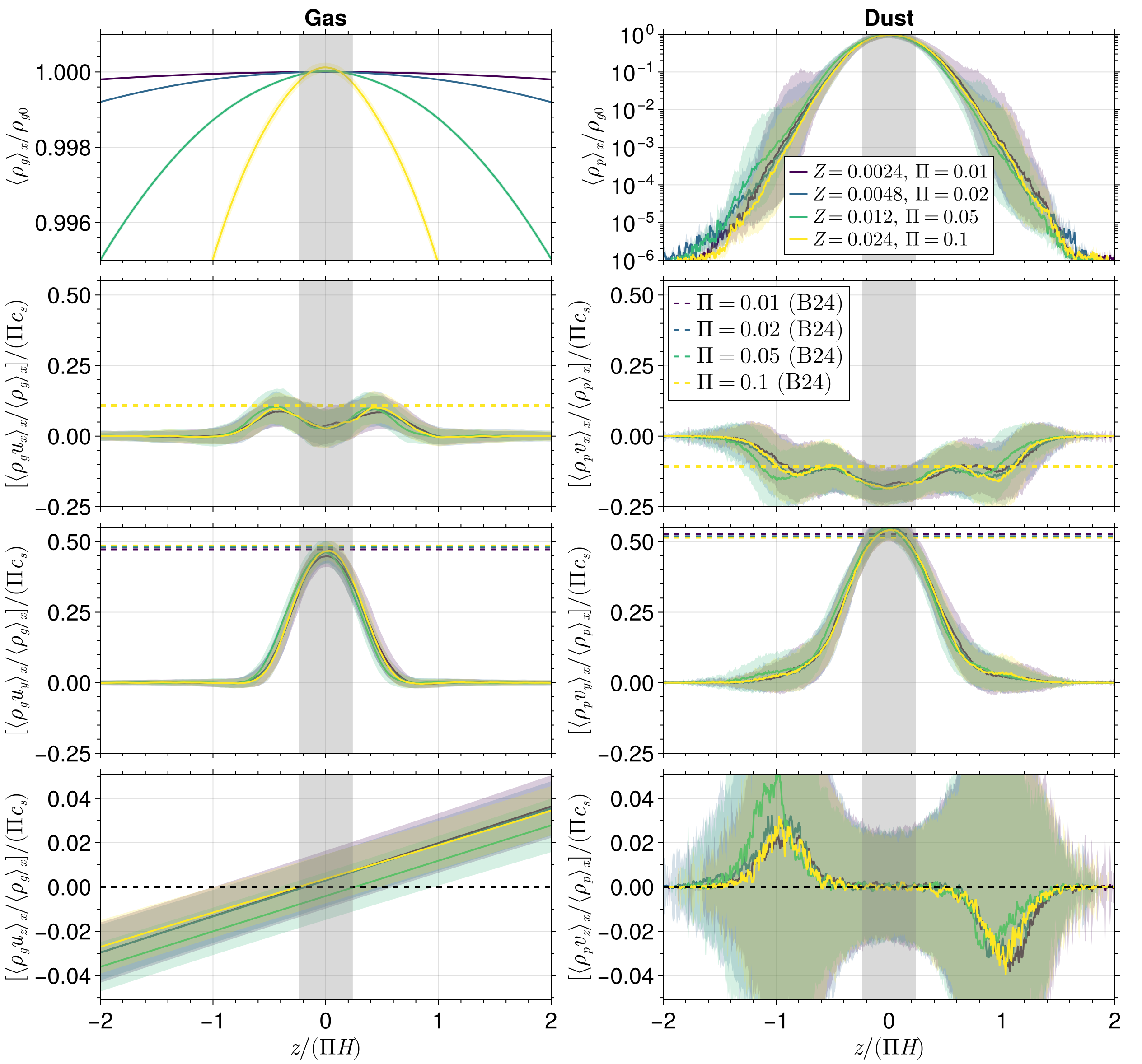}
    \caption{Vertical profiles of density and velocity fields for gas (left) and dust (right) in the saturated state. Velocities are weighted by density and normalized by $\Pi c_s$. Shaded regions around the curves indicate $1\sigma$ temporal variability, and the gray bands mark $\pm H_p$. In the middle two rows, horizontal dashed lines represent spatial and temporal averages from the AB models of \citetalias{Baronett24}. The black horizontal dashed line in the bottom panels denotes zero vertical velocity, which holds for unstratified simulations regardless of $\Pi$. Gas and dust densities follow Gaussian profiles. Radial velocities for both components peak away from the midplane and deviate from those in \citetalias{Baronett24} at $z = 0$. Azimuthal velocity profiles resemble the dust density distribution, with midplane values consistent with \citetalias{Baronett24} within the temporal variability. The gas vertical velocity increases approximately linearly with height due to the outflow boundary condition, while the dust exhibits non-zero vertical motion beyond $2H_p$, but returning back to zero as $\rho_p \rightarrow 0$. However, we caution that altitudes where the magnitude of $v_z$ begins to increase from zero are represented by only a tiny fraction of particles, and those at such heights may be sedimenting back toward the midplane without being balanced by diffusion.
    }
    \label{fig:AB_sat_vertical_profiles_actual_fields}
\end{figure*}
%%%%%%%%% Figure %%%%%%%%%%%%

Between $t/T=10$ and 50 (marked by the gray shaded regions), all runs exhibit nearly identical evolution, consistent with \citet{sekiya_two_2018}, who showed that $Z/\Pi$ controls the dust dynamics driven by the SI in stratified disks, even when values of $Z$ and $\Pi$ are different (but see the following). Yet, after $t/T\approx 50$, a bifurcation is seen, where the maximum density of the run with $\Pi=0.01$ (purple) increases sharply and reaches up to $\sim 10^3 \rho_{g0}$. The run achieves a second saturated state after $t/T \approx 80$ where the maximum density and $H_p$ level off at $\approx 300\rho_{g0}$ and $\approx 0.2 \Pi H$, respectively. We observed that a dense clump forms at the midplane after $t/T\approx 50$ and persists to the end of the simulation. For the other three simulations, $H_p$ slightly increases to $\approx 0.3 \Pi H$ and levels off there. We discuss the implication of this bifurcation on the role of $Z/\Pi$ parameter in Section \ref{sec:Discussion:caveat}. Overall, based on these observations, we define the saturated state as occurring between $t/T = 10$ and $50$ and perform time-averages of all reported quantities over the period. As we compare our simulation results during the saturated state to those from \citetalias{Baronett24}, we note that in their AB models, the saturated state ranges from $5T$ to $10T$  (see their Section 3.1 and Figure 1). 

\subsection{Vertical Profiles of Density and Velocity of the Gas and Dust}\label{sec:results:actual_fields}
In this subsection, we examine the vertical profiles of density and velocity for both gas and dust. We also compare their midplane values to those from unstratified simulations in \citetalias{Baronett24}.

Figure \ref{fig:AB_sat_vertical_profiles_actual_fields} shows the time-averaged vertical profiles of density and density-weighted velocity of the gas (left) and dust (right). Shaded regions denote the $1\sigma$ temporal variability for each quantity at each $\Pi$ value, while the gray band indicates the region within $\pm H_p$. 

The gas and dust density fields exhibit Gaussian profiles. The gas density remains nearly constant throughout the vertical domain because $L_z/2 = 4\Pi H < H$, even for the largest $\Pi$ value considered. The midplane dust-to-gas density ratio is approximately unity, consistent with the total dust-to-gas ratio in the unstratified AB models of \citetalias{Baronett24}.

The radial velocity profiles of both gas and dust (second row in Figure~\ref{fig:AB_sat_vertical_profiles_actual_fields}) exhibit a bimodal structure, with peaks located at $z \approx \pm 0.5\Pi H$ (see also Figure 4 of \citealt{Sengupta_Umurhan23}). Near the midplane, the gas radial velocity approaches zero, whereas the dust radial velocity becomes increasingly negative. Compared to the temporal and spatial averages from \citetalias{Baronett24} (dashed lines), our simulations show radial velocities that differ by approximately 60–70\%. However, as we will show in Section~\ref{sec:results:COM}, the midplane radial velocity in the COM frame is nearly indistinguishable between unstratified and stratified simulations.

The azimuthal velocities (third row in Figure~\ref{fig:AB_sat_vertical_profiles_actual_fields}) show simpler vertical dependence, which resemble the dust density profile. Compared to the average values of azimuthal velocity from \citetalias{Baronett24}\footnote{We add $\Pi c_s$ to their azimuthal velocities to account for the opposite, but consistent, treatment of the radial pressure gradient: while we apply an inward acceleration to particles, they apply an outward acceleration to the gas. As a result, in our setup (without dust feedback), particles rotate at super-Keplerian speed $(v_y > 0)$ with Keplerian gas $(u_y = 0)$, whereas in theirs, particles are Keplerian $(v_y = 0)$ and the gas is sub-Keplerian $(u_y < 0)$.},  we find a close match between the midplane values in our stratified simulations and those in their unstratified simulations.

The bottom panels show the vertical velocities of gas and dust. For the gas, the vertical velocity increases approximately linearly with height, with slight positive (negative) values for $z>0$ ($z<0$) arising from the outflow boundary conditions \citep{Li18}. Notably, the velocity crosses zero slightly away from the midplane rather than at $z=0$. We find that this offset stems from our radial domain size $(L_x)$ that is not large enough to capture sufficient number of corrugations of the dust layer. Indeed, doubling $L_x$ results in the gas velocity being zero at the midplane (see Appendix~\ref{sec:appendixA}). For the dust, the vertical velocity remains near zero within $\approx 2H_p$, becoming positive below the midplane and negative above it beyond $2H_p$, before returning to zero as $\rho_p \rightarrow 0$ with increasing $|z|$ (cf. top right panel of Figure~\ref{fig:AB_sat_vertical_profiles_actual_fields}). However, we caution that an insufficient number of particles exist at altitudes where the magnitude of $v_z$ begins to increase from zero, and those at such heights may be sedimenting back toward the midplane without being balanced by diffusion.

\subsection{COM Velocities and the Role of Vertical Gravity}\label{sec:results:COM}
In addition to examining gas and dust velocities individually, we calculate the radial and azimuthal velocities of the COM and compare these velocities to those from unstratified simulations. The vertical profiles of the COM velocities are calculated by
\begin{equation}\label{eq:U_com}
    U_{x,\rm{CM}}(z)=\frac{\langle \rho_g u_x \rangle_x + \langle \rho_p v_x \rangle_x}{\langle \rho_{g} \rangle_x+\langle \rho_p \rangle_x},
\end{equation}
and 
\begin{equation}\label{eq:V_com}
    U_{y,\rm{CM}}(z)=\frac{\langle \rho_g u_y \rangle_x + \langle \rho_p v_y \rangle_x}{\langle \rho_{g} \rangle_x+\langle \rho_p \rangle_x},
\end{equation}
where we weight $u_x$ and $u_y$ by $\rho_g$ and $v_x$ and $v_y$ by $\rho_p$. Note that according to the steady-state equilibrium solution without vertical stratification, the COM radial velocity is zero \citep{Nakagawa1986,YG05}. 

\begin{figure}
    \includegraphics[width=\columnwidth]{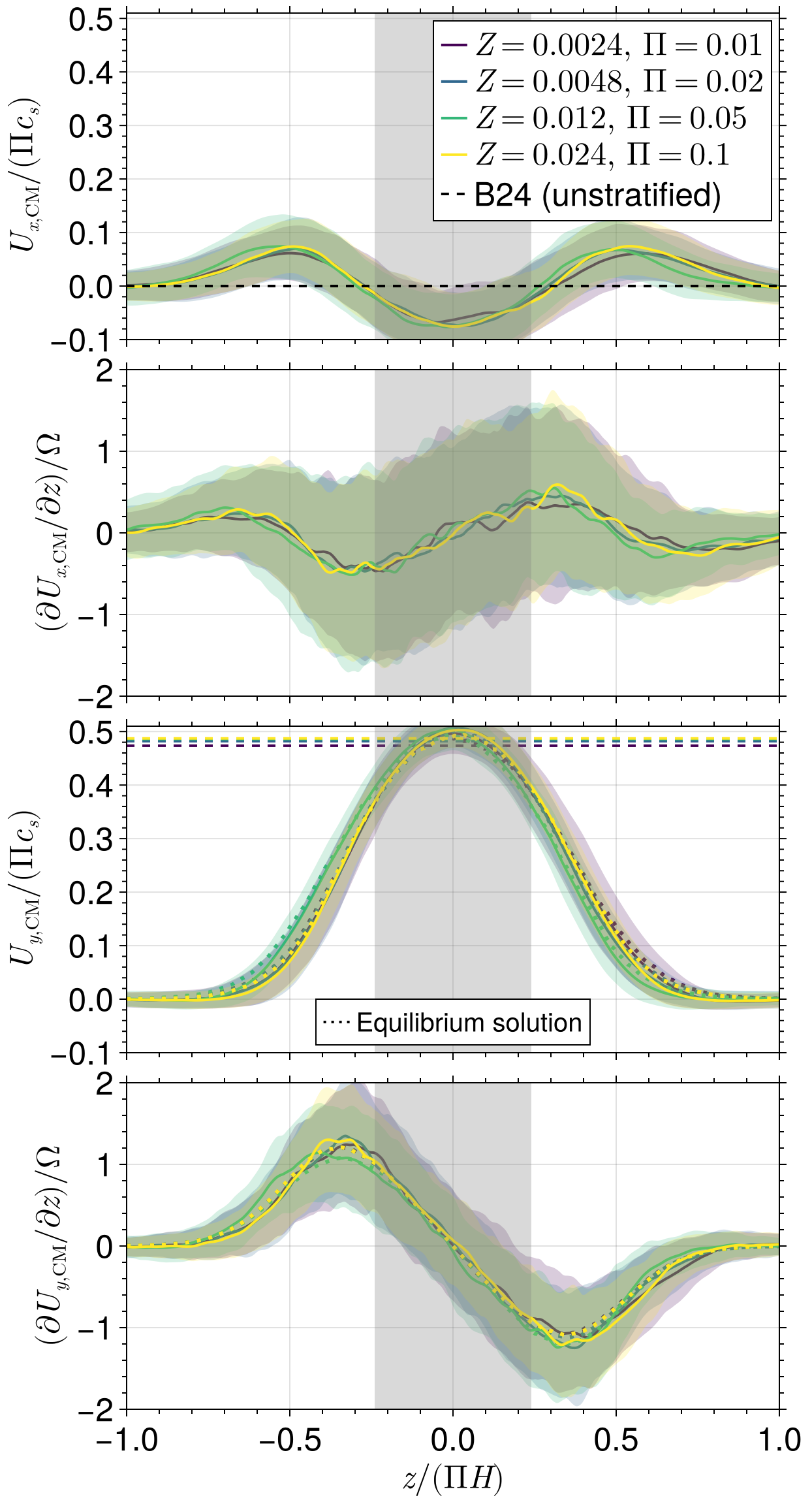}
    \caption{Time-averaged COM velocities (Equations~\ref{eq:U_com}-\ref{eq:V_com}) and their vertical gradients as functions of $z$. The top two panels show the radial component, while the bottom two panels show the azimuthal component. Shaded regions represent $1\sigma$ temporal variability, and the gray bands indicate $\pm H_p$. In the first and third panels, horizontal lines mark the COM velocities from \citetalias{Baronett24}, where the radial velocity is zero regardless of $\Pi$. The dotted curves in the third and bottom panels show the vertical profiles of the corresponding quantities for each $\Pi$ as predicted by the NSH equilibrium (Equations.~\ref{eq:Uycm_NSH}–\ref{eq:dUycm_dz_NSH}); in the bottom panel, the profiles overlap.  In our simulations, the radial velocity fluctuates around zero and deviates from the midplane value in \citetalias{Baronett24}, showing that vertical gravity redistributes momentum vertically. In contrast, the azimuthal velocity at the midplane closely matches that from \citetalias{Baronett24}. The vertical gradients of both velocity components peak near $z \approx \pm H_p$, with the azimuthal component exhibiting a larger amplitude than the radial component.}
    
    \label{fig:AB_sat_vertical_gradients}
\end{figure}

Figure~\ref{fig:AB_sat_vertical_gradients} shows $U_{x,\rm{CM}}(z)$ and its vertical gradient, $\partial U_{x,\rm{CM}}/\partial z$, in the top two panels, and the corresponding azimuthal component in the bottom two panels. The horizontal dashed lines in the top and third panels indicate the corresponding COM velocities from the unstratified simulations of \citetalias{Baronett24}. In the top panel, a single line is used because the COM radial velocity in their simulations is virtually zero for all $\Pi$, consistent with the equilibrium solution.

First, we focus on the radial and azimuthal velocity profiles. As can be seen from the top panel, $U_{x,\rm{CM}}(z)$ exhibits a bimodal shape, resembling the radial velocity profile of the gas (second row, left column in Figure~\ref{fig:AB_sat_vertical_profiles_actual_fields}). It varies around zero expected from a horizontal equilibrium. This vertical variation of $U_{x,\rm{CM}}(z)$, although not large enough to depart substantially from zero, indicates that vertical gravity acts to redistribute momentum in the vertical direction. 

The azimuthal component (third panel) behaves similarly to the corresponding gas or dust velocity profiles (third row in Figure~\ref{fig:AB_sat_vertical_profiles_actual_fields}), with its midplane value agreeing with that from \citetalias{Baronett24} within the $1\sigma$ temporal variability. This profile shape is consistent with the NSH equilibrium solution (Eq.~8 in \citealt{YJ07}):
\begin{equation}\label{eq:Uycm_NSH}
U_{y,\textrm{CM}}^{\textrm{NSH}} = -\frac{\Pi c_s}{1+\rho_p(z)/\rho_{g0}},
\end{equation}
where we assume a vertically constant gas density, $\rho_g(z) \approx \rho_{g0}$ (see the top left panel of Figure~\ref{fig:AB_sat_vertical_profiles_actual_fields}). We calculate the theoretical profile using the dust density distribution $\rho_p(z)$ (top right in Figure~\ref{fig:AB_sat_vertical_profiles_actual_fields}) and add $\Pi c_s$ to account for the  inward acceleration applied to the dust in our simulations. The resulting theoretical profiles, shown as dotted curves in the third panel, agree closely with the profiles measured in our simulations.

%%%%%%%%% Figure %%%%%%%%%%%%
%%%%%%%%% Figure %%%%%%%%%%%%

Second, the vertical gradients of both the radial (second panel) and azimuthal (bottom panel) COM velocities peak away from the midplane and diminish toward it, with the azimuthal velocity gradient exhibiting the larger amplitude. Such non-zero vertical gradients are commonly regarded as a source for various shear-driven instabilities within the dust layer. We return to this point in Section~\ref{sec:Discussion:turb}, where we provide a more detailed discussion of shear-driven instabilities. In addition, we compare $\partial U_{y,\textrm{CM}}/\partial z$ measured in our simulations to its theoretical profile, which is obtained from Equation~\ref{eq:Uycm_NSH}:
\begin{equation}\label{eq:dUycm_dz_NSH}
     \frac{\partial U_{y,\textrm{CM}}^{\textrm{NSH}}}{\partial z}\bigg/\Omega=\frac{-2\epsilon\xi}{[z/(\Pi H)]}\left[\frac{e^{-\xi/2}}{1+\epsilon e^{-\xi}}\right]^2
\end{equation}
where $\xi \equiv z^2/(2H_p^2)$ and we assume $\rho_p(z)$ follows a Gaussian profile: $\rho_p(z)=\rho_p(z=0)e^{-\xi}$. Using time-averaged values of $\epsilon$ and $H_p$ in the saturated state, we plot the theoretical gradient as dotted curves in the bottom panel. These curves show close agreement with the profiles obtained from our simulations.

\begin{figure}
    \includegraphics[width=\columnwidth]{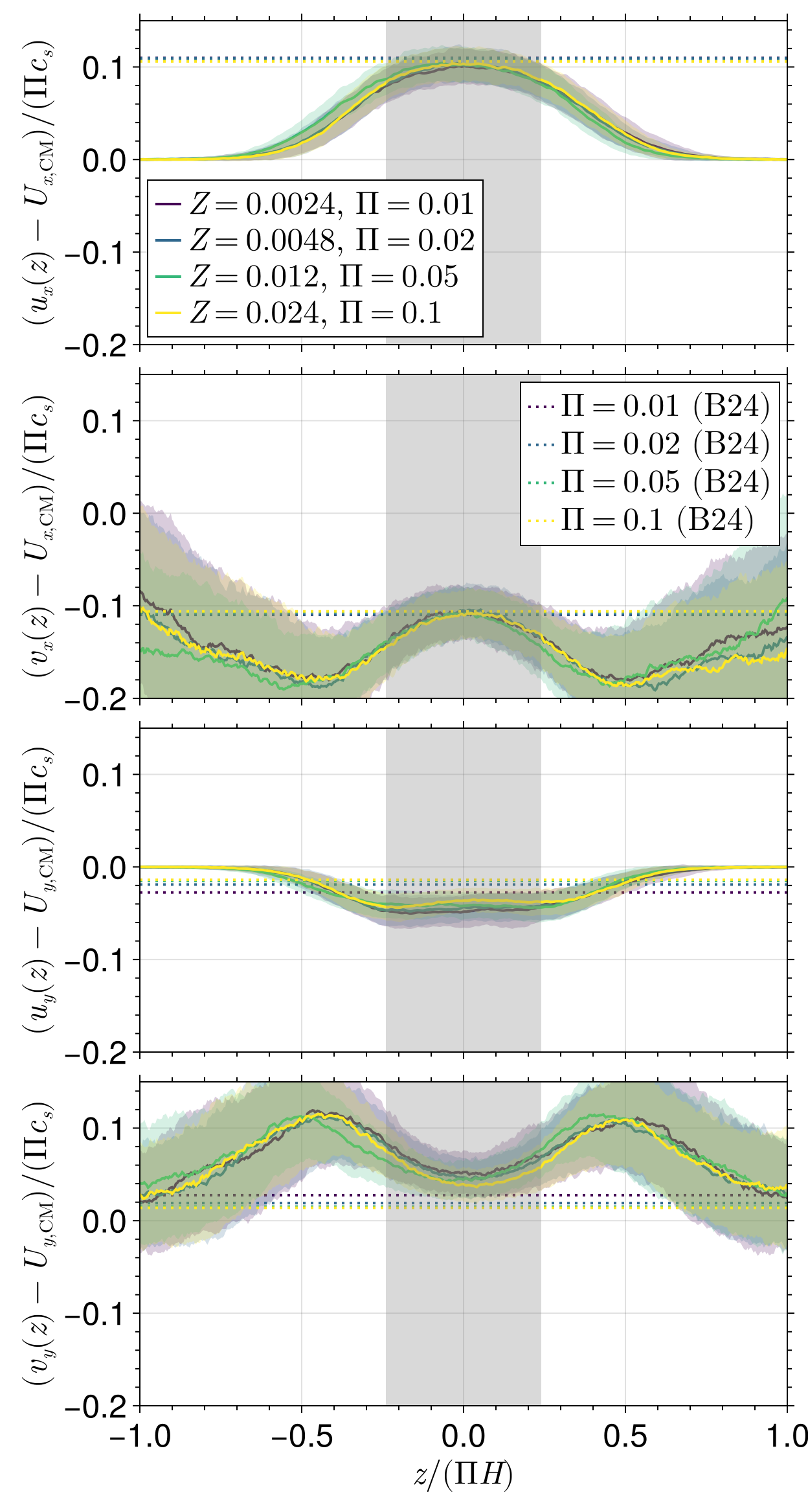}
    \caption{Similar to Figure~\ref{fig:AB_sat_vertical_profiles_actual_fields}, but showing velocities in the COM frame for the gas and dust. The top two panels display the radial velocities of gas and dust, respectively; the bottom two panels show the azimuthal velocities. Horizontal dotted lines mark the corresponding values from \citetalias{Baronett24}. Note that $u_x,~u_y,~v_x$ and $v_y$ are density-weighted. At the midplane, the azimuthal velocities differ by $\sim 2.5\%$ of $\Pi c_s$ and about $1\sigma$ temporal variability, whereas the radial velocities agree much more closely, showing that stratified and unstratified simulations remain consistent in the COM frame.
    }
    \label{fig:AB_sat_velocities_COMframe}
\end{figure}

Since the COM exhibits non-zero radial motion in the presence of vertical stratification, we calculate the radial and azimuthal velocities of the gas and dust in the COM frame and compare them to the corresponding values from the unstratified simulations of \citetalias{Baronett24}. Figure~\ref{fig:AB_sat_velocities_COMframe} shows the radial velocities of gas and dust in the COM frame in the top two panels, respectively, and the corresponding azimuthal velocities in the bottom two panels. All velocities are density-weighted. In each panel, the horizontal dotted lines mark the corresponding values from \citetalias{Baronett24}.

%%%%%%%%% Figure %%%%%%%%%%%%
%%%%%%%%% Figure %%%%%%%%%%%%

At the midplane, the radial velocities of both gas and dust in the COM frame agree much more closely with those from unstratified simulations, in contrast to the larger discrepancies seen in Figure~\ref{fig:AB_sat_vertical_profiles_actual_fields}. For the azimuthal velocities of gas and dust, the difference between stratified and unstratified simulations in the COM frame is about $2-3\%$ of $\Pi c_s$, similar to that shown in Figure~\ref{fig:AB_sat_vertical_profiles_actual_fields}.

In summary, vertical gravity acts to redistribute momentum—particularly the radial component—along the vertical direction, shaping the mean flow $(U_{x,\rm{CM}}, U_{y,\rm{CM}})$ at each $z$ (Figure~\ref{fig:AB_sat_vertical_gradients}). This redistribution likely arises from vertical momentum flux into and out of the dust layer driven by vertical gravity—an effect absent in unstratified models. However, unstratified and stratified models remain consistent in the COM frame near the midplane. In what follows, we turn to the dispersion quantities of the gas and dust to investigate streaming turbulence in our stratified simulations.

%%%%%%%%% Figure %%%%%%%%%%%%
\begin{figure*}
    \includegraphics[width=\textwidth]{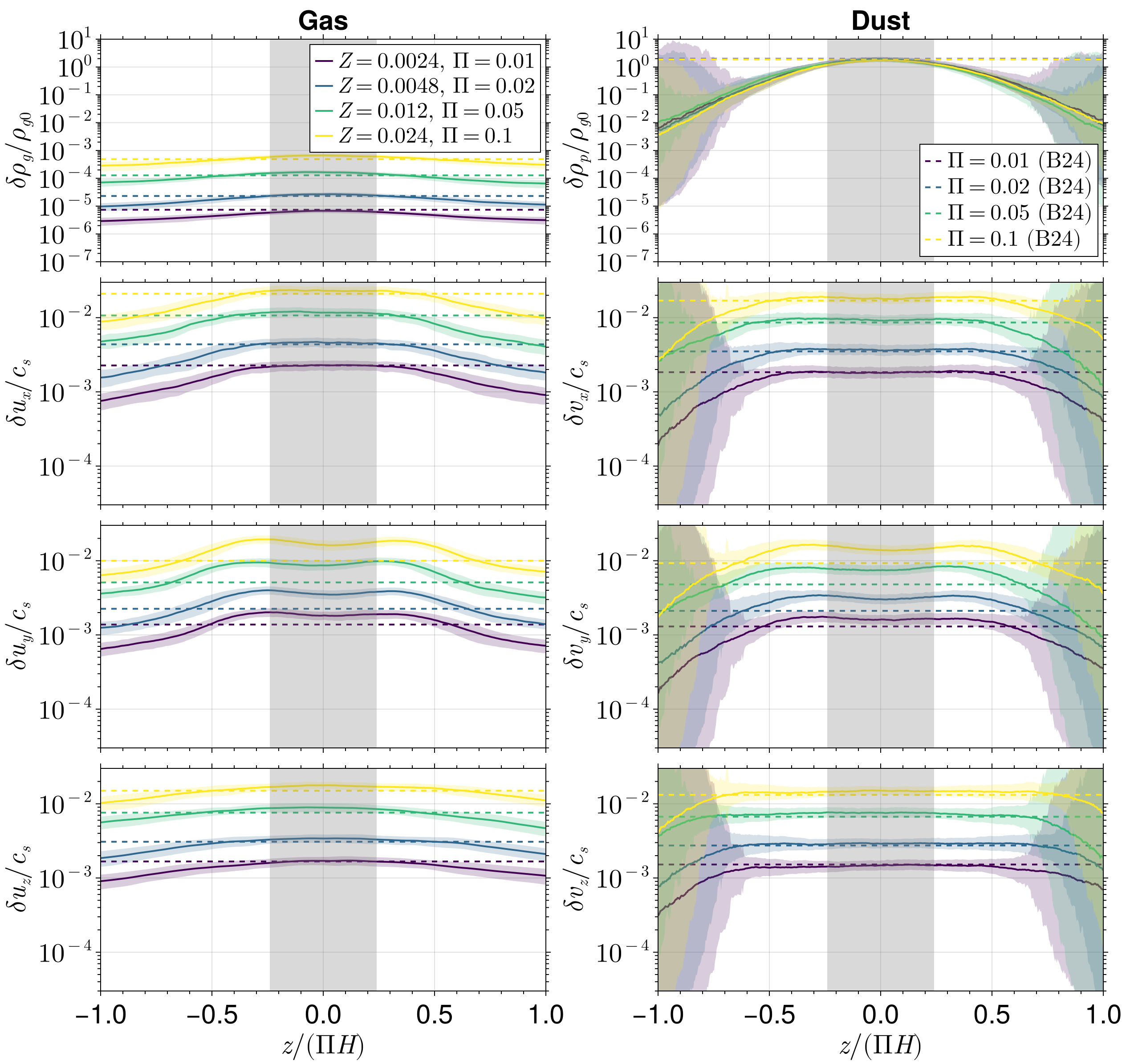}
    \caption{Similar to Figure~\ref{fig:AB_sat_vertical_profiles_actual_fields} but for the dispersions (Equations~\ref{eq:deltarhop},~\ref{eq:deltau},~\ref{eq:deltav}). Note that unlike Figure~\ref{fig:AB_sat_vertical_profiles_actual_fields}, we show the dispersion profile within $\pm \Pi H$, and velocities are in units of $c_s$. Near the midplane, the dispersions measured in the unstratified and stratified cases are nearly identical.
    }  
    \label{fig:AB_sat_vertical_profiles}
\end{figure*}
%%%%%%%%% Figure %%%%%%%%%%%%

\subsection{Dispersion Properties of the Gas and Dust}\label{sec:results:dispersion}
In this subsection, we report the dispersions of densities and velocities of the gas and dust particles for the saturated state. We measure the gas density dispersion ($\delta \rho_g$) using an equation equivalent to Equation (\ref{eq:deltarhop}) but with gas density replacing dust density. For the velocity dispersion of dust particles, we use the weighted radial average similar to Equation (\ref{eq:deltau}):
\begin{equation}\label{eq:deltav}
    \delta v_{x,y,z}(z) \equiv \sqrt{\frac{\langle \rho_p v_{x,y,z}^2 \rangle_x}{\langle \rho_p \rangle_x} - \left(\frac{\langle \rho_p v_{x,y,z} \rangle_x}{\langle \rho_p \rangle_x}\right)^2}.
\end{equation}
We first present the vertical profiles of the dispersions in the following section.

\subsubsection{Vertical profiles}\label{sec:results:dispersion:vertical_profiles}
Figure \ref{fig:AB_sat_vertical_profiles} shows the time-averaged vertical profiles of gas (left) and dust (right) dispersions. The panels from top to bottom present the density, radial, azimuthal, and vertical velocity dispersions, respectively. Each color corresponds to a simulation with different $Z$ and $\Pi$ values. The shaded region in each panel denotes $-H_p\leq z \leq H_p$, with $H_p=0.24\Pi H$ being time-averaged over the saturated state. For comparison with \citetalias{Baronett24}, we mark time-average dispersions measured in their AB models by horizontal dashed lines. 

In general, the dispersions peak at the midplane and decrease with increasing $|z|$, as expected due to stratification, because dust--gas-driven dynamics is dominant near the midplane. The only exception is the azimuthal velocity dispersion of both gas and dust, which reaches its maximum around $z \approx \pm H_p$, a feature also observed during the sedimentation phase (Figure \ref{fig:AB_sed_dispersions}). Notably, the peaks of $\delta u_y(z)$ and $\delta v_y(z)$ occur at nearly the same heights as the peak in the vertical gradient of the COM azimuthal velocity (see bottom panel of Figure~\ref{fig:AB_sat_vertical_gradients}). This suggests that the vertical shear contributes more significantly to the azimuthal velocity dispersion than to the other components. As the gradient vanishes near the midplane, the dust--gas relative motion likely becomes the dominant driver of the dispersion, resulting in the bimodal shape observed in Figure~\ref{fig:AB_sat_vertical_profiles}.

More importantly, the dispersions measured in our stratified simulations are remarkably similar to those measured in unstratified simulations (\citetalias{Baronett24}; horizontal lines) within $\pm H_p$. Outside the region, the deviation between the two cases becomes significant as $|z|$ increases. Among all components, the azimuthal velocity dispersions exhibit the largest deviation between the stratified and unstratified cases. As we explained above, this likely results from the vertical shear of the azimuthal velocity—a feature absent in unstratified simulations. Overall, Figure \ref{fig:AB_sat_vertical_profiles} highlights a strong connection between unstratified simulations and the midplane dynamics of stratified simulations. In the following, we focus on the dispersions at the midplane $(z = 0)$ and present a more quantitative comparison between unstratified and stratified cases.

%%%%%%%%% Figure %%%%%%%%%%%%
\begin{figure*}
    \includegraphics[width=\textwidth]{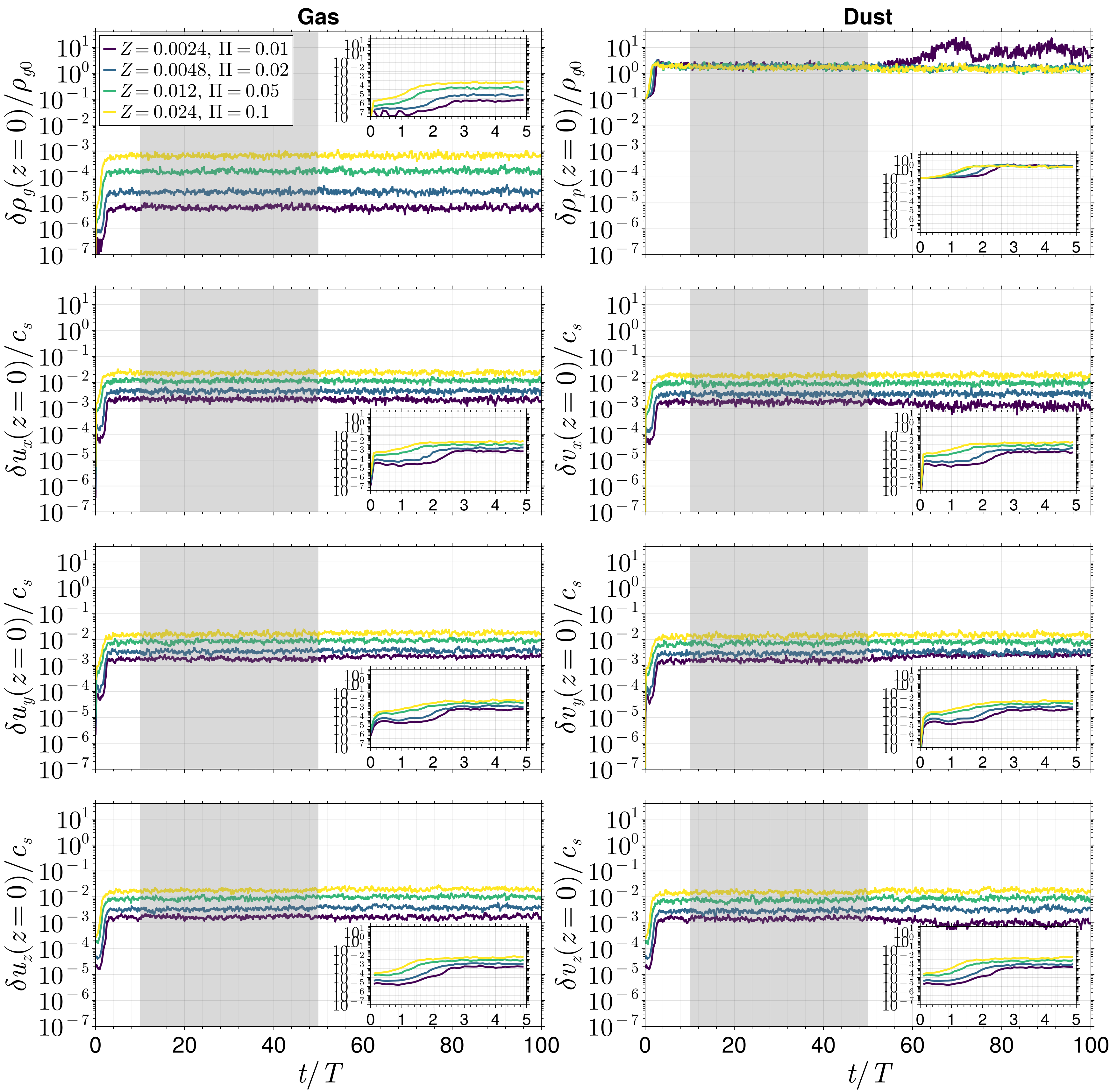}
    \caption{Time evolution of density and velocity dispersions at the midplane in simulations with $\tau_s=0.1$ and $Z/\Pi=0.24$. The left and right panels correspond to the gas and dust quantities, respectively. In order from top to bottom, the panels show the dispersions of density, radial, azimuthal, and vertical velocities. Each panel includes an inset to zoom in on the interval $t/T = 0$ to 5. Larger $\Pi$ values result in higher dispersions during the saturated state denoted by the shaded region, except for $\delta \rho_p (z=0)$. As shown in the insets, higher $\Pi$ also results in earlier saturation of the dispersions (but, see text for further discussion).}
    \label{fig:AB_stdtimeevol}
\end{figure*}
%%%%%%%%% Figure %%%%%%%%%%%%

 %%%%%%%%% Figure %%%%%%%%%%%%
\begin{figure*}
    \includegraphics[width=\textwidth]{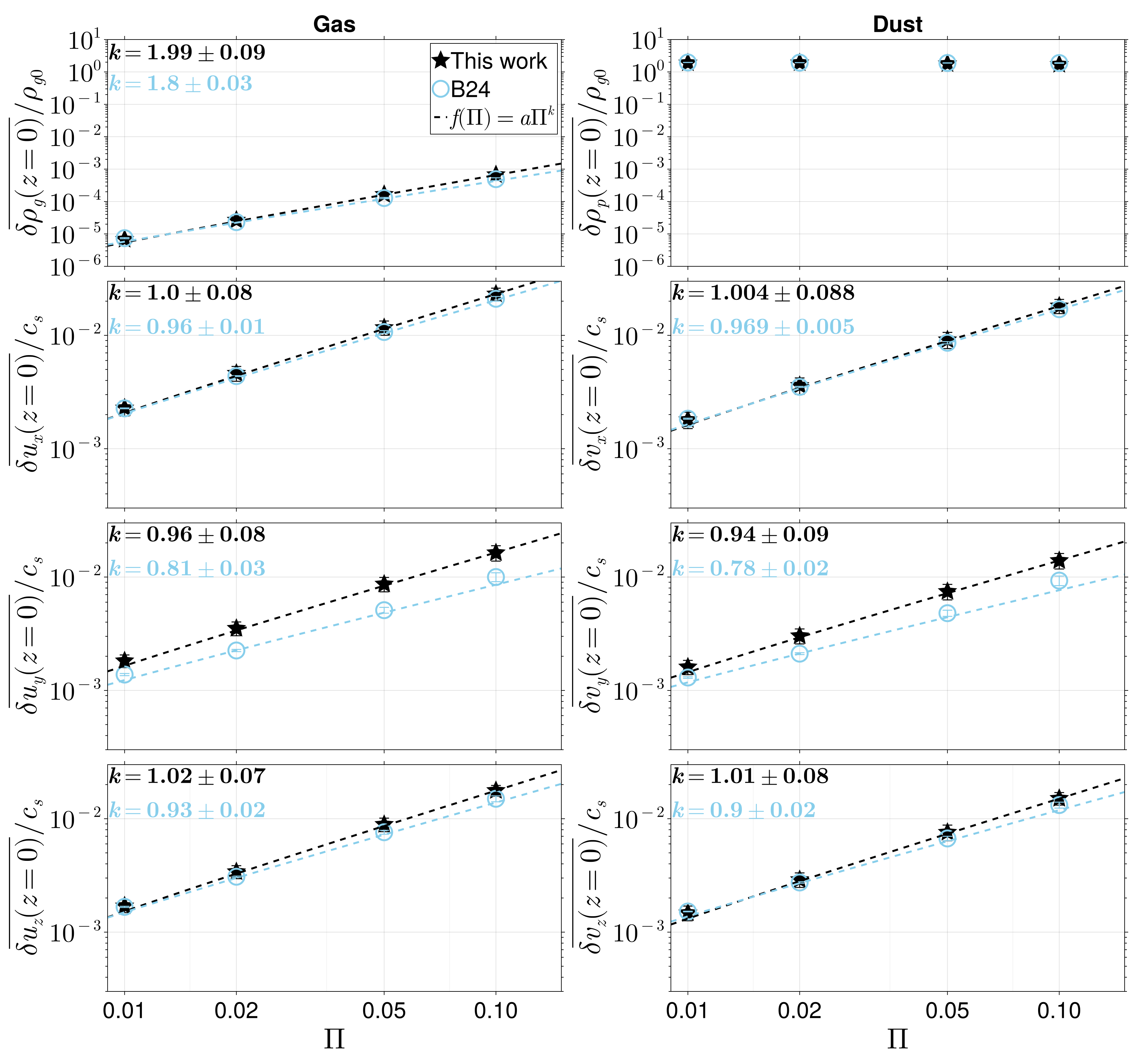}
    \caption{Similar to Figure \ref{fig:AB_stdtimeevol}, but for time-averaged midplane dispersions as functions of $\Pi$. Black and light-blue markers represent the time-averaged values from our simulations and those from \citetalias{Baronett24}, respectively. The time-averaging is performed over $t/T = 10$ to $50$ (corresponding to the gray-shaded regions in Figure \ref{fig:AB_stdtimeevol}) in our simulations and over $t/T = 5$ to $10$ in the \citetalias{Baronett24} simulations. The dashed lines in each panel (except the top right) represents the best-fit power law (see Equation \ref{eq:powerlaw}) to our results and theirs with the resulting $k$ values (i.e., the slope in log-log space) shown in bold text. The figure illustrates that, except for dust density, all quantities scale with $\Pi$, and there is a strikingly close match between stratified and unstratified simulations.}
    \label{fig:AB_stdtimeavg}
\end{figure*}
%%%%%%%%% Figure %%%%%%%%%%%%

\subsubsection{Temporal evolution and averages of midplane dispersions}\label{sec:results:dispersion:temporal_evolution}

Figure \ref{fig:AB_stdtimeevol} shows the midplane density and velocity dispersions as functions of time. To obtain the temporal evolutions shown in the figure, we calculate the dispersions as functions of $z$ and take the average over two grid cells straddling the midplane at any given time.

All quantities shown in Figure \ref{fig:AB_stdtimeevol} reach saturation after the sedimentation phase ($t/T \approx 3$). While the dust density dispersion (top right panel) is independent of $\Pi$, the other quantities exhibit a clear dependence on $\Pi$.  We will revisit this dependence more quantitatively in Figure \ref{fig:AB_stdtimeavg}. The independence of the dust density dispersion from $\Pi$ arises because we scale $Z$ proportionally with $\Pi$ to maintain a constant $Z/\Pi$ ratio, ensuring a similar $\epsilon$ across different $\Pi$ values \citep{sekiya_two_2018}.

Focusing on the early phase ($t/T$ = 0 to 5; see the insets), all dispersions exhibit similar evolution and trends with $\Pi$. Specifically, the higher the value of $\Pi$, the earlier the dispersions start growing, followed by earlier saturation. Furthermore, the timing of this growth and saturation roughly coincides with the onset of dust voids, occurring at $t/T\approx 2.0,~\approx 1.5,~\approx1.0$, and $\approx 1.0$ for $\Pi=0.01,~0.02,~0.05$, and 0.1, respectively, as shown in Figure \ref{fig:AB_sed_snapshots}. We reiterate that the earlier emergence of dust voids at higher $\Pi$ may be due to the larger initial velocity perturbations present in those runs, as seen in Figure \ref{fig:AB_stdtimeevol}. \citetalias{Baronett24} similarly observed that dispersions begin growing earlier at higher $\Pi$ in their AB models (see their Figure 1). 

Figure \ref{fig:AB_stdtimeavg} presents the time-averaged dispersions of the quantities shown in Figure \ref{fig:AB_stdtimeevol}, averaged over the saturated state ($t/T = 10$ to 50). For comparison, the time-averaged values from \citetalias{Baronett24} are also shown in each panel as light-blue circles, while our data points are marked by black stars. Table \ref{tab:timeavg} lists the time-averages.

The gas density dispersions at the midplane range from the order of $\sim 10^{-5}\rho_{g0}$ to that of $\sim 10^{-3}\rho_{g0}$, while the velocity dispersions of the gas and dust span from $\sim 10^{-3}c_s$ to $10^{-2}c_s$. This indicates that the gas remains incompressible even at our highest $\Pi$ value. Additionally, turbulence appears to be nearly isotropic in both unstratified and stratified cases (see Table \ref{tab:timeavg} and Table 2 in \citetalias{Baronett24}). 

We emphasize that the measured dispersions are nearly identical between our stratified simulations and \citetalias{Baronett24}'s unstratified AB simulations. Specifically, in both cases, the dust density dispersion is independent of $\Pi$, whereas the dispersions of gas density, gas velocities, and dust velocities scale with $\Pi$ in a similar manner. The azimuthal velocity dispersions exhibit the largest deviation between the two cases, with the highest ratio of our values to those of \citetalias{Baronett24} reaching $\approx 1.7$ at $\Pi = 0.05$. We speculate that some of the energy in the azimuthal component—generated by vertical shear away from the midplane (see Figure \ref{fig:AB_sat_vertical_gradients})— propagates toward the midplane, leading to slightly larger dispersions compared to the unstratified case, where this effect is absent. Nonetheless, the close match between the unstratified and stratified cases demonstrates that unstratified simulations accurately capture the dust--gas dynamics at the midplane of stratified disks for the parameter values we explore (e.g., for $\tau_s=0.1$).
 
The dashed lines in each panel of Figure \ref{fig:AB_stdtimeavg} represent the best power-law fit to our data points (black) and those from \citetalias{Baronett24} (light-blue), expressed as:
\begin{equation}\label{eq:powerlaw}
f(\Pi) = a\Pi^k.    
\end{equation}
We perform linear fitting in logarithmic space. The resulting $k$ value is provided in each panel with its uncertainty. We do not include a best-fit for the dust density dispersion (top right), as it is independent of $\Pi$. The resulting parameters from both this work and \citetalias{Baronett24} are listed in Table \ref{tab:powerlaw}. 

%%%%%%%%% Figure %%%%%%%%%%%%
\begin{figure*}[ht!]
    \centering
    \includegraphics[width=\textwidth]{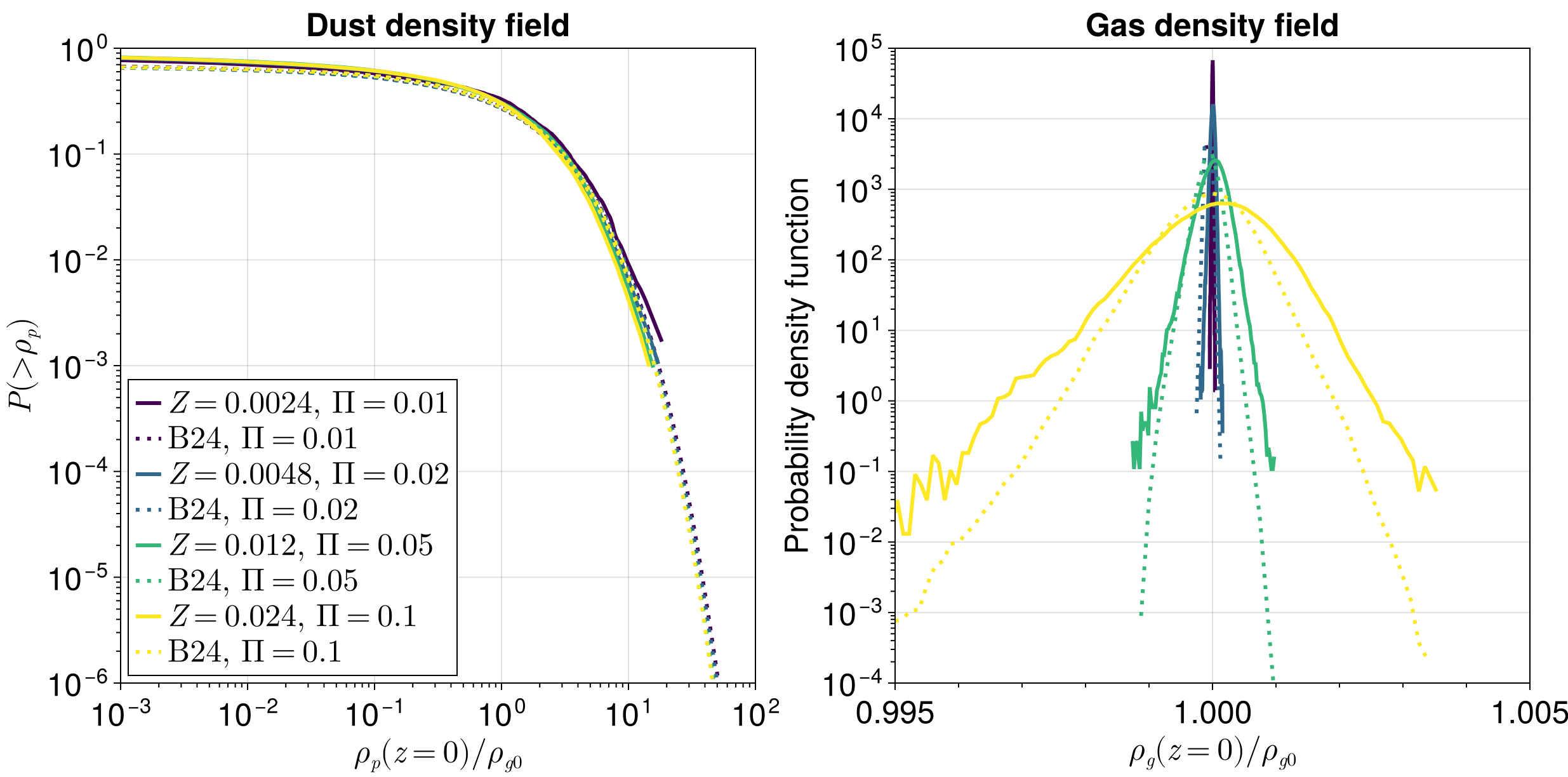}
    \caption{Time-averaged cumulative distribution functions of the midplane dust density (left) and probability density functions of the midplane gas density (right) for models with $\tau_s = 0.1$ and $Z/\Pi = 0.24$ (solid lines). Distributions from \citetalias{Baronett24} are shown as dotted lines. The time-averaging is performed over $t/T = 10$ to $50$ in our models and $t/T = 5$ to $10$ in the models by \citetalias{Baronett24}. Each color corresponds to a different $\Pi$ value. The dust and gas density fields from both unstratified and stratified simulations exhibit similar trends with $\Pi$: the dust density field remains independent of $\Pi$, while the gas density distributions become broader as $\Pi$ increases. Both the unstratified and stratified cases exhibit nearly identical dust density distributions. However, the gas density distribution is broader in the stratified case, especially for $\Pi=0.1$. }
    \label{fig:AB_cdf}
\end{figure*}
%%%%%%%%% Figure %%%%%%%%%%%%

For all dispersions to which we apply power-law fitting, the $k$ values in unstratified and stratified cases roughly agree within the uncertainties, except for the azimuthal velocity dispersions. Moreover, all velocity dispersions have $k \approx 1$, indicating a linear relationship between the velocities and $\Pi$. This aligns with the understanding that, in the linear regime of the SI, velocities scale linearly with $\Pi$ \citep{YG05,YJ07}. By contrast, the gas density dispersion (top left) exhibits a super-linear relationship with $\Pi$, with $k \approx 2$. 
This dependence aligns with the scaling relation between density and velocity with an isothermal equation of state, $\overrightarrow{\nabla}\rho_g/\rho_g \propto (\vec{u} \cdot \overrightarrow{\nabla})\vec{u}/c_s^2$. In deriving this relation, we assume that gravity, the centrifugal force, and the large-scale pressure gradient cancel out, as we are considering perturbations. As a result, the residual advection terms are balanced by the residual pressure gradient in the steady-state perturbation equations. Thus, since the gas velocity dispersions scale linearly with $\Pi$, $\delta \rho_g(z=0)$ should scale quadratically with $\Pi$.

%%%%%%%  Table 2%%%%%%%%%%%%%%%%%
\begin{deluxetable*}{cccccccccccc}
\tablecaption{Time-averages at Saturation}\label{tab:timeavg}
\tablehead{
\colhead{$\Pi$} & \colhead{$\rho_{p,\rm{max}}$} & \colhead{$\delta \rho_g (z=0)$} & \colhead{$\delta u_x (z=0)$} & \colhead{$\delta u_y (z=0)$} & \colhead{$\delta u_z (z=0)$} & \colhead{$\delta \rho_p (z=0)$} & \colhead{$\delta v_x (z=0)$} & \colhead{$\delta v_y (z=0)$} & \colhead{$\delta v_z (z=0)$} \\
%2nd row
%\colhead{} & 
\colhead{} & 
\colhead{$(\rho_{g0})$} & 
\colhead{$(10^{-4}\rho_{g0})$} &
\colhead{$(10^{-3}c_s)$} &
\colhead{$(10^{-3}c_s)$} &
\colhead{$(10^{-3}c_s)$} &
\colhead{$(\rho_{g0})$} &
\colhead{$(10^{-3}c_s)$} &
\colhead{$(10^{-3}c_s)$} &
\colhead{$(10^{-3}c_s)$} \\  
%3rd row
\colhead{(1)} &
\colhead{(2)} &
\colhead{(3)} &
\colhead{(4)} &
\colhead{(5)} &
\colhead{(6)} &
\colhead{(7)} &
\colhead{(8)} &
\colhead{(9)} &
\colhead{(10)} &
%\colhead{(11)} &
}
\startdata
 0.01 & $46.65 \pm 10.82$ &  $0.07 \pm 0.01$ &  $2.29 \pm 0.35$ & $1.82 \pm 0.23$ &  $1.70\pm 0.22$ & $1.90\pm 0.35$ & $1.80 \pm 0.29$ & $1.61 \pm 0.23$ &  $1.48 \pm 0.22$ \\
0.02 & $42.51 \pm 9.40$ & $0.27 \pm 0.04$ & $4.63 \pm 0.69$ & $ 3.53 \pm 0.49$ & $3.41 \pm 0.45$ & $1.91 \pm 0.31$ & $3.31 \pm 0.60$ &  $3.03 \pm 0.46$ & $2.90 \pm 0.43$ \\
0.05 & $37.12 \pm 8.26$ & $1.65 \pm 0.27$ & $11.69 \pm 1.68$ &  $8.66\pm 1.26$ & $8.90 \pm 1.22$ &  $1.79 \pm 0.31$ &  $9.13\pm 1.48$ &  $7.45 \pm 1.16$ &  $7.59 \pm 1.20$ \\ 
0.1 & $33.06 \pm 6.26$ & $6.70 \pm 1.05$ & $23.13 \pm 2.93$ & $16.34 \pm 2.53$ & $17.72\pm 1.94$ &  $1.71 \pm 0.25$ & $18.14 \pm 2.60$ &  $13.95 \pm 2.17$ & $15.10 \pm 1.91$
\enddata
\tablecomments{Columns: (1) dimensionless parameter for the radial pressure gradient (Equation \ref{eq:Pi}; see Table \ref{tab:simlist} for the corresponding $Z$ value for each $\Pi$); (2) maximum dust density; (3) dispersion of gas density at the midplane; (4) - (6) dispersion of radial, azimuthal, and vertical velocities of the gas at the midplane, respectively (Equation \ref{eq:deltau}) (7) dispersion of dust density at the midplane (Equation \ref{eq:deltarhop}); (8)-(10) dispersion of radial, azimuthal, and vertical velocities of dust at the midplane, respectively (Equation \ref{eq:deltav}). See Section \ref{sec:results:saturation} for how we defined the saturated state. Time-averaged values of dust scale height ($H_p$, Equation \ref{eq:Hp}) and the midplane density ratio ($\epsilon$, Equation \ref{eq:eps}) are $\approx 0.24 \Pi H$ and $\approx 1$ in all of the runs listed above. 
}    
\end{deluxetable*}
%%%%%%%  Table 2%%%%%%%%%%%%%%%%%

\subsection{Gas and Dust Density Distributions}\label{sec:results:densityfield}
In this subsection, we examine density distributions of the gas and dust at the midplane. In addition, we compare these results to unstratified AB models in \citetalias{Baronett24}.

We present time-averaged midplane density distributions of dust and the gas in Figure \ref{fig:AB_cdf}. The distributions are time-averaged during the saturated state. The left panel shows cumulative dust density distributions (i.e., the fraction of grid cells with $\geq \rho_p$), while the right panel displays the probability density functions  of midplane gas density. We compute probability density functions instead of cumulative distribution functions for the gas because the gas density field spans too narrow a range for the cumulative form to be informative. Distributions from our simulations are represented by solid curves, and those from \citetalias{Baronett24} by dotted curves. For both linestyles, each color corresponds to a different value of $\Pi$. We discuss each density field separately below. 

The cumulative dust density distributions are nearly identical and independent of $\Pi$ in both the stratified and unstratified cases. This aligns with our finding in Section \ref{sec:results:dispersion} that the dust density dispersion does not depend on $\Pi$. However, we observe that the maximum dust density, time-averaged over the saturated state (shaded region in the top panel of Figure \ref{fig:AB_DmaxHpEps}), increases as $\Pi$ decreases (see Table \ref{tab:timeavg}), with a relative difference of $\approx 41\%$ between $\Pi$ = 0.01 and $\Pi$ = 0.1. This trend is also seen in the simulations of \citetalias{Baronett24}. Overall, the distributions are in strong agreement with each other for the range of probabilities shared by both works (e.g., $10^{-3} < P(>\rho_p) < 1$).\footnote{The extension of the cumulative distribution of the unstratified model (dotted; \citetalias{Baronett24}) to $P(>\rho_p)$ of $10^{-6}$ is due to the difference in the number of grid cells considered in the cumulative distribution functions between their and our simulations. While we present $\rho_p$ averaged over two horizontal slabs of cells around the midplane, which consists of 1024 cells out of $1024\times2048$ (radial-vertical), they included all grid cells in their simulations ($2048^2$) because their entire simulation domain had no vertical stratification.}

We now examine the gas density distribution shown in the right panel of Figure \ref{fig:AB_cdf}. Both unstratified (dotted) and stratified (solid) cases clearly demonstrate that the distribution becomes broader with increasing $\Pi$, consistent with the observed increase in gas density dispersion with $\Pi$ (Figures \ref{fig:AB_stdtimeevol}-\ref{fig:AB_stdtimeavg}). However, the time-averaged distributions in the stratified cases are slightly broader than those in the unstratified cases, especially when $\Pi=0.1$, and the differences between the two cases appear to increase as $\Pi$ increases. This is further supported by the result that the gas density dispersion has a slightly steeper power-law with $\Pi$ in the stratified cases ($k\approx 2.0$) than in the unstratified cases $(k\approx 1.8)$ (see Figure \ref{fig:AB_stdtimeavg} and Table \ref{tab:powerlaw}). We speculate that the broader distribution in the vertically stratified case is likely due to the gas from regions adjacent to the midplane being mixed into the midplane by turbulence.

%%%%%%%%% Figure %%%%%%%%%%%%
\begin{figure}
    \centering
    \includegraphics[width=\columnwidth]{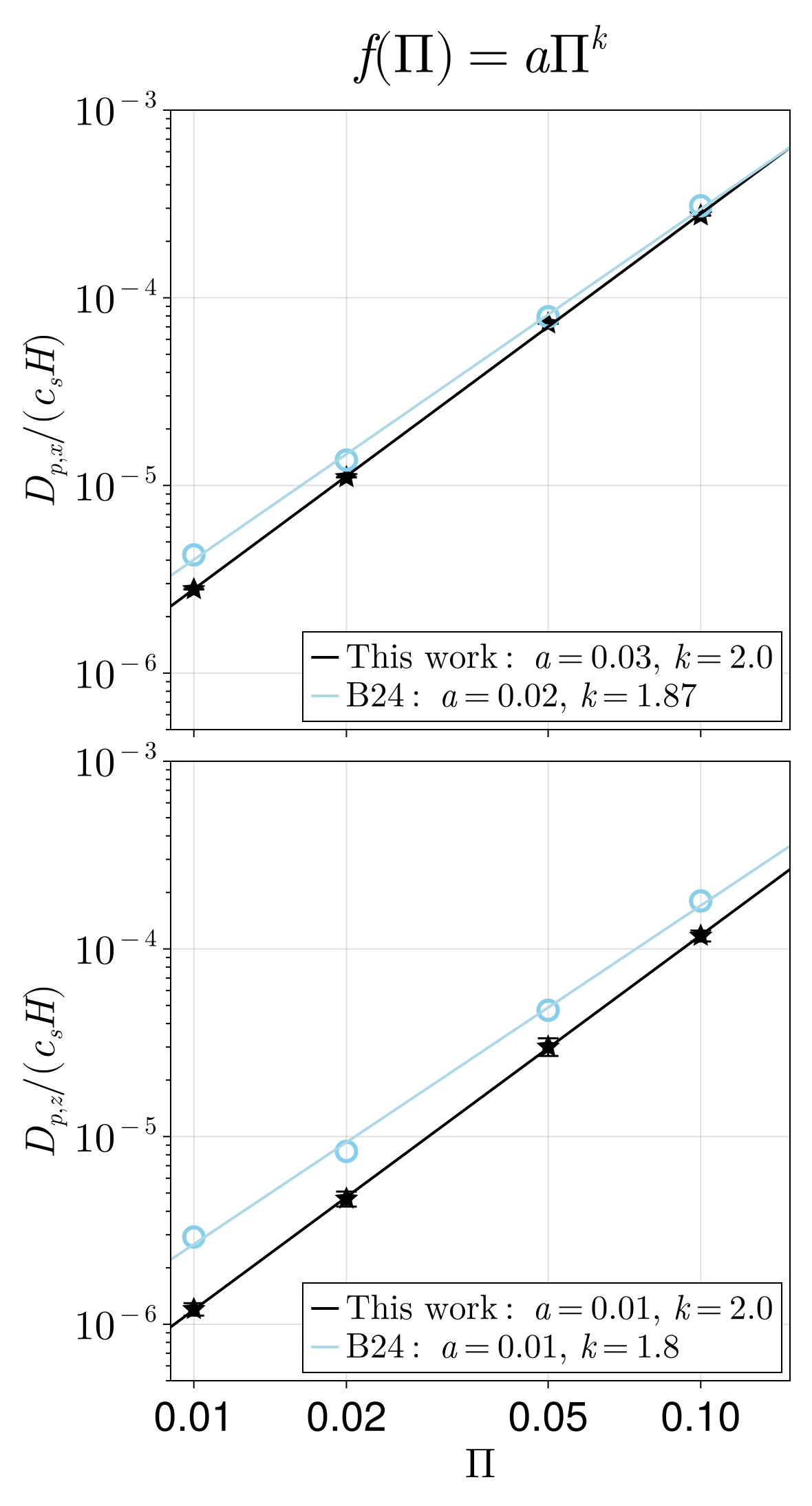}
    \caption{Radial (top) and vertical (bottom) diffusion coefficients as a function of $\Pi$ for simulations with $\tau_s = 0.1$ and $Z/\Pi=0.24$. In each panel, black and light-blue markers represent the diffusion coefficients from this work and \citetalias{Baronett24}, respectively. Solid lines indicate the best-fitting power-law models, defined by Equation (\ref{eq:powerlaw}) (see figure title), with the best-fitting parameters shown in the legends. Despite minor differences in the data points and fitting lines, the unstratified and stratified models yield similar diffusion coefficients and exhibit consistent trends with $\Pi$.}
    \label{fig:AB_dcoeff}
\end{figure}
%%%%%%%%% Figure %%%%%%%%%%%%

%%%%%%%  Table %%%%%%%%%%%%%%%%%

\begin{deluxetable}{lcc}[t!]
\label{tab:powerlaw}
\tablecaption{Best-fit Power-law Parameters}
\tablehead{
\colhead{Quantity} & \colhead{This work} & \colhead{\citetalias{Baronett24}} \\ 
\colhead{} & \colhead{$a\quad\quad k$} & \colhead{$a\quad\quad k$}
%\multicolumn{2}{c}{$a$} &
%\multicolumn{2}{c}{$k$}  
} 
\decimals
\startdata
$\delta \rho_g(z=0)/\rho_{g0}$ & 0.07 (2) \ 1.99 (9) & 0.027 (4) \ 1.80 (3) \\
$\delta u_x (z=0)/c_s$ & 0.24 (6) \ 1.00 (8) & 0.19 (1) \ 0.96 (1) \\ 
$\delta u_y (z=0)/c_s$ & 0.15 (4) \ 0.96 (8) & 0.05 (1) \ 0.81 (3) \\
$\delta u_z (z=0)/c_s$ & 0.19 (4) \ 1.02 (7) & 0.12 (1) \ 0.93 (2) \\
$\delta v_x (z=0)/c_s$ & 0.18 (6) \ 1.00 (9) & 0.158 (2) \ 0.969 (5) \\
$\delta v_y (z=0)/c_s$ & 0.12 (4) \ 0.94 (8) & 0.046 (5) \ 0.78 (2) \\
$\delta v_z (z=0)/c_s$ & 0.16 (4) \ 1.01 (8) & 0.09 (1) \ 0.90 (1) \\
$D_{p,x}/(c_sH)$ & 0.028 (2) \ 2.00 (2)  & 0.022 (3) \ 1.87 (4) \\
$D_{p,z}/(c_sH)$ & 0.0117 (5) \ 2.00 (1) & 0.011 (2) \ 1.80 (6) \\
\enddata
\tablecomments{The best-fit power-law parameters $a$ and $k$ (Equation~\ref{eq:powerlaw}) for the measured quantities in this work and \citetalias{Baronett24}. From top to bottom, each row corresponds to: gas density dispersion (computed similarly to Equation~\ref{eq:deltarhop}); radial, azimuthal, and vertical velocity dispersions of the gas (Equation~\ref{eq:deltau}); radial, azimuthal, and vertical velocity dispersions of dust (Equation~\ref{eq:deltav}); and radial and vertical diffusion coefficients of dust (Equations~\ref{eq:Dpx}–\ref{eq:Dpz}). We show in parentheses the uncertainty in the least  significant digit of each parameter.
}
\end{deluxetable}
%%%%%%%  Table %%%%%%%%%%%%%%%%%

\subsection{Diffusion Coefficients of Dust}\label{sec:results:diffusion}
In this subsection, we compute the radial and vertical diffusion coefficients of the dust and compare them to those in \citetalias{Baronett24}. To calculate the radial diffusion coefficient, we track the radial displacement of positions of particles at time $t$ from their positions at an initial time $t_0$ and compute the variance $\sigma^2_x(t)$ of the $x_i(t)-x_i(t_0)$ distribution.  We set $t_0=10T$ and calculate $\sigma^2_x$ every $0.1T$ up to $t=50T$. Then, the radial diffusion coefficient can be obtained by the best-fit of 
\begin{equation}\label{eq:Dpx}
    D_{p,x}=\frac{1}{2}\frac{\partial \sigma^2_x(t)}{\partial t}
\end{equation}
as in \citet{JY07}. 

For the vertical diffusion coefficient ($D_{p,z}$), we follow the approach described by \citet{Lim24a}\footnote{We omit the $(5/\Pi)^2$ term in their Equation 23 because external turbulence is not included in our simulations.} to find
\begin{equation}\label{eq:Dpz}
D_{p,z} = (1+\epsilon)\tau_s\left(\frac{H_p}{H}\right)^2c_sH.
\end{equation}
To derive this equation from their Equation 23, we assumed that $[D_{p,z}/(c_sH)] / \tau_s \ll 1$. The $(1+\epsilon)$ term in Equation (\ref{eq:Dpz}) accounts for the mass-loading effects of dust on the gas, which increase the inertia of the dust--gas mixture \citep{ShiChiang2013,LaibePrice2014,LinYoudin2017,ChenLin2018, YangZhu20}. This implies that stronger diffusion is required when $\epsilon \gg 1$ compared to when $\epsilon \ll 1$ to maintain the same ratio of $H_p / H$. We use $H_p/H$ averaged over the saturated state (see the middle panel of Figure \ref{fig:AB_DmaxHpEps}) to calculate $D_{p,z}$.

\citetalias{Baronett24} followed \citet{Yang2009} (see their Section 4.1) to calculate both  $D_{p,x}$  and  $D_{p,z}$. The method they used is essentially the same as Equation (\ref{eq:Dpx}). The equation is the analytical solution to the diffusion equation with the Gaussian distribution of the radial displacement of dust particles. However, in \citet{Yang2009} and \citetalias{Baronett24}, the Gaussian distribution is expressed as  $\propto e^{-x^2/\sigma_x^2}$  (see Equation 28 in \citealt{Yang2009}), leading to  $D_{p,x} = (1/4)\partial \sigma_x^2 / \partial t$. By contrast, Equation (\ref{eq:Dpx}) is derived using the form $\propto e^{-x^2/(2\sigma_x^2)}$. To ensure consistency, we multiply the $D_{p,x}$ and $D_{p,z}$ values reported in \citetalias{Baronett24} by 2.

We present $D_{p,x}$ and $D_{p,z}$ as functions of $\Pi$ in the upper and lower panels of Figure \ref{fig:AB_dcoeff}, respectively. The black and light-blue markers represent results from our simulations and those from \citetalias{Baronett24}, respectively. The best-fit power-law models (Equation \ref{eq:powerlaw}) for $D_{p,x}$ and $D_{p,z}$ are shown as black lines for our data and light-blue lines for the data from \citetalias{Baronett24}. The corresponding best-fit parameters ($a$ and $k$) from Equation~\ref{eq:powerlaw} (also see title of Figure~\ref{fig:AB_dcoeff}) are displayed in the figure legends for reference. 

In general, $D_{p,x}$ and $D_{p,z}$ increase as $\Pi$ increases in both stratified (black) and unstratified (light-blue) cases. In addition, the radial diffusion is consistently stronger than the vertical diffusion across all $\Pi$ values. Specifically, unstratified and stratified cases show $D_{p,x}/D_{p,z} \approx 1.5$ and $\approx 2.4$, respectively, with little variation across $\Pi$. Stronger radial diffusion compared to vertical diffusion has been reported in previous simulations of unstratified AB models by \citet{JY07}. Furthermore, \citet{LiYoudin21} found that $D_{p,x}$ is consistently larger than $D_{p,z}$ across a wide range of stopping times ($\tau_s = 10^{-3}$ to 1) in their stratified simulations. It is important to note, however, that both studies only considered $\Pi = 0.05$.

We now focus on the comparison of the diffusion coefficients between unstratified and stratified cases. While the two cases yield similar values for both $D_{p,x}$ and $D_{p,z}$, the vertical diffusion coefficient exhibits larger deviations. Specifically, the ratio of $D_{p,x}$ from \citetalias{Baronett24} to our results is highest $(\approx 1.5)$ at $\Pi = 0.01$ and decreases to $\approx 1$ at higher $\Pi$. By contrast, the ratio for $D_{p,z}$ ranges from $\approx 2.4$ at $\Pi = 0.01$ to $\approx 1.5$ at $\Pi = 0.1$. Given that dust particles undergo a random walk in both directions in the unstratified case, the closer match in $D_{p,x}$ between the two cases suggests that radial diffusion in the stratified case can also be described by a random walk. However, the presence of gravitational pull toward the midplane in the stratified case prevents dust particles from diffusing freely in $z$. In fact, the vertical motion of dust particles can be described as a damped harmonic oscillator stochastically forced by gas drag \citep{Carballido06, youdin_particle_2007}. Based on this, we speculate that the weaker vertical diffusion observed in stratified cases is related to the influence of vertical gravity on the stochastic forcing by the gas. Despite the influence of vertical gravity, we reiterate that the small differences in both $D_{p,x}$ and $D_{p,z}$ between the unstratified and stratified simulations highlight a strong connection between the two cases.

Regarding the power-law relation between the diffusion coefficients and $\Pi$, the stratified case exhibits a steeper power-law slope compared to the unstratified case for both diffusion coefficients. Nevertheless, the fact that $k$ is close to 2 for both $D_{p,x}$ and $D_{p,z}$ in both cases indicates that the characteristic length and velocity scales of turbulence scale with $\Pi$ in approximately the same way independent of whether or not vertical stratification is present.

%%%%%%%%% Figure %%%%%%%%%%%%
\begin{figure*}
    \centering
    \includegraphics[width=\textwidth]{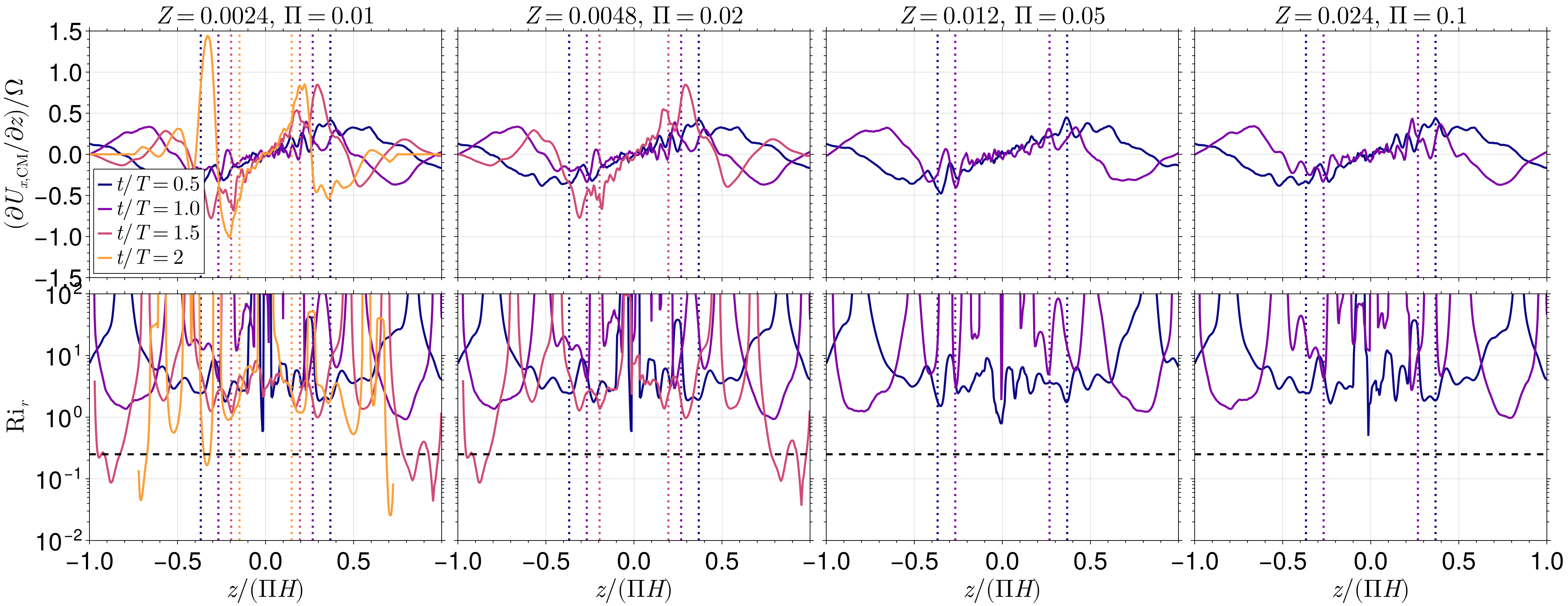}
    \caption{The vertical gradient of radial velocity (upper row; Equation \ref{eq:U_com}) and the radial Richardson number (lower row; Equation \ref{eq:Ri}) as a function of $z$ across different $\Pi$ values during the sedimentation phase. In order from left to right, panels correspond to $\Pi=0.01,~0.02,~0.05$, and 0.1. For each $\Pi$, we include snapshots before the dust layer becomes turbulent. We mark $\pm H_p$ at a given time using dotted lines in each panel, with the dotted lines matching the color of the corresponding profile at that time. The magnitude of the vertical gradient of radial velocity vanishes at the midplane but increases with height up to $\approx \pm H_p$ in all snapshots and $\Pi$ values. The Richardson number is above  $\rm{Ri_{r,\rm{crit}}}$ (0.25; horizontal dashed lines in the lower panels) within the dust layer making it stable to KHI.
    }
    \label{fig:Discussion:sed-dUdz-Ri}
\end{figure*}
%%%%%%%%% Figure %%%%%%%%%%%%

%%%%%%%%% Figure %%%%%%%%%%%%
\begin{figure*}
    \centering
    \includegraphics[width=\textwidth]{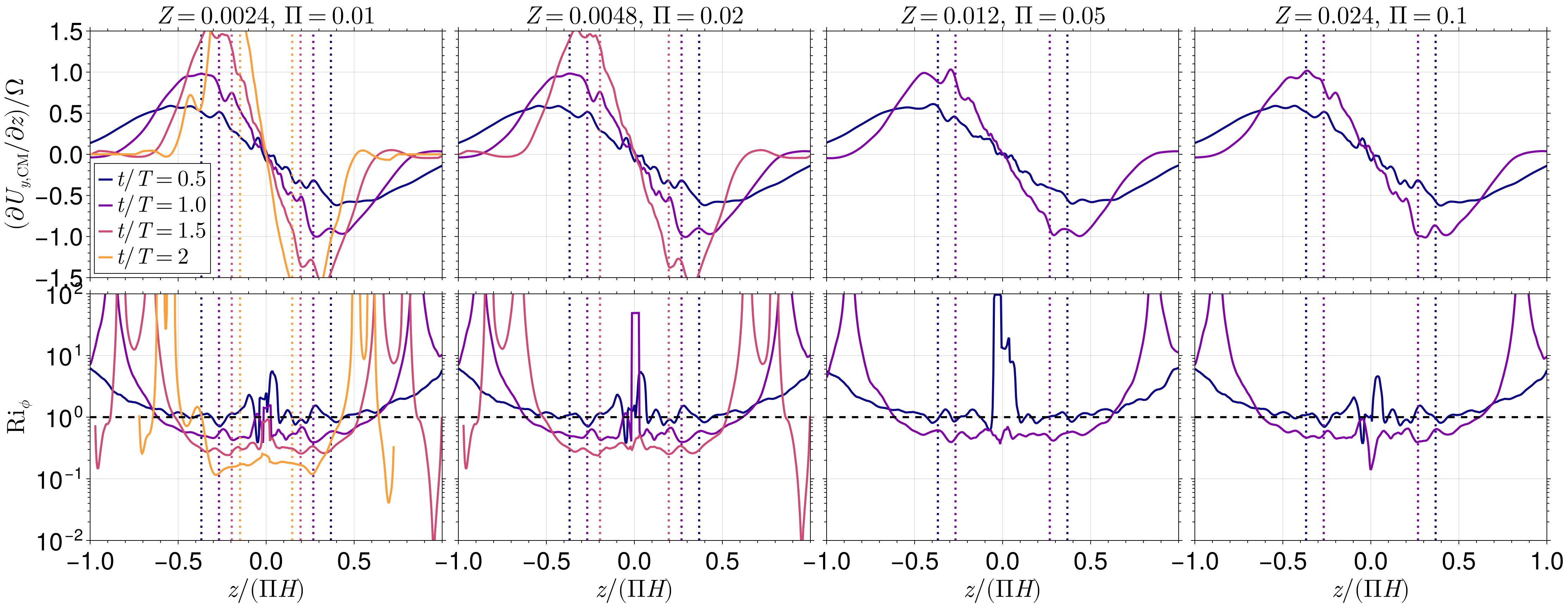}
    \caption{Similar to Figure \ref{fig:Discussion:sed-dUdz-Ri}, but for the vertical gradient of azimuthal velocity (upper row; Equation \ref{eq:V_com}) and the azimuthal Richardson number (lower row, Equation \ref{eq:Ri_phi}). Unlike $\textrm{Ri}_{r}$, $\textrm{Ri}_{\phi}$ falls below the threshold ($\textrm{Ri}_{\phi, \textrm{crit}}$, the horizontal line in the lower panels) within the dust layer (marked by vertical lines) as dust settles toward the midplane across all simulations presented here making it unstable to SymI.
    }
    \label{fig:Discussion:sed-dVdz-Riphi}
\end{figure*}
%%%%%%%%% Figure %%%%%%%%%%%%

\section{Discussion}\label{sec:Discussion}
%We have demonstrated that unstratified models can accurately predict gas and particle dynamics near the midplane in stratified models during the saturation state. However, stratified models include a dust sedimentation phase before reaching saturation, a process absent in unstratified models. Additionally, our run with $\Pi = 0.01$ reveals a sharp increase in maximum particle density after the saturation state (see the top panel of Figure \ref{fig:AB_DmaxHpEps}). In the next two subsections, we discuss these two distinct phases: the sedimentation phase and the strong-clumping phase.

\subsection{Turbulence in Stratified Models}\label{sec:Discussion:turb}
We have demonstrated in Section \ref{sec:results:dispersion} that during the saturated state, the gas and dust velocity dispersions exhibit close agreement between unstratified and stratified models. However, this does not necessarily imply that turbulence is driven by the same mechanism in both cases.  In fact, stratification naturally induces vertical shear in radial and azimuthal velocities, which can trigger various shear-driven instabilities in a stratified dusty layer, such as the Kelvin-Helmholtz instability (KHI; \citealt{Weidenschilling80,WeidenschillingCuzzi93,Sekiya98,Johansen06,Chiang2008ApJ,Lee2010,Sengupta_Umurhan23}),
the symmetric instability (SymI; \citealt{Sengupta_Umurhan23}), 
and the vertically shearing streaming instability (VSSI; \citealt{Lin2021}). Both the SymI and VSSI require a vertical gradient in the dust-gas mixture's azimuthal velocity. The key difference is that the SymI arises from a single-fluid approximation (see Section 5.2 and Appendix C in \citealt{Sengupta_Umurhan23} for details), meaning it does not require a relative velocity between dust and gas. By contrast, the VSSI harnesses the relative velocity, particularly at small wavenumbers \citep{Lin2021}. These shear-driven instabilities are absent in unstratified models. 

Given these considerations, we investigate the role of shear-driven turbulence in our simulations during both the sedimentation and saturated states.  
We use the radial Richardson number ($\textrm{Ri}_r$) to assess the onset of the axisymmetric KHI\footnote{Non-axisymmetric KHI, which is driven by the vertical gradient in \textit{azimuthal} velocities \citep{Johansen06,Chiang2008ApJ,Lee2010}, is suppressed in our axisymmetric simulations. Throughout this section, we use the term ‘axisymmetric KHI’ to refer to the KHI associated with the vertical gradient of the COM radial velocity (i.e., $U_{x,\mathrm{CM}}$)}. in our simulations, following \citet{Sengupta_Umurhan23}:
\begin{equation}\label{eq:Ri}
    \textrm{Ri}_r= -\left(\frac{\Omega^2 z}{\rho_{g0}+\langle \rho_p \rangle_x}\right) 
    \frac{\partial \langle \rho_p \rangle_x}{\partial z} 
    \Big/ \left(\frac{\partial U_{x,\rm{CM}}}{\partial z}\right)^2.
\end{equation}
If $\textrm{Ri}_r$ falls below a critical threshold, the axisymmetric KHI is triggered, leading to overturning and mixing within the dust layer. In non-rotating flows, this threshold is 0.25 \citep{Chandrasekhar1961,Howard1961}. In differentially rotating disks, although the critical threshold for $\textrm{Ri}_r$ is neither unique nor well-defined, \citet{Sengupta_Umurhan23} showed that the axisymmetric KHI roll-up develops when $\textrm{Ri}_r < 0.25$. Therefore, we adopt $\textrm{Ri}_{r,\textrm{crit}} = 0.25$ as an approximation for the critical threshold.

For the evaluation of the SymI, we use the azimuthal Richardson number $(\textrm{Ri}_{\phi})$, which is defined by
\begin{equation}\label{eq:Ri_phi}
    \textrm{Ri}_{\phi}= -\left(\frac{\Omega^2 z}{\rho_{g0}+\langle \rho_p \rangle_x}\right) 
    \frac{\partial \langle \rho_p \rangle_x}{\partial z} 
    \Big/ \left(\frac{\partial U_{y,\rm{CM}}}{\partial z}\right)^2.
\end{equation}

\citet{Sengupta_Umurhan23} performed a linear analysis for the SymI under the single-fluid assumption (i.e., $\tau_s \rightarrow 0$) and demonstrated that the condition for the instability is $\textrm{Ri}_{\phi} < \textrm{Ri}_{\phi,\textrm{crit}}=1$ (see their Figure 24). We adopt the same condition to assess whether the SymI is present in our simulations.

As mentioned above, the VSSI requires both partial dust--gas coupling (i.e., not perfect coupling as in the case of $\tau_s$ = 0) and a vertical gradient in the rotation velocity of the dust--gas mixture, with the latter being the primary driving mechanism. While we cannot probe the possiblity of the VSSI in the sedimentation phase (because its linear analysis assumes an equilibrium dust layer where sedimentation is balanced by vertical diffusion driven by underlying turbulence) as discussed in the next subsection, we do revisit this instability in Section \ref{sec:Discussion:turb:sat}.

%%%%%%%%% Figure %%%%%%%%%%%%
\begin{figure*}
    \centering
    \includegraphics[width=\textwidth]{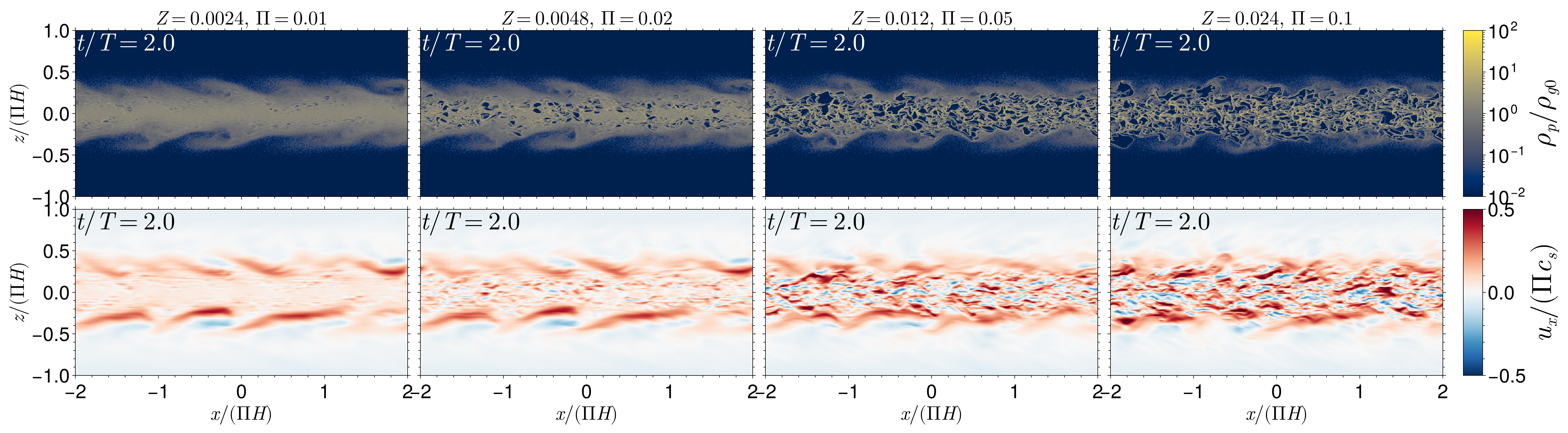}
    \caption{Snapshots of the dust density ($\rho_p$; upper panels) and radial gas velocity ($u_x$; lower panels) at $t/T=2$. In order from left to right, panels correspond to $\Pi=0.01$, 0.02, 0.05, and 0.1, respectively. For $\Pi = 0.05$ and 0.1, dust filaments are nearly fully developed, and the midplane is already turbulent. By contrast, for $\Pi = 0.01$ and 0.02, the midplane is still under development, while the radial velocity reaches its largest magnitude near the top of the dust layer, where large radial structures (as large as $\sim \Pi H$) develop. 
    }
    \label{fig:Discussion:sed-rhopux}
\end{figure*}
%%%%%%%%% Figure %%%%%%%%%%%%

\subsubsection{Sedimentation phase}\label{sec:Discussion:turb:sed}

In Figure \ref{fig:Discussion:sed-dUdz-Ri}, we present the vertical gradient of radial velocity ($\partial U_{x,\rm{CM}}/\partial z$; upper panels) and $\textrm{Ri}_r$ (lower panels) as a function of $z$ during the sedimentation phase. The panels from left to right correspond to $\Pi = 0.01,~0.02,~0.05$, and $0.1$, respectively. In each panel, we use darker to brighter colors to represent earlier to later times, and two vertical dotted lines at $\pm H_p(t)$ (cf. Figure~\ref{fig:AB_DmaxHpEps}, middle panel). For a given $\Pi$, we only include snapshots before turbulence develops within the dust layer as we are interested in the onset of the axisymmetric KHI. In the lower panels, horizontal dashed lines mark the critical KHI threshold of 0.25.

As shown in the upper panels of Figure \ref{fig:Discussion:sed-dUdz-Ri}, the magnitude of the vertical gradient of radial velocity vanishes at the midplane and increases with height, reaching a maximum around $\pm H_p(t)$, similar to Figure~\ref{fig:AB_sat_vertical_gradients} (second panel). Nevertheless, the lower panels show that $\textrm{Ri}_r$ is always above the threshold within $\pm H_p$ and even beyond except for a very localized excursion below 0.25 when $\Pi=0.01$ and $t/T=2$ (orange in the lower leftmost panel). This suggests that the dust layer is stable against the axisymmetric KHI during the sedimentation phase.

We now present the vertical profiles of the vertical gradient of azimuthal velocity (Equation \ref{eq:V_com}) and the azimuthal Richardson number (Equation \ref{eq:Ri_phi}) in Figure \ref{fig:Discussion:sed-dVdz-Riphi} to study the potential onset of the SymI. In contrast to $\textrm{Ri}_r$, the figure clearly shows that $\textrm{Ri}_{\phi}$ drops below $\textrm{Ri}_{\phi, \textrm{crit}}$ within $\pm H_p$, regardless of $\Pi$. This implies that the dust layer is susceptible to the SymI during the sedimentation phase. Similarly, \citet{Sengupta_Umurhan23} found in their axisymmetric simulations that $\textrm{Ri}_{\phi} < \textrm{Ri}_{\phi,\textrm{crit}}$ in both the $\tau_s=0.04$ and 0.1 cases, while the higher $\tau_s$ case also appears to be unstable to the the axisymmetric KHI. 

The SymI may explain the findings in Section \ref{sec:results:sedimentation}, where the locations of both dust void formation and velocity dispersion peaks shift toward the midplane as $\Pi$ increases (see Figures \ref{fig:AB_sed_snapshots}-\ref{fig:AB_sed_dispersions}). While $\textrm{Ri}_{\phi} < \textrm{Ri}_{\phi,\rm{crit}}$ regardless of $\Pi$, the SI is expected to grow faster with increasing $\Pi$, as shown by \citet{YG05} and \citet{Jacquet11} under the terminal velocity approximation. As a result, in cases with $\Pi = 0.05$ and 0.1, the SI plays a dominant role, amplifying velocities at the midplane and leading to the formation of dust voids there. By contrast, when $\Pi = 0.01$ and 0.02, the slower growth of the SI allows the SymI to dominate initially. This leads to faster velocity amplification away from the midplane, resulting in the emergence of dust voids near $\pm H_p$.

However, we note a caveat in our argument above, which arises from the assumptions made in the linear analyses of the SymI and the SI. First, it remains unclear whether the zero-stopping-time approximation is valid during sedimentation, as dust particles with $\tau_s = 0$ are not expected to settle. Moreover, the $\tau_s \to 0$ limit may not be applicable when $\tau_s = 0.1$. Second, the dependence of the SI's linear growth rate on $\Pi$ \citep{YG05,Jacquet11} is derived from unstratified models.

Nonetheless, we emphasize that the dynamics during the sedimentation phase in our simulations are distinct from those in the unstratified case. To illustrate this more clearly and reinforce our finding in Section \ref{sec:results:sedimentation}, Figure \ref{fig:Discussion:sed-rhopux} presents snapshots of the dust density ($\rho_p$; upper panels) and radial gas velocity ($u_x$; lower panels) at $t/T = 2$. At lower $\Pi$ (left two columns), the radial velocity reaches its maximum magnitude away from the midplane, forming radial structures along which dust voids begin to emerge. By contrast, at higher $\Pi$ (right two columns), the dust layer is filled with fully-developed dust filaments, giving rise to a turbulent midplane. Given that vertical gradients of radial and azimuthal velocities have their largest magnitudes near $H_p$ (see Figures \ref{fig:Discussion:sed-dUdz-Ri}-\ref{fig:Discussion:sed-dVdz-Riphi}), the velocity and dust density perturbations near the top of the dust layer in the $\Pi = 0.01$ and 0.02 cases is likely  driven by  mechanism(s) associated with the vertical shear of gas velocities. This finding underscores the need for a detailed examination of the role of shear-driven instabilities in stratified disks.

\subsubsection{Saturated state}\label{sec:Discussion:turb:sat}
\begin{figure}
    \centering
    \includegraphics[width=\columnwidth]{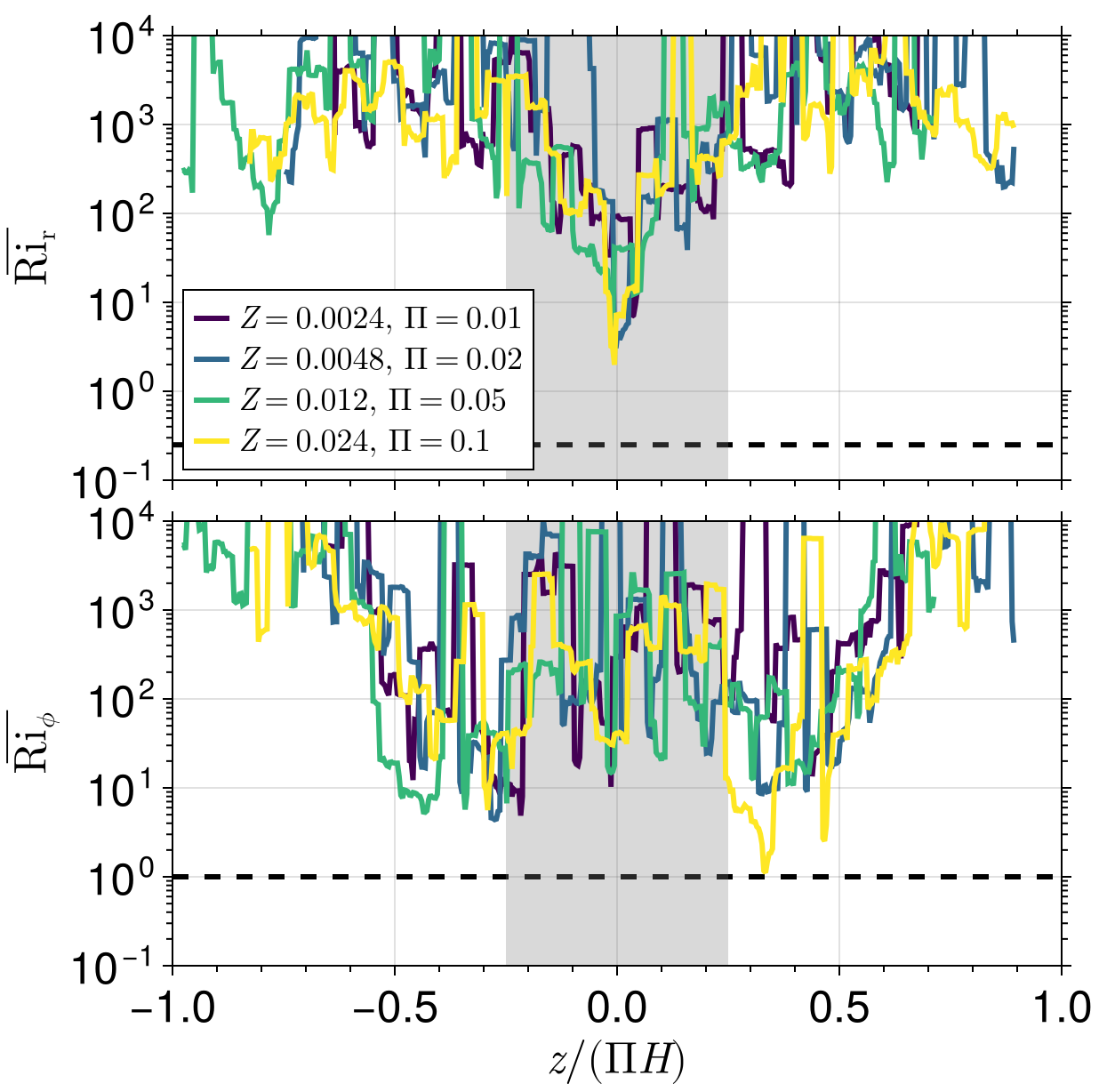}
    \caption{Time-averaged vertical profiles of $\textrm{Ri}_r$ (upper panel, Equation \ref{eq:Ri}) and $\textrm{Ri}_{\phi}$ (lower panel, Equation \ref{eq:Ri_phi}) over the saturated state.  Their critical values are denoted by horizontal lines in each panel. The gray shaded region marks $\pm H_p$.   Regardless of $\Pi$, the Richardson numbers remain well above their corresponding critical values (0.25 for $\textrm{Ri}_r$ and 1.0 for $\textrm{Ri}_{\phi}$).  
    }
    \label{fig:Discussion:sat-Ri}
\end{figure}

In this section, we carry out a similar analysis as was done for the sedimentation phase but for the saturated state. In Figure \ref{fig:Discussion:sat-Ri}, we present time-averaged Richardson numbers as functions of $z$: $\textrm{Ri}_r$ and $\textrm{Ri}_{\phi}$ in the upper and lower panels, respectively. In each panel, the shaded region indicates $\pm H_p$. The horizontal lines denote $\rm{Ri}_{r,\rm{crit}}$ or $\rm{Ri}_{\phi,\rm{crit}}$ depending on the plot.

The figure shows that both $\textrm{Ri}_r$ and $\textrm{Ri}_{\phi}$ are always above their respective critical values within a dust layer, indicating that neither the axisymmetric KHI nor SymI is present in the saturated state. Similarly, \citet{Sengupta_Umurhan23} demonstrated that for $\tau_s = 0.2$ and $\Pi = 0.05$, the system becomes stable against the axisymmetric KHI as it transitions into the saturated state. 

We examine the possibility of the VSSI during the saturated state of our simulations. As mentioned above, the VSSI grows on radial length scales of $\sim 10^{-3}H$ \citep{Lin2021}. Assuming that a mode is properly resolved with at least 16 grid cells, only our lowest $\Pi$ case resolves the $10^{-3}H$ scale, with 25 grid cells.

\begin{figure}
    \centering
    \includegraphics[width=\columnwidth]{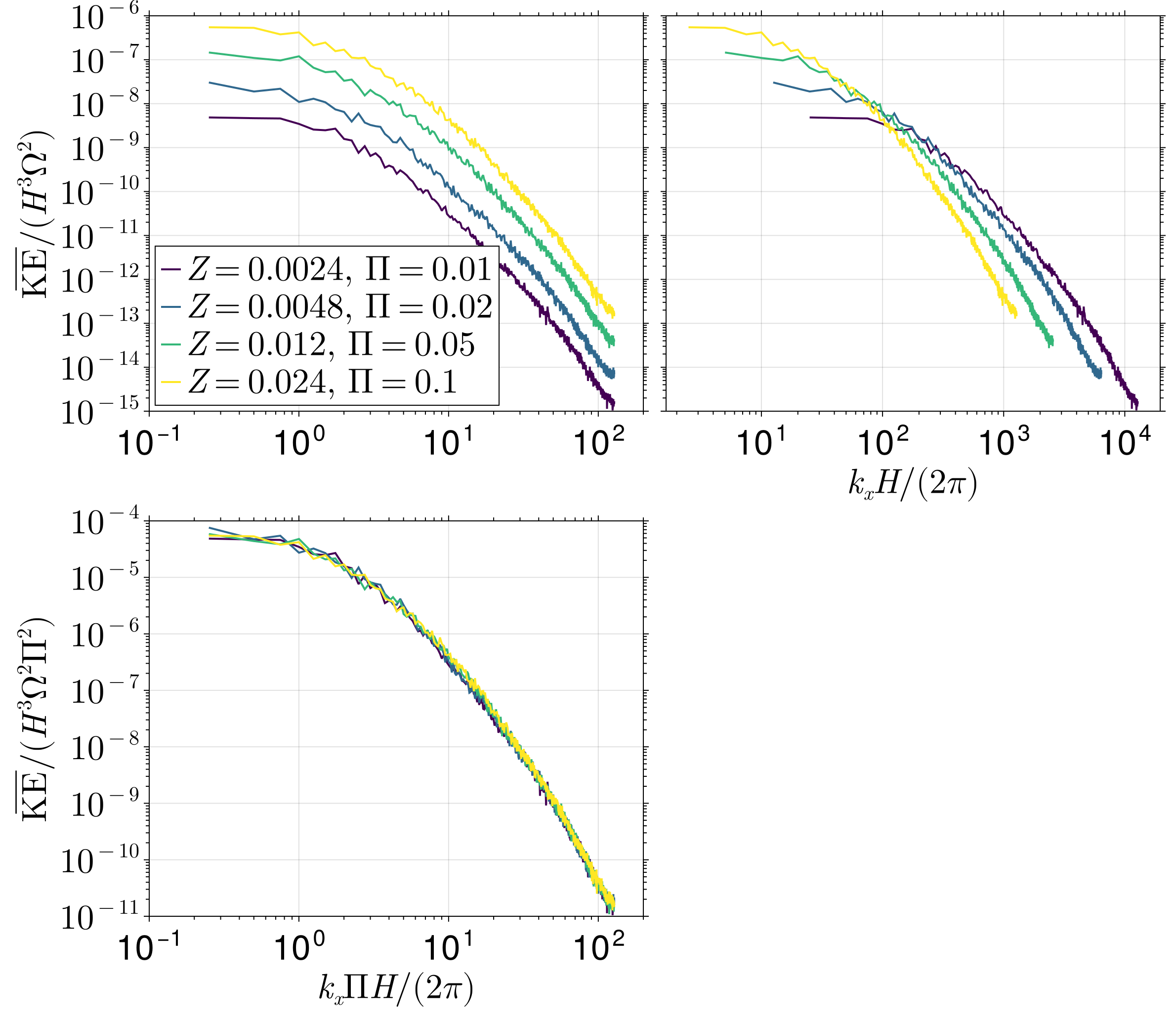}
    \caption{Time-averaged power spectrum of the specific kinetic energy ($\overline{\textrm{KE}}$; Equation \ref{eq:KE}) over the saturated state as a function of radial wave number $k_x$. All panels display the same power spectra but with different scaling on the $x$ and $y$ axes. The left two panels share the same $x$-axis $(k_x \Pi H / (2\pi))$, while in the lower left panel, $\overline{\textrm{KE}}$ is scaled by $\Pi^2$. The upper panels share the same $y$-axis (not scaled by $\Pi^2$), but in the top right panel, $k_x$ is expressed in terms of the gas scale height $H$. The power at any given $k_x\Pi H$ increases with $\Pi$ (upper left) and scales with $\Pi^2$ (lower left). Most of the kinetic energy is concentrated near $k_x \Pi H / (2\pi) \sim 1$, corresponding to a wavelength of $\Pi H$ (lower left). Given that both length and velocity scales are proportional to $\Pi H$, the SI plays a dominant role in the dynamics during the saturated state.
    }
    \label{fig:Discussion:sat-usqk}
\end{figure}

For a more quantitative evaluation, we present the power spectrum of gas kinetic energy density ($\textrm{KE}$) in Figure \ref{fig:Discussion:sat-usqk}. We compute the specific kinetic energy as follows:
\begin{equation}\label{eq:KE}
    \textrm{KE} = \frac{1}{2}\left[(\Delta u_x)^2+(\Delta u_y)^2+(\Delta u_z)^2\right],
\end{equation}
where $\Delta u_i \equiv u_i - \langle u_i \rangle_x$ for $i = x, y, z$.  After computing $\textrm{KE}$, we average it over $-0.5 \leq z/(\Pi H) \leq 0.5$ and apply a Fourier transform to obtain the specific kinetic energy  between $k_x$ and $k_x + dk_x$, where $k_x$ is the radial wave number. This calculation is performed for every snapshot during the saturated state (between $t/T=10$ and 50 with a cadence of $0.1T$), and the results are time-averaged. 

All three panels in Figure \ref{fig:Discussion:sat-usqk} display the same power spectra but with different scaling on the $x$ and $y$ axes. In the top left panel, the power spectra are plotted as a function of $k_x \Pi H / (2\pi)$ [i.e., the wave number in units of ($\Pi H$$)^{-1}$]. Although the power spectra do not exhibit a discernible peak, most of the power is concentrated near $k_x \Pi H / (2\pi) \sim 1$, corresponding to the characteristic scale of the SI ($\sim \Pi H$), which is significantly larger than that of the VSSI. 

In the top right panel, we show the power spectra as a function of $k_xH/(2\pi)$. The length scale of the VSSI corresponds to the wave number of $\sim 10^3$ in this panel. As can be seen, the $\Pi = 0.05$ and $0.1$ cases cannot resolve the characteristic length scale of the VSSI, as the corresponding wave number is too close to the cutoff wave number (i.e., the Nyquist frequency). Even for the lowest two $\Pi$ values, we do not find any indication of VSSI activity near this wave number. This is supported by Figure \ref{fig:AB_sat_vertical_profiles} showing that the velocity dispersions are nearly uniform within $-0.5 < z/(\Pi H) < 0.5$, while the linear perturbation of the VSSI is expected to be amplified near $2H_p$ ($\approx 0.5 \Pi H$ during saturation) in the case of $\tau_s=0.1$ (see Figure 8 in \citealt{Lin2021}). We note, however, that this is not a rigorous evaluation of the VSSI's presence or absence in our simulations. A more quantitative assessment would require performing a pseudo-energy decomposition \citep{Ishitsu2009,Lin2021} to identify which component (e.g., dust--gas coupling or vertical shear) contributes the most to the total energy at a given spatial scale. Such an analysis is beyond the scope of this work.

Finally, we remark on the properties of the power spectra of SI-driven turbulence. First, the two left panels in Figure \ref{fig:Discussion:sat-usqk} illustrate how the velocity scale of the SI varies with $\Pi$. The upper left panel shows that, at any given $k_x$, the power increases with $\Pi$. The bottom left panel further demonstrates that the power spectrum scales as $\Pi^2$, consistent with the scaling relation between velocity dispersion and $\Pi$ (Figure \ref{fig:AB_stdtimeavg}). Second, the upper right panel shows that the entire spectrum shifts toward higher wave numbers as $\Pi$ decreases, and the amount of the shift scales with $\Pi$ (lower left), reinforcing  the SI length scale being approximately $\Pi H$. This provides additional quantitative support for the finding in \citetalias{Baronett24} that dust filaments appear to grow larger as $\Pi$ increases (see their Figures 2 and 6). 

Summarizing the last two subsections, our analysis indicates not only the presence of the SI but also the potential role of shear-driven instabilities during the sedimentation phase. As the system transitions to the saturated state, the SI becomes the dominant mechanism for generating turbulence and governing dust--gas dynamics within the explored parameter space. However, this result should not be considered universal, as \cite{Sengupta_Umurhan23} argue the susceptibility of a dust layer to instabilities depends on dust properties such as $\tau_s$ and $Z$. In fact, these authors demonstrated that when $\tau_s$ and $Z/\Pi$ are chosen such that the SI remains weak, KHI or the SymI can outgrow the SI and become the primary driver of turbulence, with the SI growing within an already turbulent dust layer. The early presence of the SI in our simulations is likely due to $\epsilon$ reaching $\approx 1$ as early as $t/T=1$, enabling the SI to grow rapidly.

%%%%%%%%% Figure %%%%%%%%%%%%
%%%%%%%%% Figure %%%%%%%%%%%%

%%%%%%%%% Figure %%%%%%%%%%%%
%%%%%%%%% Figure %%%%%%%%%%%%

\subsection{Outlooks and Caveats}\label{sec:Discussion:caveat}
%In this work, we have demonstrated a close connection between the nonlinear saturated state of the SI in unstratified simulations and that at the midplane of stratified simulations, across a range of radial pressure gradients. Although stratified simulations exhibit unique features such as vertical settling and shear-driven instabilities, our results not only confirm that the SI in stratified simulations is fundamentally the same as in unstratified simulations, but also suggest that unstratified SI simulations can serve as a reliable predictor of SI-driven dust--gas dynamics at the midplane of stratified disks.

In this work, we have demonstrated a close connection between the nonlinear saturated state of the SI in unstratified simulations and that at the midplane of stratified simulations, across a range of radial pressure gradients. The close connection suggests that while vertical gravity acts to redistribute momentum vertically, it may play a negligible role in dust-gas dynamics driven by the SI around the midplane during the saturated state (but see several important caveats below). This suggests that the SI operates on top of the redistribution, resulting in similar behavior between unstratified models and the midplane of stratified models. In other words, if a sample volume around the midplane were extracted from a stratified simulation in its saturated state and evolved further with vertical gravity turned off, the system would likely remain statistically unchanged. This implies that the key features of SI-driven dynamics, such as velocity dispersions and filamentary structures, are primarily governed by local dynamics near the midplane and are only weakly influenced by global vertical stratification. As a result, although stratified simulations exhibit unique features such as vertical settling, redistribution of momentum, and shear-driven instabilities, our findings not only confirm that the SI in stratified simulations is fundamentally the same as in unstratified simulations, but also suggest that unstratified simulations can serve as a reliable predictor of SI-driven dust--gas dynamics at the midplane of stratified disks.

However, we have considered only a single $\tau_s$ value. The behavior of the SI (e.g., morphology of dust density fields or dust--gas-driven turbulence) is significantly influenced by $\tau_s$ in both unstratified (\citealt{JY07}; \citetalias{Baronett24}) and stratified simulations \citep{Yang2017,LiYoudin21}.  Thus, an open question remains as to whether unstratified simulations can predict stratified SI behavior for both less coupled ($\tau_s \sim 1$) and more strongly coupled ($\tau_s \lesssim 10^{-2}$) regimes. Further study is required to determine whether the strong agreement observed in this study extends across a broader range of $\tau_s$ values.

Despite the close connection between the two models in the saturated state for the parameters considered in this work, our run with $\Pi=0.01$ deviates significantly from the other three runs after $t/T=50$ (see Figure \ref{fig:AB_DmaxHpEps}) and reaches a second saturated state with a maximum dust density of $\approx 300 \rho_{g0}$. This result raises two important considerations.

First, the sharp increase in the maximum dust density, regarded as a prerequisite for planetesimal formation \citep{carrera_how_2015,Yang2017,LiYoudin21,Lim24b}, has been frequently observed in previous stratified simulations. For example, \citet{LiYoudin21} conducted 2D axisymmetric, stratified simulations at $\Pi = 0.05$ across various $\tau_s$ and $Z$ values. They proposed a threshold, $Z_{\rm{crit}}/\Pi$, beyond which the SI can concentrate dust such that their maximum density surpasses the Hill density--approximately $180\rho_{g0}$, assuming a Toomre Q of 32 (\citealt{Toomre1964}). This process, which they termed strong clumping, would directly lead to gravitational collapse into planetesimals.  Although they only studied $\Pi=0.05$, their focus on $Z_{\rm{crit}}/\Pi$ instead of $Z_{\rm{crit}}$ assumes that the SI leads to the same degree of dust clumping so long as $Z/\Pi$ remains constant, even if $Z$ and $\Pi$ change independently \citep{sekiya_two_2018}. However, despite having the same $Z/\Pi$ and exhibiting a similar behavior during the saturated state, only the $\Pi=0.01$ case in our simulations shows strong clumping. Given that strong clumping is favored in larger domain sizes \citep{Yang2017,LiYoudin21,Lim24b} and that our simulations with $\Pi=0.05$ and $L_x/(\Pi H)=8$ show strong clumping (Appendix \ref{sec:appendixA}), this inconsistency may originate from our smaller domain size ($L_x=4\Pi H$) compared to those in previous studies ($L_x=8$–$16\Pi H$).  Nevertheless, our simulation results suggest that the $Z_{\rm{crit}}/\Pi$ criterion may not accurately capture the conditions for the SI to form planetesimals.

Second, at $\tau_s = 0.1$, \citet{LiYoudin21} found in their stratified models that $\epsilon \gtrapprox 0.3$ is required for the SI to drive strong clumping (see their Figure 4). Although unstratified simulations by \citet{Schreiber_Klahr2018} at the same $\tau_s$ value and $\epsilon > 0.3$ show filamentary structures (characteristic of the SI), by contrast, their maximum dust densities appear to remain far below the Hill density criteria for strong clumping defined by \citet{LiYoudin21}. Even for $\epsilon \geq 1$, the maximum dust density in unstratified simulations remains below 100 times the initial gas density (see AB and AC models in \citealt{JY07}; see also AB models in \citetalias{Baronett24}).  \citet{Schreiber_Klahr2018} demonstrated in their radial--vertical 2D models that the unstratified SI can enhance dust concentrations beyond 100 times the initial gas density when the initial density ratio is $\sim 10$. Therefore, the SI appears to require a much higher dust-to-gas density ratio in unstratified cases than in stratified cases to achieve a comparable dust density enhancement. While a systematic comparison between unstratified and stratified simulations is needed to understand this discrepancy, the existing results imply that vertical gravity appears to be a necessary ingredient to trigger strong clumping and planetesimal formation.

Finally, both our simulations and those in \citetalias{Baronett24} consider a single dust size instead of a distribution of grain sizes. Numerical simulations of the SI with a more realistic dust size distribution have been conducted in both unstratified \citep{YangZhu2021,Matthijsse2025} and stratified \citep{BaiStone10b_stratified,Schaffer2018,Schaffer2021,Rucska_Wadsley2023} cases.  While both cases exhibit similar behaviors—such as larger dust species preferentially participating in SI clumping—further studies are required to determine whether unstratified and stratified SI maintain similar properties with a dust size distribution. 

\section{Summary}\label{sec:Summary}
In this work, we investigate the effect of radial pressure gradients (parameterized by $\Pi$; Equation \ref{eq:Pi}) on the sedimentation and nonlinear saturated states of the SI in stratified simulations for a dimensionless stopping time $\tau_s = 0.1$. Additionally, we aim to bridge the nonlinear saturated states of unstratified (without vertical gravity) and stratified SI across various radial pressure gradients.  

To achieve this, we perform 2D axisymmetric (radial-vertical) SI simulations with vertical stratification, considering four different values of $Z$ (dust-to-gas column density ratio; Equation \ref{eq:Z}) and $\Pi$ each (see Table \ref{tab:simlist}), while maintaining a constant ratio of $Z/\Pi$. This parameter choice allows us to maintain a dust-to-gas midplane density ratio of $\approx 1$, comparable to the density ratio in the AB models of \citetalias{Baronett24}. We then systematically compare the results from our stratified simulations to their simulations. We summarize our main results as follows.

\begin{enumerate}
    \item Dust filaments emerge around the midplane as dust settles (Figure \ref{fig:AB_sed_snapshots}) and exhibit morphological similarities to those in unstratified simulations (\citealt{JY07}; \citetalias{Baronett24}). 
    \item In the saturated state of the SI, the midplane values of densities and velocities agree well with those in the unstratified AB models of \citetalias{Baronett24}, except for the radial velocities (Figure~\ref{fig:AB_sat_vertical_profiles_actual_fields})
    \item We find that the COM exhibits non-zero radial motion along the vertical direction (Figure~\ref{fig:AB_sat_vertical_gradients}), an effect absent in unstratified models. This shows that vertical gravity redistributes radial momentum vertically.
    \item By considering the COM frame to account for the effect of vertical gravity, we find that the unstratified and stratified models remain consistent near the midplane (Figure~\ref{fig:AB_sat_velocities_COMframe}).
    \item Within one dust scale height, the dispersions of gas and dust density and velocity are in close agreement with those from the AB unstratified models of \citetalias{Baronett24} (Figure~\ref{fig:AB_sat_vertical_profiles}). Furthermore, except for the dust density, all dispersions increase with increasing $\Pi$ at any given height.
    \item We use the midplane values of the dispersions to quantitatively compare our results with those from \citetalias{Baronett24} and find similar scaling relations between the dispersions and $\Pi$ (Figure~\ref{fig:AB_stdtimeavg}).
    \item While the gas density distributions from our simulations are slightly broader than that in \citetalias{Baronett24}, particularly for $\Pi \geq 0.05$, unstratified and stratified simulations exhibit similar density distributions for both the gas and dust (Figure \ref{fig:AB_cdf}).  
    \item We find that both the radial and vertical diffusion coefficients of dust particles follow a power law of $\propto \Pi^2$. Compared to the unstratified case, vertical diffusion is slightly lower (by a factor of $\approx 2$) in the stratified case, whereas the radial diffusion coefficients remain nearly identical across all $\Pi$ values (Figure \ref{fig:AB_dcoeff}). 
\end{enumerate}
In conclusion, our results demonstrate that for $\tau_s = 0.1$ and a range of radial pressure gradients, dust--gas dynamics driven by the SI exhibit striking similarities between unstratified simulations and near the midplane in stratified simulations. This clearly demonstrates that, apart from extra features that emerge as a result of vertical settling and shear-driven instabilities, the SI in stratified simulations behaves similarly to that in unstratified simulations—at least for the parameters explored here. This similarity holds before the onset of strong clumping, and it remains unclear how the saturated state of the SI determines strong clumping.

%%%%%%%%%%%%%%%%%%%%%%%%%%%%%%%%%%%%%%%%%
% Acknowledgements 
%%%%%%%%%%%%%%%%%%%%%%%%%%%%%%%%%%%%%%%%%
\section*{Acknowledgments} 
J.L., J.B.S, O.M.U, C-C. Y., and W.L. acknowledge support from NASA under the Theoretical and Computational Astrophysical Networks (TCAN) grant \# 80NSSC21K0497.
J.L. and J.B.S also acknowledge support from NASA under Exoplanets Research Program (XRP) Grant \# 80NSSC22K0267 and J.L. acknowledges support from NASA under the Future Investigators in NASA Earth and Space Science and Technology grant \# 80NSSC22K1322. 
SAB acknowledges support from the University of Nevada, Las Vegas, Foundation Board of Trustees Fellowship.
C.C.Y.\ is also grateful for the support from NASA via the Emerging Worlds program (grant \#80NSSC23K0653) and the Astrophysics Theory Program (grant \#80NSSC24K0133).
D.S. was partially supported by grant \# 80NSSC24K1282 from NASA under the Emerging World (EW) program. O.M.U and D.S. acknowledge support from NASA under the Emerging Worlds (EW) program grant \# 80NSSC25K7022. W.L. acknowledges support from NASA under the Emerging Worlds (EW) program grant \#80NSSC22K1419 and NSF via grant AST-2007422. The computations were performed using Stampede3 at the Texas Advanced Computing Center using XSEDE/ACCESS grant TG-AST120062 and on Pleiades at the NASA High-End Computing (HEC) Program through the NASA Advanced Supercomputing (NAS) Division at Ames Research Center. J.L. acknowledges the use of ChatGPT, which has been used on rare occasions to improve the clarity of the writing. The outputs have been checked for their correctness and no original content has been created.

\software{Julia \citep{bezanson2017julia}, Makie.jl \citep{DanischKrumbiegel2021}, Athena \citep{Stone08,stone_implementation_2010,BaiStone10a}, ChatGPT \citep{2023arXiv230308774O}}
          
\appendix
\counterwithin{figure}{section}
\section{Convergence Tests on Domain size and Grid Resolution}\label{sec:appendixA}

\begin{figure*}[h!]
    \includegraphics[width=\textwidth]{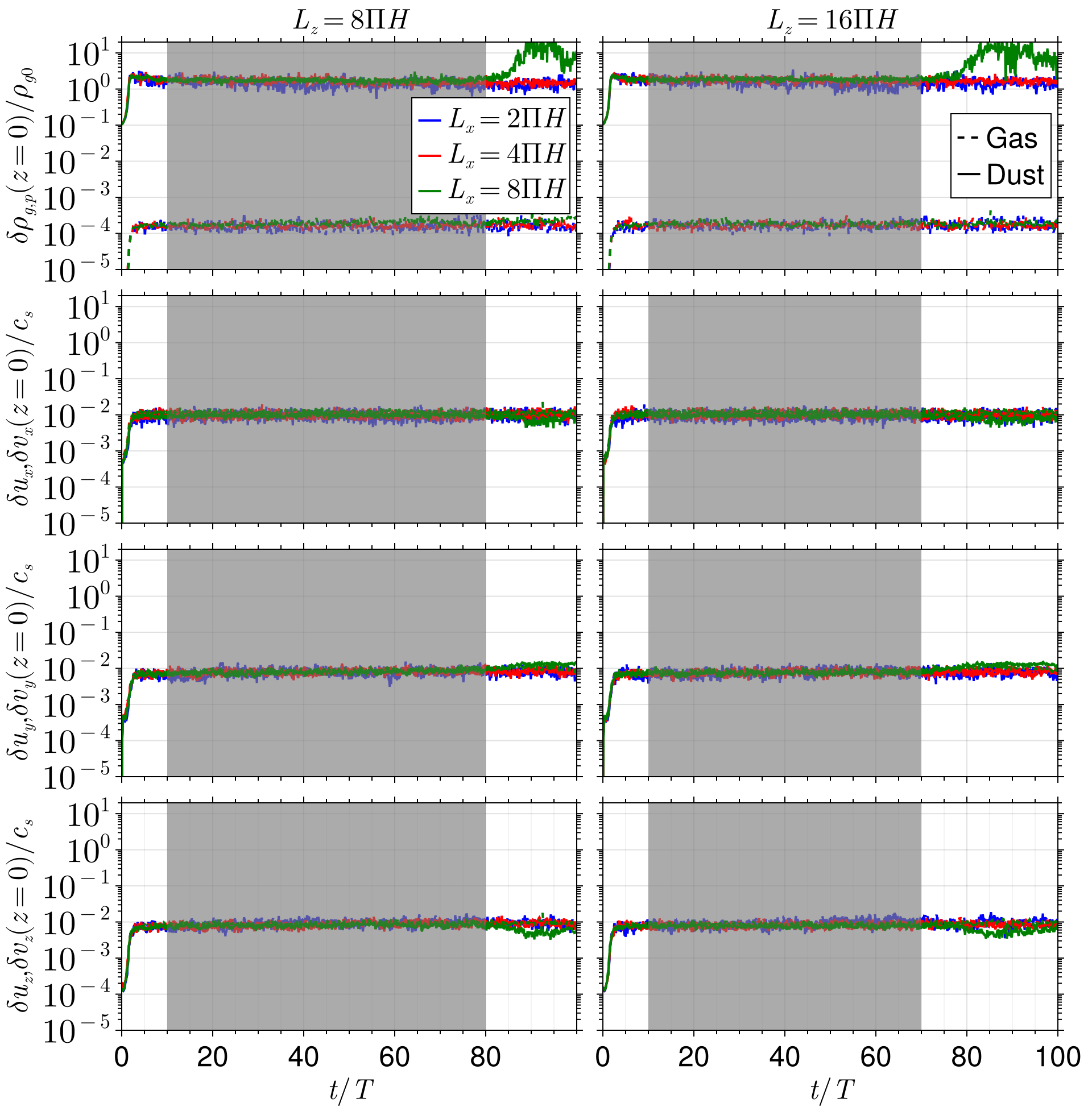}
    \caption{Similar to Figure \ref{fig:AB_stdtimeevol} but for different domain sizes ($L_x$ and $L_z$) at $Z=0.012$ and $\Pi=0.05$ with $N_{\Pi H}=256$. Left and right columns correspond to $L_z=8\Pi H$ and $16\Pi H$, respectively. In order from top to bottom, rows correspond to density, radial-, azimuthal-, and vertical-velocity dispersions. Different colors denote different $L_x$ values, with solid and dashed curves representing the dust and the gas, respectively. The gray shaded regions indicate the saturated state, spanning from $t/T=10$ to 80 for $L_z=8\Pi H$ and from $t/T=10$ to 70 for $L_z=16\Pi H$. While runs with $L_x=8\Pi H$ show a sharp increase in $\sigma_{\rho_p}(z=0)$ during the final $20-30$ orbits depending on $L_z$, the dispersions do not significantly depend on domain size during the saturated state.}
    \label{fig:Boxsize-deviations}
\end{figure*}

\begin{figure}[h!]
    \includegraphics[width=\columnwidth]{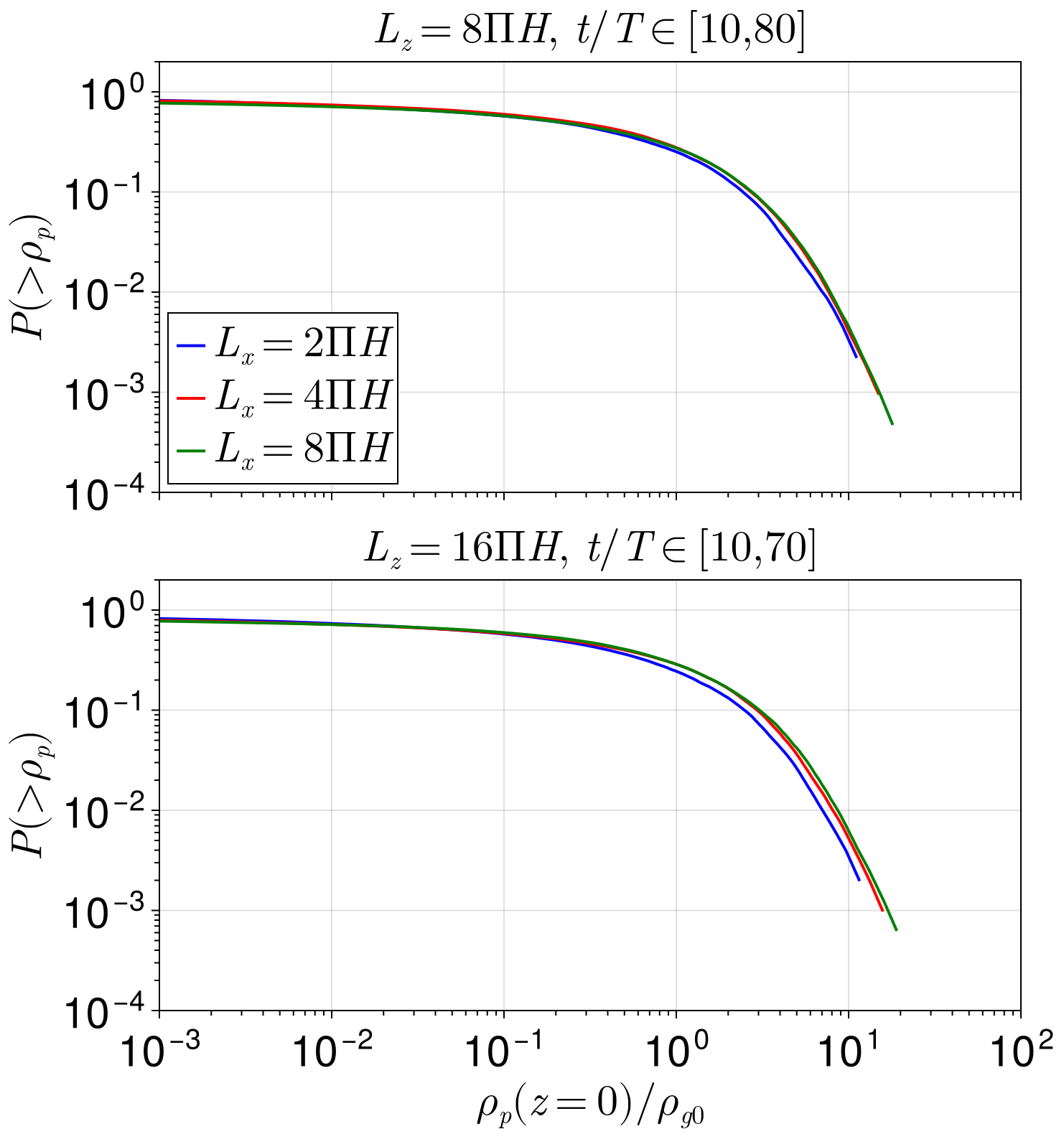}
    \caption{Similar to Figure \ref{fig:AB_cdf} but for different domain sizes ($L_x$ and $L_z$) at $Z=0.012$ and $\Pi=0.05$ with $N_{\Pi H}=256$. Upper and lower panels correspond to $L_z=8\Pi H$ and $16\Pi H$, respectively. Different colors denote different $L_x$ values. The title of each panel indicates the time period over which the distributions are averaged. The comparison shows convergence of the dust density distributions across various domain sizes, except for the maximum densities, which increase with $L_x$.}
    \label{fig:Boxsize-CDF}
\end{figure}

%%%%%%%%% Figure %%%%%%%%%%%%
\begin{figure*}[]
    \centering
    \includegraphics[width=\textwidth]{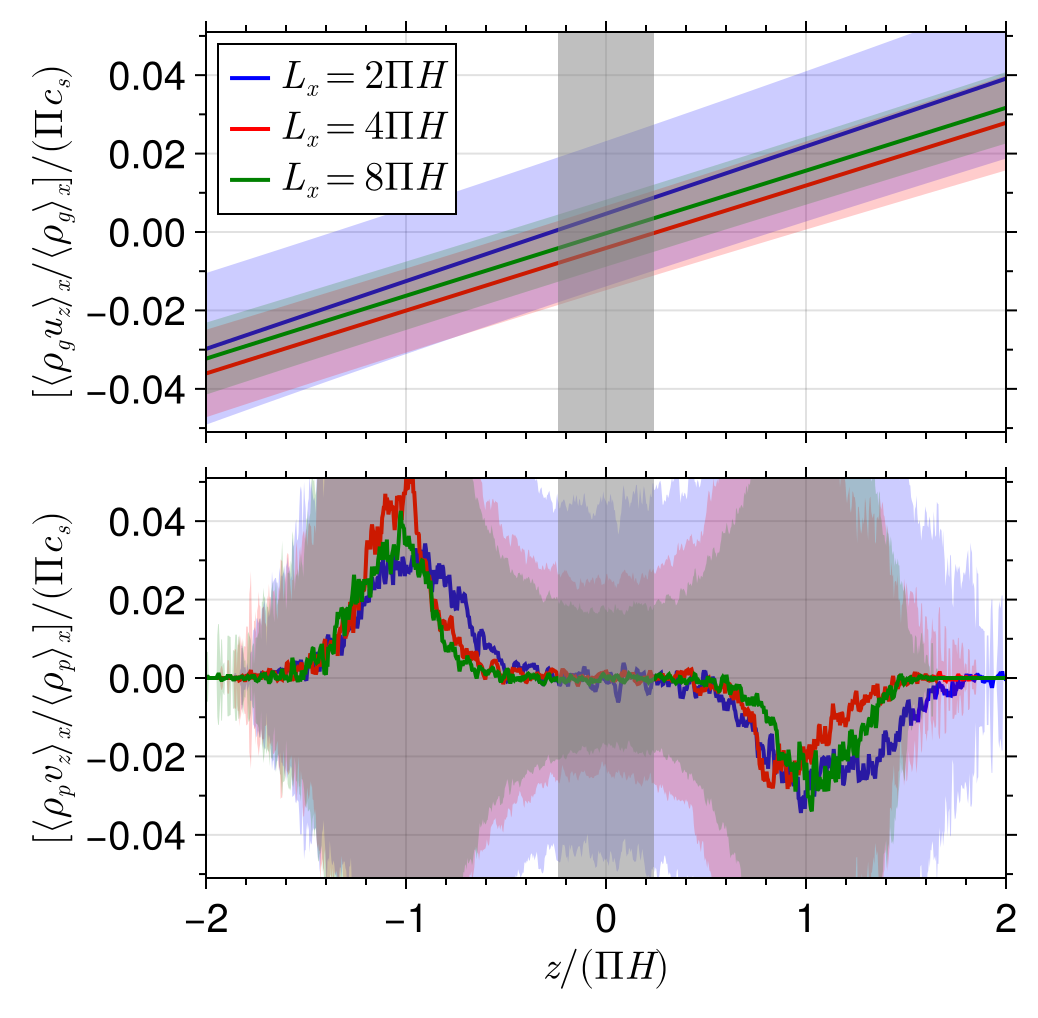}
    \caption{Similar to Figure \ref{fig:AB_sat_vertical_profiles_actual_fields}, but showing only vertical velocities of gas (top) and dust (bottom) for simulations with varying radial domain sizes at fixed $\Pi = 0.05$ and $L_z/(\Pi H)= 8$. Blue, red, and green curves correspond to $L_x/(\Pi H) = 2$, 4 (fiducial), and 8, respectively. All of the runs have time-averaged $H_p$ of $\approx 0.24\Pi H$, which is indicated by the gray vertical band. In the gas, the $L_x/(\Pi H) = 8$ run shows zero vertical velocity at the midplane, while the smaller-domain runs reach zero velocity away from the midplane. All runs exhibit non-zero dust vertical velocity outside of $\approx \pm 2H_p$, with the radial domain size having negligible effect on the dust velocity.
    }
    \label{fig:AppendixB-uzvz-Lx}
\end{figure*}
%%%%%%%%% Figure %%%%%%%%%%%%

%%%%%%%%% Figure %%%%%%%%%%%%
\begin{figure*}
    \centering
    \includegraphics[width=\textwidth]{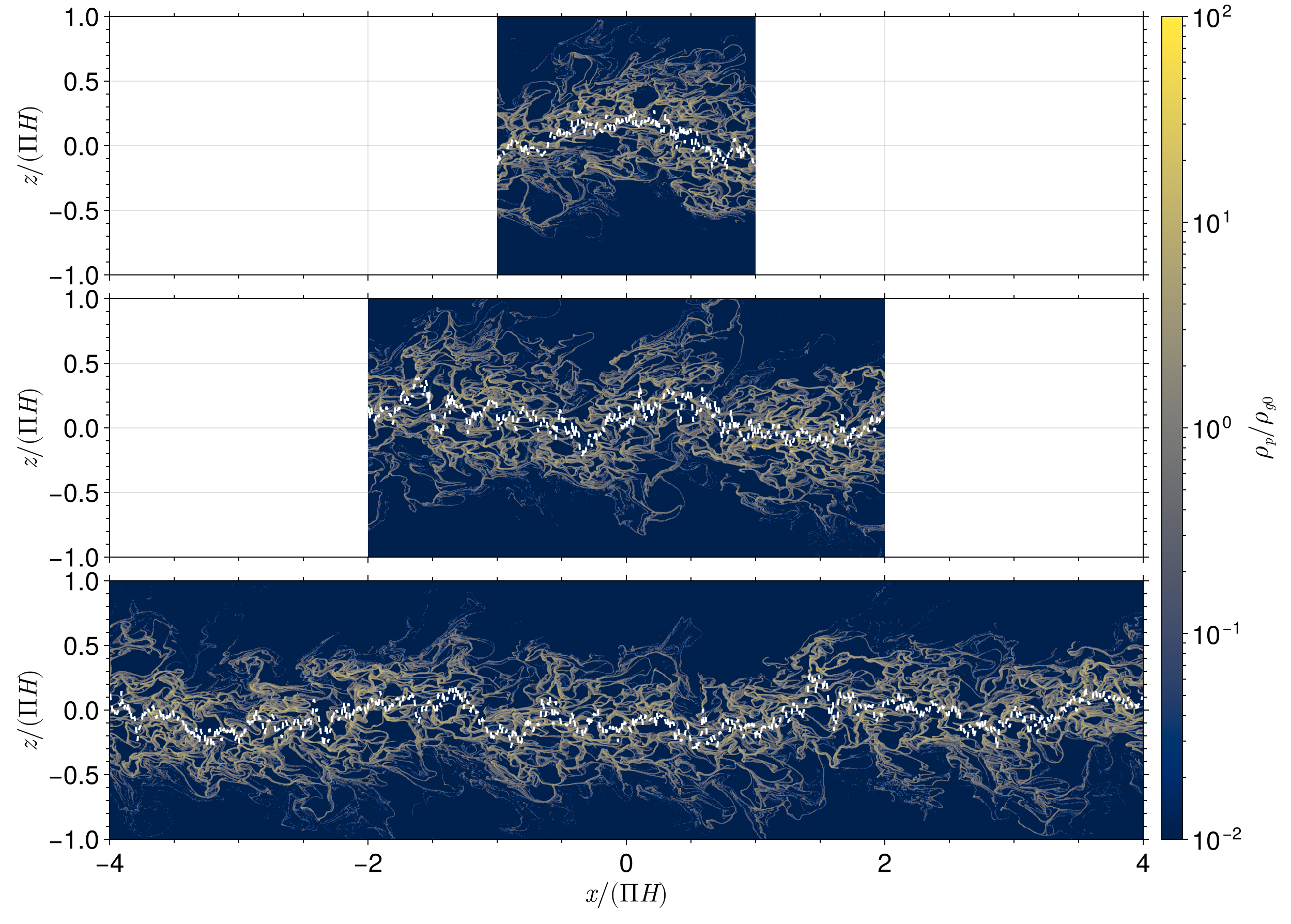}
    \caption{Snapshot of $\rho_p$ at $t/T = 50$ for the simulations shown in Figure \ref{fig:AppendixB-uzvz-Lx}. From top to bottom, panels correspond to $L_x/(\Pi H) = 2$, 4, and 8, respectively. Only the central vertical region ($L_z/(\Pi H) \in [-1, 1]$) is displayed, although the full vertical extent is $8\Pi H$. White dashed curves show the mean vertical position of dust as a function of $x$. As $L_x$ increases, more wavelengths of the undulating dust layer are captured, with $L_x \leq 4\Pi H$ appearing to contain only a single mode or less.
    }
    \label{fig:AppendixB-rhop-snapshot-t50}
\end{figure*}
%%%%%%%%% Figure %%%%%%%%%%%%

\begin{figure*}[h!]
    \includegraphics[width=\textwidth]{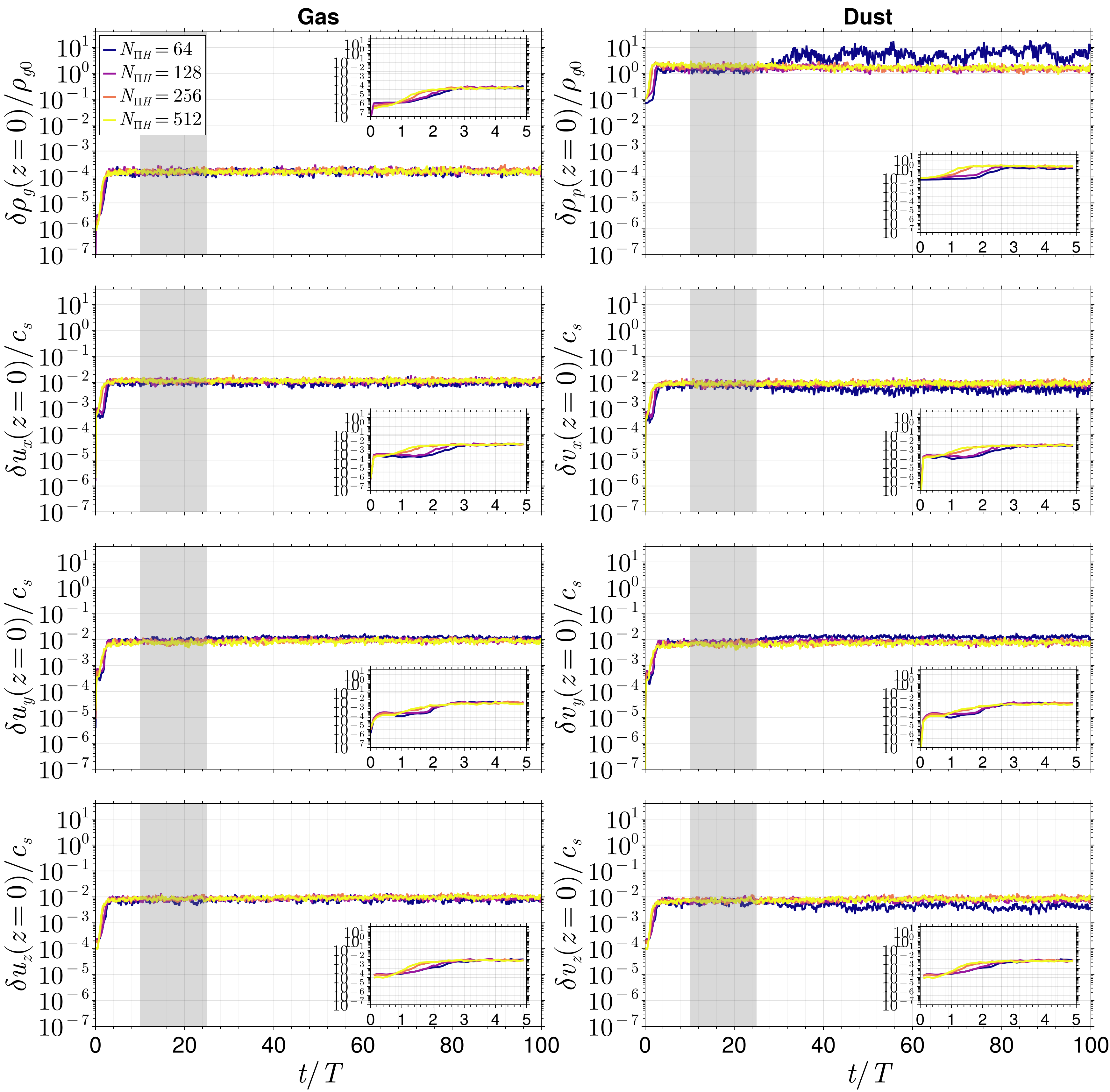}
    \caption{Similar to Figure \ref{fig:AB_stdtimeevol} but for different grid resolutions ($N_{\Pi H}$) at $Z=0.012$ and $\Pi=0.05$. Runs shown here have $L_x=4\Pi H$ and $L_z=8\Pi H$. Different colors denote different $N_{\Pi H}$ values. The gray shaded regions indicate the saturated state, spanning from $t/T=10$ to 25. The dispersions are well converged with grid resolution during the saturated state.}
    \label{fig:resolution-deviations}
\end{figure*}

\begin{figure}[h!]
    \includegraphics[width=\columnwidth]{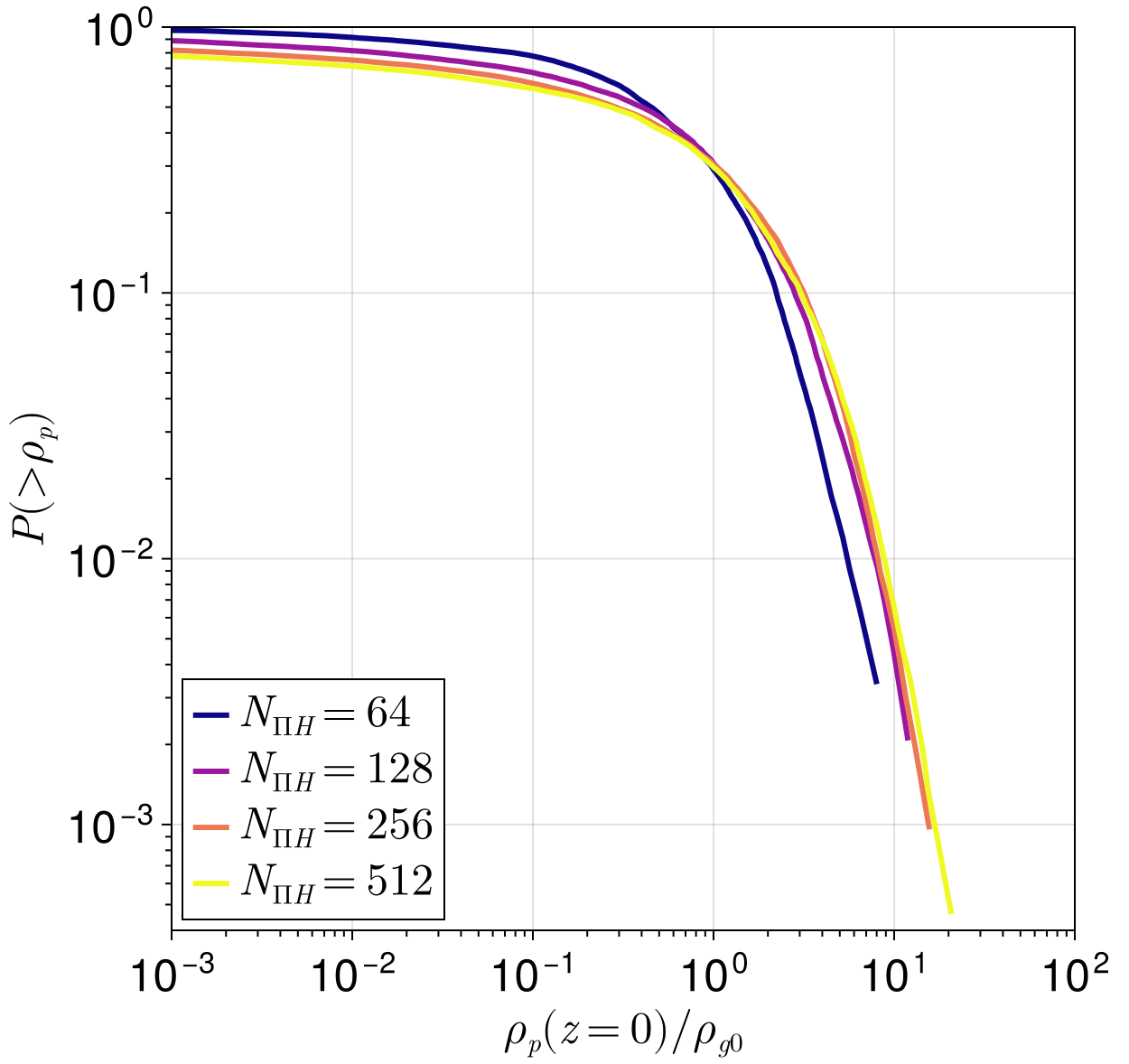}
    \caption{Similar to Figure \ref{fig:AB_cdf} but for different grid resolutions ($N_{\Pi H}$) at $Z=0.012$ and $\Pi=0.05$. In these models, $L_x=4\Pi H$ and $L_z=8\Pi H$. The distributions are time-averaged between $t/T=10$ and 25 (see Figure \ref{fig:resolution-deviations}). When $N_{\Pi H}\gtrsim 128$, the distributions show satisfactory convergence with grid resolution. }
    \label{fig:resolution-CDF}
\end{figure}

Since numerical simulations of the SI are known to be sensitive to domain size and grid resolution \citep{YangJohansen14,Li18}, we perform a convergence test on both. For this test, we maintain $n_p = 1$ and $\Pi = 0.05$ across all simulations presented in this section and perform diagnostics similar to those in Figures \ref{fig:AB_stdtimeevol} and \ref{fig:AB_cdf}. We note that maintaining $n_p = 1$ leads to differences in the total number of particles in simulations with varying domain sizes or grid resolutions.
 
We present the result from the domain size test first. We examine three different $L_x / (\Pi H)$ values $(2,~4,~8)$ at either $L_z / (\Pi H) = 8$ or $16$, using the fiducial grid resolution $(N_{\Pi H}=256)$. In Figure \ref{fig:Boxsize-deviations}, we show the density and velocity dispersions of the gas and dust at the midplane (see Equations \ref{eq:deltarhop},~\ref{eq:deltau},~\ref{eq:deltav} for their definitions) across various domain sizes. Left and right panels correspond to the case of $L_z/(\Pi H)=8$ and 16, respectively. From top to bottom, density, radial, azimuthal, and vertical velocity dispersions are shown. In each panel, solid and dashed curves denote the dust and the gas, respectively. At each $L_z$, we consider $L_x/(\Pi H)$ values of 2 (blue), 4 (red), and 8 (green). We note that when $L_x/(\Pi H)=8$, the dust density dispersions deviate from those for $L_x/(\Pi H)=2$ and 4 by sharply increasing, due to strong clumping, after $t/T \approx 80$ for $L_z/(\Pi H)=8$ and after $t/T \approx 70$ for $L_z/(\Pi H)=16$. Therefore, we determine the saturated state to be $t/T \in [10,80]$ and $t/T \in [10,70]$ for $L_z/(\Pi H)=8$ and 16, respectively, denoted by the gray shaded regions. The dispersions all reach similar values at saturation across different domain sizes. 

We present the time-averaged cumulative distribution functions for the midplane dust density in Figure \ref{fig:Boxsize-CDF}. The top and bottom panels correspond to $L_z/(\Pi H) = 8$ and 16, respectively. Blue, red, and green curves represent $L_x/(\Pi H) = 2$, 4, and 8, respectively. The distribution functions appear to be nearly independent of domain size. The smaller $P(>\rho_p)$ at the maximum $\rho_p$ for larger $L_x$ models is simply due to the greater number of grid cells compared to smaller $L_x$ models. Because of this, the maximum dust density increases with increasing $L_x$ for both $L_z$ values, with a factor of $\approx 2$ difference between $L_x/(\Pi H) = 2$ (blue) and $L_x/(\Pi H) = 8$ (green). This trend may explain the late strong dust clumping observed in Figure \ref{fig:Boxsize-deviations} when $L_x/(\Pi H) = 8$.  The increase in maximum dust density in larger $L_x$ is consistent with previous studies showing that SI-induced clumping tends to be stronger in larger radial domain sizes \citep{YangJohansen14,Yang2017,Li18}.

Despite the satisfactory convergence of the dispersions and dust density distribution, we find that $L_x$ slightly affects the vertical profile of gas vertical velocity. To show this, Figure \ref{fig:AppendixB-uzvz-Lx} presents density-weighted vertical velocities of gas (top) and dust (bottom) as functions of $z$ for different $L_x$ values but with the same $L_z$ of $8\Pi H$. In the gas profiles, the largest $L_x$ case shows the vertical velocity crossing zero at the midplane, unlike our fiducial case of $L_x=4\Pi H$ (see Figure~\ref{fig:AB_sat_vertical_profiles_actual_fields}) and the smallest $L_x$ case. For dust, the radial domain size has relatively negligible influence, with the different $L_x$ values yielding similar profiles of the velocity. 

The dependence of the gas vertical velocity on $L_x$ arises from the fact that the larger $L_x$ captures more wavelengths of the undulating dust layer. To illustrate this, Figure \ref{fig:AppendixB-rhop-snapshot-t50} presents snapshots of $\rho_p$ at $t/T=50$ for the three different $L_x$ values. In each panel, the white dashed curves show the mean vertical position of dust as functions of $x$, calculated by $\int \rho_p zdz/\int \rho_p dz$. As the figure shows, a larger radial domain accommodates more radial modes of longer wavelengths. Fewer modes being captured within the radial domain can cause the mean vertical dust position to deviate from zero. Indeed, we find that time-averaged mean vertical position of particles ($\overline{z_p}$) over $t/T = 10$–50 is close to zero for $L_x = 8\Pi H$ ($\approx 8 \times 10^{-4} \Pi H$), but shifts to $\approx 0.03\Pi H$ and $-0.02\Pi H$ for $L_x = 2\Pi H$ and $4\Pi H$, respectively. These results indicate that $L_x > 4\Pi H$ is required to adequately resolve the vertical undulations of the dust layer for $\tau_s = 0.1$, consistent with previous studies \citep{YangJohansen14,Li18}.

While the radial domain length ($L_x$) influences gas and dust vertical velocities both near and away from the midplane, it is unlikely to affect our comparison with \citetalias{Baronett24}. This is because we find satisfactory convergence in dispersions of densities and velocities and dust density distributions across different $L_x$ and $L_z$ values (Figure~\ref{fig:Boxsize-deviations}). Moreover, although dust particles exhibit non-zero vertical velocity beyond $\approx 2H_p$ (Figure~ \ref{fig:AB_sat_vertical_profiles_actual_fields}), suggesting that their vertical motion is not in equilibrium in those regions, this likely has minimal impact on our midplane-focused analysis, as only a few percent of particles reside beyond $2H_p$. Also, we caution that such small number of particles are insufficient to represent the altitude where the magnitude of $v_z$ begins to increase from zero, and those at such heights may be sedimenting back toward the midplane without being balanced by diffusion.

We now examine the resolution convergence at $Z = 0.012$ and $\Pi = 0.05$. For this test, we explore four different grid resolutions: $N_{\Pi H} = 64,~128,~256,~512$, while keeping the fiducial domain size $(L_x, L_z) = (4,8)\Pi H$ and the number of particles per cell $(n_p=1)$. Figure \ref{fig:resolution-deviations} depicts the density and velocity dispersions at the midplane, similar to Figure \ref{fig:AB_stdtimeevol}. The left and right panels correspond to the dispersions of the gas and dust, respectively. From top to bottom, each panel shows density and the three velocity components.  In each panel, lighter colors represent higher grid resolution runs. The gray-shaded region denotes the saturated state for the runs shown in the figure. The saturated state ends at $t/T = 25$ due to a sharp increase in dust density dispersion in the lowest-resolution case beyond this point. We find that the dispersions are converged over the saturated state. In addition, models with higher resolutions show faster growth of the dispersions before reaching the saturation (see the insets). The same behavior was also seen by \citetalias{Baronett24} (see their Figure A1). 

We present the time-averaged cumulative distribution function for the midplane dust density across different grid resolutions in Figure \ref{fig:resolution-CDF}. Except for the lowest-resolution case, the distributions for each resolution nearly overlap. The higher maximum dust density at higher grid resolutions, previously reported by \citealt{BaiStone10a} and \citetalias{Baronett24}, is attributed to the ability of higher grid resolutions to resolve smaller regions of the computational domain. However, we note that the maximum density is not a reliable indicator of convergence, as it is determined by only a tiny fraction of grid cells. Overall, based on the overlapping distributions, we conclude that the resolution convergence occurs for $N_{\Pi H}\gtrsim 128$ for our parameter choices. 

In summary, we find satisfactory convergence in both domain size and grid resolution during the saturated state of the SI, provided that $N_{\Pi H} \gtrsim 128$. Therefore, the analysis conducted with our fiducial domain size ($L_x=4\Pi H$ and $L_z=8\Pi H$) and grid resolution ($N_{\Pi H} = 256$) should be representative of the saturated state.

\bibliography{main}{}

\begin{thebibliography}{}
\expandafter\ifx\csname natexlab\endcsname\relax\def\natexlab#1{#1}\fi
\providecommand{\url}[1]{\href{#1}{#1}}
\providecommand{\dodoi}[1]{doi:~\href{http://doi.org/#1}{\nolinkurl{#1}}}
\providecommand{\doeprint}[1]{\href{http://ascl.net/#1}{\nolinkurl{http://ascl.net/#1}}}
\providecommand{\doarXiv}[1]{\href{https://arxiv.org/abs/#1}{\nolinkurl{https://arxiv.org/abs/#1}}}

\bibitem[{C.~P. Abod {et~al.}(2019)Abod, Simon, Li, Armitage, Youdin, \& Kretke}]{abod_mass_2019}
Abod, C.~P., Simon, J.~B., Li, R., {et~al.} 2019, \bibinfo{title}{The {Mass} and {Size} {Distribution} of {Planetesimals} {Formed} by the {Streaming} {Instability}. {II}. {The} {Effect} of the {Radial} {Gas} {Pressure} {Gradient},} The Astrophysical Journal, 883, 192, \dodoi{10.3847/1538-4357/ab40a3}

\bibitem[{ {ALMA Partnership} {et~al.}(2015){ALMA Partnership}, {Brogan}, {P{\'e}rez}, {Hunter}, {Dent}, {Hales}, {Hills}, {Corder}, {Fomalont}, {Vlahakis}, {Asaki}, {Barkats}, {Hirota}, {Hodge}, {Impellizzeri}, {Kneissl}, {Liuzzo}, {Lucas}, {Marcelino}, {Matsushita}, {Nakanishi}, {Phillips}, {Richards}, {Toledo}, {Aladro}, {Broguiere}, {Cortes}, {Cortes}, {Espada}, {Galarza}, {Garcia-Appadoo}, {Guzman-Ramirez}, {Humphreys}, {Jung}, {Kameno}, {Laing}, {Leon}, {Marconi}, {Mignano}, {Nikolic}, {Nyman}, {Radiszcz}, {Remijan}, {Rod{\'o}n}, {Sawada}, {Takahashi}, {Tilanus}, {Vila Vilaro}, {Watson}, {Wiklind}, {Akiyama}, {Chapillon}, {de Gregorio-Monsalvo}, {Di Francesco}, {Gueth}, {Kawamura}, {Lee}, {Nguyen Luong}, {Mangum}, {Pietu}, {Sanhueza}, {Saigo}, {Takakuwa}, {Ubach}, {van Kempen}, {Wootten}, {Castro-Carrizo}, {Francke}, {Gallardo}, {Garcia}, {Gonzalez}, {Hill}, {Kaminski}, {Kurono}, {Liu}, {Lopez}, {Morales}, {Plarre}, {Schieven}, {Testi}, {Videla}, {Villard}, {Andreani}, {Hibbard}, \&
  {Tatematsu}}]{ALMA2015}
{ALMA Partnership}, {Brogan}, C.~L., {P{\'e}rez}, L.~M., {et~al.} 2015, \bibinfo{title}{{The 2014 ALMA Long Baseline Campaign: First Results from High Angular Resolution Observations toward the HL Tau Region},} \apjl, 808, L3, \dodoi{10.1088/2041-8205/808/1/L3}

\bibitem[{S.~M. Andrews {et~al.}(2018)Andrews, Huang, Pérez, Isella, Dullemond, Kurtovic, Guzmán, Carpenter, Wilner, Zhang, Zhu, Birnstiel, Bai, Benisty, Hughes, Öberg, \& Ricci}]{andrews_disk_2018}
Andrews, S.~M., Huang, J., Pérez, L.~M., {et~al.} 2018, \bibinfo{title}{The {Disk} {Substructures} at {High} {Angular} {Resolution} {Project} ({DSHARP}). {I}. {Motivation}, {Sample}, {Calibration}, and {Overview},} The Astrophysical Journal, 869, L41, \dodoi{10.3847/2041-8213/aaf741}

\bibitem[{B.~A. {Ayliffe} \& M.~R. {Bate}(2012){Ayliffe} \& {Bate}}]{AyliffeBate2012}
{Ayliffe}, B.~A., \& {Bate}, M.~R. 2012, \bibinfo{title}{{The growth and hydrodynamic collapse of a protoplanet envelope},} \mnras, 427, 2597, \dodoi{10.1111/j.1365-2966.2012.21979.x}

\bibitem[{X.-N. {Bai} \& J.~M. {Stone}(2010{\natexlab{a}}){Bai} \& {Stone}}]{BaiStone10c_pressure_gradient}
{Bai}, X.-N., \& {Stone}, J.~M. 2010{\natexlab{a}}, \bibinfo{title}{{The Effect of the Radial Pressure Gradient in Protoplanetary Disks on Planetesimal Formation},} \apjl, 722, L220, \dodoi{10.1088/2041-8205/722/2/L220}

\bibitem[{X.-N. {Bai} \& J.~M. {Stone}(2010{\natexlab{b}}){Bai} \& {Stone}}]{BaiStone10a}
{Bai}, X.-N., \& {Stone}, J.~M. 2010{\natexlab{b}}, \bibinfo{title}{{Particle-gas Dynamics with Athena: Method and Convergence},} \apjs, 190, 297, \dodoi{10.1088/0067-0049/190/2/297}

\bibitem[{X.-N. {Bai} \& J.~M. {Stone}(2010{\natexlab{c}}){Bai} \& {Stone}}]{BaiStone10b_stratified}
{Bai}, X.-N., \& {Stone}, J.~M. 2010{\natexlab{c}}, \bibinfo{title}{{Dynamics of Solids in the Midplane of Protoplanetary Disks: Implications for Planetesimal Formation},} \apj, 722, 1437, \dodoi{10.1088/0004-637X/722/2/1437}

\bibitem[{S.~A. {Baronett} {et~al.}(2024){Baronett}, {Yang}, \& {Zhu}}]{Baronett24}
{Baronett}, S.~A., {Yang}, C.-C., \& {Zhu}, Z. 2024, \bibinfo{title}{{Dust-gas dynamics driven by the streaming instability with various pressure gradients},} \mnras, 529, 275, \dodoi{10.1093/mnras/stae272}

\bibitem[{J. Bezanson {et~al.}(2017)Bezanson, Edelman, Karpinski, \& Shah}]{bezanson2017julia}
Bezanson, J., Edelman, A., Karpinski, S., \& Shah, V.~B. 2017, \bibinfo{title}{Julia: A fresh approach to numerical computing,} SIAM review, 59, 65.
\newblock \url{https://doi.org/10.1137/141000671}

\bibitem[{B. {Bitsch} {et~al.}(2015){Bitsch}, {Lambrechts}, \& {Johansen}}]{Bitsch+15}
{Bitsch}, B., {Lambrechts}, M., \& {Johansen}, A. 2015, \bibinfo{title}{{The growth of planets by pebble accretion in evolving protoplanetary discs},} \aap, 582, A112, \dodoi{10.1051/0004-6361/201526463}

\bibitem[{J. Blum \& G. Wurm(2008)Blum \& Wurm}]{Blum08}
Blum, J., \& Wurm, G. 2008, \bibinfo{title}{The growth mechanisms of macroscopic bodies in protoplanetary disks,} Annual Review of Astronomy and Astrophysics, 46, 21, \dodoi{10.1146/annurev.astro.46.060407.145152}

\bibitem[{A. {Carballido} {et~al.}(2006){Carballido}, {Fromang}, \& {Papaloizou}}]{Carballido06}
{Carballido}, A., {Fromang}, S., \& {Papaloizou}, J. 2006, \bibinfo{title}{{Mid-plane sedimentation of large solid bodies in turbulent protoplanetary discs},} \mnras, 373, 1633, \dodoi{10.1111/j.1365-2966.2006.11118.x}

\bibitem[{D. Carrera {et~al.}(2015)Carrera, Johansen, \& Davies}]{carrera_how_2015}
Carrera, D., Johansen, A., \& Davies, M.~B. 2015, \bibinfo{title}{How to form planetesimals from mm-sized chondrules and chondrule aggregates,} Astronomy and Astrophysics, 579, \dodoi{10.1051/0004-6361/201425120}

\bibitem[{D. Carrera \& J.~B. Simon(2022)Carrera \& Simon}]{Carrera2022}
Carrera, D., \& Simon, J.~B. 2022, \bibinfo{title}{The Streaming Instability Cannot Form Planetesimals from Millimeter-size Grains in Pressure Bumps,} The Astrophysical Journal Letters, 933, L10, \dodoi{10.3847/2041-8213/ac6b3e}

\bibitem[{D. Carrera {et~al.}(2021)Carrera, Simon, Li, Kretke, \& Klahr}]{Carrera2021}
Carrera, D., Simon, J.~B., Li, R., Kretke, K.~A., \& Klahr, H. 2021, \bibinfo{title}{Protoplanetary Disk Rings as Sites for Planetesimal Formation,} The Astronomical Journal, 161, 96, \dodoi{10.3847/1538-3881/abd4d9}

\bibitem[{S. {Chandrasekhar}(1961){Chandrasekhar}}]{Chandrasekhar1961}
{Chandrasekhar}, S. 1961, {Hydrodynamic and hydromagnetic stability}

\bibitem[{J.-W. {Chen} \& M.-K. {Lin}(2018){Chen} \& {Lin}}]{ChenLin2018}
{Chen}, J.-W., \& {Lin}, M.-K. 2018, \bibinfo{title}{{Dusty disc-planet interaction with dust-free simulations},} \mnras, 478, 2737, \dodoi{10.1093/mnras/sty1166}

\bibitem[{E. {Chiang}(2008){Chiang}}]{Chiang2008ApJ}
{Chiang}, E. 2008, \bibinfo{title}{{Vertical Shearing Instabilities in Radially Shearing Disks: The Dustiest Layers of the Protoplanetary Nebula},} \apj, 675, 1549, \dodoi{10.1086/527354}

\bibitem[{S. Danisch \& J. Krumbiegel(2021)Danisch \& Krumbiegel}]{DanischKrumbiegel2021}
Danisch, S., \& Krumbiegel, J. 2021, \bibinfo{title}{{Makie.jl}: Flexible high-performance data visualization for {Julia},} Journal of Open Source Software, 6, 3349, \dodoi{10.21105/joss.03349}

\bibitem[{K. Gerbig \& R. Li(2023)Gerbig \& Li}]{gerbig_planetesimal_2023}
Gerbig, K., \& Li, R. 2023, \bibinfo{title}{Planetesimal {Initial} {Mass} {Functions} {Following} {Diffusion}-regulated {Gravitational} {Collapse},} The Astrophysical Journal, 949, 81, \dodoi{10.3847/1538-4357/acca1a}

\bibitem[{K. Gerbig {et~al.}(2020)Gerbig, Murray-Clay, Klahr, \& Baehr}]{Gerbig20}
Gerbig, K., Murray-Clay, R.~A., Klahr, H., \& Baehr, H. 2020, \bibinfo{title}{Requirements for {Gravitational} {Collapse} in {Planetesimal} {Formation}—{The} {Impact} of {Scales} {Set} by {Kelvin}–{Helmholtz} and {Nonlinear} {Streaming} {Instability},} The Astrophysical Journal, 895, 91, \dodoi{10.3847/1538-4357/ab8d37}

\bibitem[{P. {Goldreich} \& D. {Lynden-Bell}(1965){Goldreich} \& {Lynden-Bell}}]{GoldreichLynden-Bell1965}
{Goldreich}, P., \& {Lynden-Bell}, D. 1965, \bibinfo{title}{{II. Spiral arms as sheared gravitational instabilities},} \mnras, 130, 125, \dodoi{10.1093/mnras/130.2.125}

\bibitem[{D.~A. Gole {et~al.}(2020)Gole, Simon, Li, Youdin, \& Armitage}]{Gole20}
Gole, D.~A., Simon, J.~B., Li, R., Youdin, A.~N., \& Armitage, P.~J. 2020, \bibinfo{title}{Turbulence {Regulates} the {Rate} of {Planetesimal} {Formation} via {Gravitational} {Collapse},} The Astrophysical Journal, 904, 132, \dodoi{10.3847/1538-4357/abc334}

\bibitem[{R. {Greenberg} {et~al.}(1978){Greenberg}, {Wacker}, {Hartmann}, \& {Chapman}}]{Greenberg1978}
{Greenberg}, R., {Wacker}, J.~F., {Hartmann}, W.~K., \& {Chapman}, C.~R. 1978, \bibinfo{title}{{Planetesimals to planets: Numerical simulation of collisional evolution},} \icarus, 35, 1, \dodoi{10.1016/0019-1035(78)90057-X}

\bibitem[{C. Güttler {et~al.}(2010)Güttler, Blum, Zsom, Ormel, \& Dullemond}]{Guttler10}
Güttler, C., Blum, J., Zsom, A., Ormel, C.~W., \& Dullemond, C.~P. 2010, \bibinfo{title}{The outcome of protoplanetary dust growth: pebbles, boulders, or planetesimals?: {II}. {Introducing} the bouncing barrier,} Astronomy and Astrophysics, 513, \dodoi{10.1051/0004-6361/200912852}

\bibitem[{J.~F. Hawley {et~al.}(1995)Hawley, Gammie, \& Balbus}]{Hawley95}
Hawley, J.~F., Gammie, C.~F., \& Balbus, S.~A. 1995, \bibinfo{title}{Hawley+ 1995,} The Astrophysical Journal, 440, 742, \dodoi{10.1086/175311}

\bibitem[{L.~N. {Howard}(1961){Howard}}]{Howard1961}
{Howard}, L.~N. 1961, \bibinfo{title}{{Note on a paper of John W. Miles},} Journal of Fluid Mechanics, 10, 509, \dodoi{10.1017/S0022112061000317}

\bibitem[{N. {Ishitsu} {et~al.}(2009){Ishitsu}, {Inutsuka}, \& {Sekiya}}]{Ishitsu2009}
{Ishitsu}, N., {Inutsuka}, S.-i., \& {Sekiya}, M. 2009, \bibinfo{title}{{Two-fluid Instability of Dust and Gas in the Dust Layer of a Protoplanetary Disk},} arXiv e-prints, arXiv:0905.4404, \dodoi{10.48550/arXiv.0905.4404}

\bibitem[{E. {Jacquet} {et~al.}(2011){Jacquet}, {Balbus}, \& {Latter}}]{Jacquet11}
{Jacquet}, E., {Balbus}, S., \& {Latter}, H. 2011, \bibinfo{title}{{On linear dust-gas streaming instabilities in protoplanetary discs},} \mnras, 415, 3591, \dodoi{10.1111/j.1365-2966.2011.18971.x}

\bibitem[{A. {Johansen} {et~al.}(2006){Johansen}, {Klahr}, \& {Henning}}]{Johansen06}
{Johansen}, A., {Klahr}, H., \& {Henning}, T. 2006, \bibinfo{title}{{Gravoturbulent Formation of Planetesimals},} \apj, 636, 1121, \dodoi{10.1086/498078}

\bibitem[{A. {Johansen} {et~al.}(2015){Johansen}, {Mac Low}, {Lacerda}, \& {Bizzarro}}]{Johansen15}
{Johansen}, A., {Mac Low}, M.-M., {Lacerda}, P., \& {Bizzarro}, M. 2015, \bibinfo{title}{{Growth of asteroids, planetary embryos, and Kuiper belt objects by chondrule accretion},} Science Advances, 1, 1500109, \dodoi{10.1126/sciadv.1500109}

\bibitem[{A. Johansen {et~al.}(2007)Johansen, Oishi, Low, Klahr, Henning, \& Youdin}]{johansen_rapid_2007}
Johansen, A., Oishi, J.~S., Low, M. M.~M., {et~al.} 2007, \bibinfo{title}{Rapid planetesimal formation in turbulent circumstellar disks,} Nature, 448, 1022, \dodoi{10.1038/nature06086}

\bibitem[{A. Johansen \& A. Youdin(2007)Johansen \& Youdin}]{JY07}
Johansen, A., \& Youdin, A. 2007, \bibinfo{title}{{PROTOPLANETARY} {DISK} {TURBULENCE} {DRIVEN} {BY} {THE} {STREAMING} {INSTABILITY}: {NONLINEAR} {SATURATION} {AND} {PARTICLE} {CONCENTRATION},} The Astrophysical Journal, 662

\bibitem[{A. Johansen {et~al.}(2009)Johansen, Youdin, \& Mac~Low}]{Johansen09b}
Johansen, A., Youdin, A., \& Mac~Low, M.~M. 2009, \bibinfo{title}{Particle clumping and planetesimal formation depend strongly on metallicity,} Astrophysical Journal, 704, \dodoi{10.1088/0004-637X/704/2/L75}

\bibitem[{H. Klahr \& A. Schreiber(2020)Klahr \& Schreiber}]{klahr_turbulence_2020}
Klahr, H., \& Schreiber, A. 2020, \bibinfo{title}{Turbulence {Sets} the {Length} {Scale} for {Planetesimal} {Formation}: {Local} {2D} {Simulations} of {Streaming} {Instability} and {Planetesimal} {Formation},} The Astrophysical Journal, 901, 54, \dodoi{10.3847/1538-4357/abac58}

\bibitem[{E. {Kokubo} \& S. {Ida}(1998){Kokubo} \& {Ida}}]{KokuboIda1998}
{Kokubo}, E., \& {Ida}, S. 1998, \bibinfo{title}{{Oligarchic Growth of Protoplanets},} \icarus, 131, 171, \dodoi{10.1006/icar.1997.5840}

\bibitem[{G. {Laibe} \& D.~J. {Price}(2014){Laibe} \& {Price}}]{LaibePrice2014}
{Laibe}, G., \& {Price}, D.~J. 2014, \bibinfo{title}{{Dusty gas with one fluid},} \mnras, 440, 2136, \dodoi{10.1093/mnras/stu355}

\bibitem[{M. {Lambrechts} \& A. {Johansen}(2012){Lambrechts} \& {Johansen}}]{LambrechtsJohansen2012}
{Lambrechts}, M., \& {Johansen}, A. 2012, \bibinfo{title}{{Rapid growth of gas-giant cores by pebble accretion},} \aap, 544, A32, \dodoi{10.1051/0004-6361/201219127}

\bibitem[{C.~J. {Law} {et~al.}(2021){Law}, {Teague}, {Loomis}, {Bae}, {{\"O}berg}, {Czekala}, {Andrews}, {Aikawa}, {Alarc{\'o}n}, {Bergin}, {Bergner}, {Booth}, {Bosman}, {Calahan}, {Cataldi}, {Cleeves}, {Furuya}, {Guzm{\'a}n}, {Huang}, {Ilee}, {Le Gal}, {Liu}, {Long}, {M{\'e}nard}, {Nomura}, {P{\'e}rez}, {Qi}, {Schwarz}, {Soto}, {Tsukagoshi}, {Yamato}, {van't Hoff}, {Walsh}, {Wilner}, \& {Zhang}}]{MAPSLaw21}
{Law}, C.~J., {Teague}, R., {Loomis}, R.~A., {et~al.} 2021, \bibinfo{title}{{Molecules with ALMA at Planet-forming Scales (MAPS). IV. Emission Surfaces and Vertical Distribution of Molecules},} \apjs, 257, 4, \dodoi{10.3847/1538-4365/ac1439}

\bibitem[{A.~T. {Lee} {et~al.}(2010){Lee}, {Chiang}, {Asay-Davis}, \& {Barranco}}]{Lee2010}
{Lee}, A.~T., {Chiang}, E., {Asay-Davis}, X., \& {Barranco}, J. 2010, \bibinfo{title}{{Forming Planetesimals by Gravitational Instability. I. The Role of the Richardson Number in Triggering the Kelvin-Helmholtz Instability},} \apj, 718, 1367, \dodoi{10.1088/0004-637X/718/2/1367}

\bibitem[{R. Li \& A.~N. Youdin(2021)Li \& Youdin}]{LiYoudin21}
Li, R., \& Youdin, A.~N. 2021, \bibinfo{title}{Thresholds for {Particle} {Clumping} by the {Streaming} {Instability},} The Astrophysical Journal, 919, 107, \dodoi{10.3847/1538-4357/ac0e9f}

\bibitem[{R. Li {et~al.}(2018)Li, Youdin, \& Simon}]{Li18}
Li, R., Youdin, A.~N., \& Simon, J.~B. 2018, \bibinfo{title}{On the {Numerical} {Robustness} of the {Streaming} {Instability}: {Particle} {Concentration} and {Gas} {Dynamics} in {Protoplanetary} {Disks},} The Astrophysical Journal, 862, 14, \dodoi{10.3847/1538-4357/aaca99}

\bibitem[{R. Li {et~al.}(2019)Li, Youdin, \& Simon}]{Li_demographics_2019}
Li, R., Youdin, A.~N., \& Simon, J.~B. 2019, \bibinfo{title}{Demographics of {Planetesimals} {Formed} by the {Streaming} {Instability},} The Astrophysical Journal, 885, 69, \dodoi{10.3847/1538-4357/ab480d}

\bibitem[{J. {Lim} {et~al.}(2025){Lim}, {Simon}, {Li}, {Carrera}, {Baronett}, {Youdin}, {Lyra}, \& {Yang}}]{Lim24b}
{Lim}, J., {Simon}, J.~B., {Li}, R., {et~al.} 2025, \bibinfo{title}{{Probing Conditions for Strong Clumping by the Streaming Instability: Small Dust Grains and Low Dust-to-gas Density Ratio},} \apj, 981, 160, \dodoi{10.3847/1538-4357/adb311}

\bibitem[{J. {Lim} {et~al.}(2024){Lim}, {Simon}, {Li}, {Armitage}, {Carrera}, {Lyra}, {Rea}, {Yang}, \& {Youdin}}]{Lim24a}
{Lim}, J., {Simon}, J.~B., {Li}, R., {et~al.} 2024, \bibinfo{title}{{Streaming Instability and Turbulence: Conditions for Planetesimal Formation},} \apj, 969, 130, \dodoi{10.3847/1538-4357/ad47a2}

\bibitem[{M.-K. {Lin}(2021){Lin}}]{Lin2021}
{Lin}, M.-K. 2021, \bibinfo{title}{{Stratified and Vertically Shearing Streaming Instabilities in Protoplanetary Disks},} \apj, 907, 64, \dodoi{10.3847/1538-4357/abcd9b}

\bibitem[{M.-K. {Lin} \& A.~N. {Youdin}(2017){Lin} \& {Youdin}}]{LinYoudin2017}
{Lin}, M.-K., \& {Youdin}, A.~N. 2017, \bibinfo{title}{{A Thermodynamic View of Dusty Protoplanetary Disks},} \apj, 849, 129, \dodoi{10.3847/1538-4357/aa92cd}

\bibitem[{W. {Lyra} {et~al.}(2024){Lyra}, {Yang}, {Simon}, {Umurhan}, \& {Youdin}}]{Lyra24}
{Lyra}, W., {Yang}, C.-C., {Simon}, J.~B., {Umurhan}, O.~M., \& {Youdin}, A.~N. 2024, \bibinfo{title}{{Rapid Protoplanet Formation in Vortices: Three-dimensional Local Simulations with Self-gravity},} \apjl, 970, L19, \dodoi{10.3847/2041-8213/ad5af6}

\bibitem[{J. {Matthijsse} {et~al.}(2025){Matthijsse}, {Aly}, \& {Paardekooper}}]{Matthijsse2025}
{Matthijsse}, J., {Aly}, H., \& {Paardekooper}, S.-J. 2025, \bibinfo{title}{{Polydisperse formation of planetesimals: The dust size distribution in clumps},} \aap, 695, A158, \dodoi{10.1051/0004-6361/202453072}

\bibitem[{Y. {Nakagawa} {et~al.}(1986){Nakagawa}, {Sekiya}, \& {Hayashi}}]{Nakagawa1986}
{Nakagawa}, Y., {Sekiya}, M., \& {Hayashi}, C. 1986, \bibinfo{title}{{Settling and growth of dust particles in a laminar phase of a low-mass solar nebula},} \icarus, 67, 375, \dodoi{10.1016/0019-1035(86)90121-1}

\bibitem[{D. {Nesvorn{\'y}} {et~al.}(2021){Nesvorn{\'y}}, {Li}, {Simon}, {Youdin}, {Richardson}, {Marschall}, \& {Grundy}}]{Nesvorny2021}
{Nesvorn{\'y}}, D., {Li}, R., {Simon}, J.~B., {et~al.} 2021, \bibinfo{title}{{Binary Planetesimal Formation from Gravitationally Collapsing Pebble Clouds},} \psj, 2, 27, \dodoi{10.3847/PSJ/abd858}

\bibitem[{D. {Nesvorn{\'y}} {et~al.}(2019){Nesvorn{\'y}}, {Li}, {Youdin}, {Simon}, \& {Grundy}}]{Nesvorny2019}
{Nesvorn{\'y}}, D., {Li}, R., {Youdin}, A.~N., {Simon}, J.~B., \& {Grundy}, W.~M. 2019, \bibinfo{title}{{Trans-Neptunian binaries as evidence for planetesimal formation by the streaming instability},} Nature Astronomy, 3, 808, \dodoi{10.1038/s41550-019-0806-z}

\bibitem[{ {OpenAI} {et~al.}(2023){OpenAI}, {Achiam}, {Adler}, {Agarwal}, {Ahmad}, {Akkaya}, {Leoni Aleman}, {Almeida}, {Altenschmidt}, {Altman}, {Anadkat}, {Avila}, {Babuschkin}, {Balaji}, {Balcom}, {Baltescu}, {Bao}, {Bavarian}, {Belgum}, {Bello}, {Berdine}, {Bernadett-Shapiro}, {Berner}, {Bogdonoff}, {Boiko}, {Boyd}, {Brakman}, {Brockman}, {Brooks}, {Brundage}, {Button}, {Cai}, {Campbell}, {Cann}, {Carey}, {Carlson}, {Carmichael}, {Chan}, {Chang}, {Chantzis}, {Chen}, {Chen}, {Chen}, {Chen}, {Chen}, {Chess}, {Cho}, {Chu}, {Chung}, {Cummings}, {Currier}, {Dai}, {Decareaux}, {Degry}, {Deutsch}, {Deville}, {Dhar}, {Dohan}, {Dowling}, {Dunning}, {Ecoffet}, {Eleti}, {Eloundou}, {Farhi}, {Fedus}, {Felix}, {Posada Fishman}, {Forte}, {Fulford}, {Gao}, {Georges}, {Gibson}, {Goel}, {Gogineni}, {Goh}, {Gontijo-Lopes}, {Gordon}, {Grafstein}, {Gray}, {Greene}, {Gross}, {Gu}, {Guo}, {Hallacy}, {Han}, {Harris}, {He}, {Heaton}, {Heidecke}, {Hesse}, {Hickey}, {Hickey}, {Hoeschele}, {Houghton}, {Hsu}, {Hu}, {Hu}, {Huizinga},
  {Jain}, {Jain}, {Jang}, {Jiang}, {Jiang}, {Jin}, {Jin}, {Jomoto}, {Jonn}, {Jun}, {Kaftan}, {Kaiser}, {Kamali}, {Kanitscheider}, {Shirish Keskar}, {Khan}, {Kilpatrick}, {Kim}, {Kim}, {Kim}, {Hendrik Kirchner}, {Kiros}, {Knight}, {Kokotajlo}, {Kondraciuk}, {Kondrich}, {Konstantinidis}, {Kosic}, {Krueger}, {Kuo}, {Lampe}, {Lan}, {Lee}, {Leike}, {Leung}, {Levy}, {Li}, {Lim}, {Lin}, {Lin}, {Litwin}, {Lopez}, {Lowe}, {Lue}, {Makanju}, {Malfacini}, {Manning}, {Markov}, {Markovski}, {Martin}, {Mayer}, {Mayne}, {McGrew}, {McKinney}, {McLeavey}, {McMillan}, {McNeil}, {Medina}, {Mehta}, {Menick}, {Metz}, {Mishchenko}, {Mishkin}, {Monaco}, {Morikawa}, {Mossing}, {Mu}, {Murati}, {Murk}, {M{\'e}ly}, {Nair}, {Nakano}, {Nayak}, {Neelakantan}, {Ngo}, {Noh}, {Ouyang}, {O'Keefe}, {Pachocki}, {Paino}, {Palermo}, {Pantuliano}, {Parascandolo}, {Parish}, {Parparita}, {Passos}, {Pavlov}, {Peng}, {Perelman}, {de Avila Belbute Peres}, {Petrov}, {Ponde de Oliveira Pinto}, {Michael}, {Pokorny}, {Pokrass}, {Pong}, {Powell}, {Power},
  {Power}, {Proehl}, {Puri}, \& {Radford}}]{2023arXiv230308774O}
{OpenAI}, {Achiam}, J., {Adler}, S., {et~al.} 2023, \bibinfo{title}{{GPT-4 Technical Report},} arXiv e-prints, arXiv:2303.08774, \dodoi{10.48550/arXiv.2303.08774}

\bibitem[{C.~W. {Ormel} \& H.~H. {Klahr}(2010){Ormel} \& {Klahr}}]{OrmelKlahr2010}
{Ormel}, C.~W., \& {Klahr}, H.~H. 2010, \bibinfo{title}{{The effect of gas drag on the growth of protoplanets. Analytical expressions for the accretion of small bodies in laminar disks},} \aap, 520, A43, \dodoi{10.1051/0004-6361/201014903}

\bibitem[{J.~B. {Pollack} {et~al.}(1996){Pollack}, {Hubickyj}, {Bodenheimer}, {Lissauer}, {Podolak}, \& {Greenzweig}}]{Pollack+96}
{Pollack}, J.~B., {Hubickyj}, O., {Bodenheimer}, P., {et~al.} 1996, \bibinfo{title}{{Formation of the Giant Planets by Concurrent Accretion of Solids and Gas},} \icarus, 124, 62, \dodoi{10.1006/icar.1996.0190}

\bibitem[{S.~N. {Raymond} {et~al.}(2009){Raymond}, {O'Brien}, {Morbidelli}, \& {Kaib}}]{Raymond2009}
{Raymond}, S.~N., {O'Brien}, D.~P., {Morbidelli}, A., \& {Kaib}, N.~A. 2009, \bibinfo{title}{{Building the terrestrial planets: Constrained accretion in the inner Solar System},} \icarus, 203, 644, \dodoi{10.1016/j.icarus.2009.05.016}

\bibitem[{J.~J. {Rucska} \& J.~W. {Wadsley}(2023){Rucska} \& {Wadsley}}]{Rucska_Wadsley2023}
{Rucska}, J.~J., \& {Wadsley}, J.~W. 2023, \bibinfo{title}{{Planetesimal formation via the streaming instability with multiple grain sizes},} \mnras, 526, 1757, \dodoi{10.1093/mnras/stad2855}

\bibitem[{U. {Sch{\"a}fer} {et~al.}(2024){Sch{\"a}fer}, {Johansen}, {Haugb{\o}lle}, \& {Nordlund}}]{Schafer2024}
{Sch{\"a}fer}, U., {Johansen}, A., {Haugb{\o}lle}, T., \& {Nordlund}, {\r{A}}. 2024, \bibinfo{title}{{Thousands of planetesimals: Simulating the streaming instability in very large computational domains},} \aap, 691, A258, \dodoi{10.1051/0004-6361/202450639}

\bibitem[{U. {Sch{\"a}fer} {et~al.}(2017){Sch{\"a}fer}, {Yang}, \& {Johansen}}]{Schafer17}
{Sch{\"a}fer}, U., {Yang}, C.-C., \& {Johansen}, A. 2017, \bibinfo{title}{{Initial mass function of planetesimals formed by the streaming instability},} \aap, 597, A69, \dodoi{10.1051/0004-6361/201629561}

\bibitem[{N. {Schaffer} {et~al.}(2021){Schaffer}, {Johansen}, \& {Lambrechts}}]{Schaffer2021}
{Schaffer}, N., {Johansen}, A., \& {Lambrechts}, M. 2021, \bibinfo{title}{{Streaming instability of multiple particle species. II. Numerical convergence with increasing particle number},} \aap, 653, A14, \dodoi{10.1051/0004-6361/202140690}

\bibitem[{N. {Schaffer} {et~al.}(2018){Schaffer}, {Yang}, \& {Johansen}}]{Schaffer2018}
{Schaffer}, N., {Yang}, C.-C., \& {Johansen}, A. 2018, \bibinfo{title}{{Streaming instability of multiple particle species in protoplanetary disks},} \aap, 618, A75, \dodoi{10.1051/0004-6361/201832783}

\bibitem[{A. {Schreiber} \& H. {Klahr}(2018){Schreiber} \& {Klahr}}]{Schreiber_Klahr2018}
{Schreiber}, A., \& {Klahr}, H. 2018, \bibinfo{title}{{Azimuthal and Vertical Streaming Instability at High Dust-to-gas Ratios and on the Scales of Planetesimal Formation},} \apj, 861, 47, \dodoi{10.3847/1538-4357/aac3d4}

\bibitem[{U. Schäfer \& A. Johansen(2022)Schäfer \& Johansen}]{schafer_coexistence_2022}
Schäfer, U., \& Johansen, A. 2022, \bibinfo{title}{The coexistence of the streaming instability and the vertical shear instability in protoplanetary disks: {Planetesimal} formation thresholds explored in two-dimensional global models,} Astronomy and Astrophysics, 666, \dodoi{10.1051/0004-6361/202243655}

\bibitem[{M. Sekiya(1998)Sekiya}]{Sekiya98}
Sekiya, M. 1998, \bibinfo{title}{Quasi-{Equilibrium} {Density} {Distributions} of {Small} {Dust} {Aggregations} in the {Sola} r {Nebula},} ICARUS, 133, 298

\bibitem[{M. Sekiya \& I.~K. Onishi(2018)Sekiya \& Onishi}]{sekiya_two_2018}
Sekiya, M., \& Onishi, I.~K. 2018, \bibinfo{title}{Two {Key} {Parameters} {Controlling} {Particle} {Clumping} {Caused} by {Streaming} {Instability} in the {Dead}-zone {Dust} {Layer} of a {Protoplanetary} {Disk},} The Astrophysical Journal, 860, 140, \dodoi{10.3847/1538-4357/aac4a7}

\bibitem[{D. {Sengupta} \& O.~M. {Umurhan}(2023){Sengupta} \& {Umurhan}}]{Sengupta_Umurhan23}
{Sengupta}, D., \& {Umurhan}, O.~M. 2023, \bibinfo{title}{{Turbulence in Particle-laden Midplane Layers of Planet-forming Disks},} \apj, 942, 74, \dodoi{10.3847/1538-4357/ac9411}

\bibitem[{J.-M. {Shi} \& E. {Chiang}(2013){Shi} \& {Chiang}}]{ShiChiang2013}
{Shi}, J.-M., \& {Chiang}, E. 2013, \bibinfo{title}{{From Dust to Planetesimals: Criteria for Gravitational Instability of Small Particles in Gas},} \apj, 764, 20, \dodoi{10.1088/0004-637X/764/1/20}

\bibitem[{J.~B. Simon {et~al.}(2016)Simon, Armitage, Li, \& Youdin}]{simon_mass_2016}
Simon, J.~B., Armitage, P.~J., Li, R., \& Youdin, A.~N. 2016, \bibinfo{title}{{THE} {MASS} {AND} {SIZE} {DISTRIBUTION} {OF} {PLANETESIMALS} {FORMED} {BY} {THE} {STREAMING} {INSTABILITY}. {I}. {THE} {ROLE} {OF} {SELF}-{GRAVITY},} The Astrophysical Journal, 822, 55, \dodoi{10.3847/0004-637X/822/1/55}

\bibitem[{J. {Stadler} {et~al.}(2025){Stadler}, {Benisty}, {Winter}, {Izquierdo}, {Longarini}, {Galloway-Sprietsma}, {Curone}, {Andrews}, {Bae}, {Facchini}, {Rosotti}, {Teague}, {Barraza-Alfaro}, {Cataldi}, {Cuello}, {Czekala}, {Fasano}, {Flock}, {Fukagawa}, {Garg}, {Hall}, {Hammond}, {Hilder}, {Huang}, {Ilee}, {Kanagawa}, {Lesur}, {Lodato}, {Loomis}, {Menard}, {Orihara}, {Pinte}, {Price}, {Yen}, {Wafflard-Fernandez}, {Wilner}, {W{\"o}lfer}, {Yoshida}, \& {Zawadzki}}]{exoALMAStadler25}
{Stadler}, J., {Benisty}, M., {Winter}, A.~J., {et~al.} 2025, \bibinfo{title}{{exoALMA. VI. Rotating under Pressure: Rotation Curves, Azimuthal Velocity Substructures, and Gas Pressure Variations},} \apjl, 984, L11, \dodoi{10.3847/2041-8213/adb152}

\bibitem[{J.~M. Stone \& T.~A. Gardiner(2010)Stone \& Gardiner}]{stone_implementation_2010}
Stone, J.~M., \& Gardiner, T.~A. 2010, \bibinfo{title}{Implementation of the shearing box approximation in {Athena},} Astrophysical Journal, Supplement Series, 189, 142, \dodoi{10.1088/0067-0049/189/1/142}

\bibitem[{J.~M. Stone {et~al.}(2008)Stone, Gardiner, Teuben, Hawley, \& Simon}]{Stone08}
Stone, J.~M., Gardiner, T.~A., Teuben, P., Hawley, J.~F., \& Simon, J.~B. 2008, \bibinfo{title}{{ATHENA}: {A} {NEW} {CODE} {FOR} {ASTROPHYSICAL} {MHD},} The Astrophysical Journal Supplement Series, 178, 137, \dodoi{10.1086/588755}

\bibitem[{A. {Toomre}(1964){Toomre}}]{Toomre1964}
{Toomre}, A. 1964, \bibinfo{title}{{On the gravitational stability of a disk of stars.},} \apj, 139, 1217, \dodoi{10.1086/147861}

\bibitem[{S.~J. Weidenschilling(1980)Weidenschilling}]{Weidenschilling80}
Weidenschilling, S.~J. 1980, \bibinfo{title}{Dust to {Planetesimals}" {Settling} and {Coagulation} in the {Solar} {Nebula},} ICARUS, 44, 172

\bibitem[{S.~J. {Weidenschilling} \& J.~N. {Cuzzi}(1993){Weidenschilling} \& {Cuzzi}}]{WeidenschillingCuzzi93}
{Weidenschilling}, S.~J., \& {Cuzzi}, J.~N. 1993, in Protostars and Planets III, ed. E.~H. {Levy} \& J.~I. {Lunine}, 1031

\bibitem[{Z. Xu \& X.-N. Bai(2022)Xu \& Bai}]{xu_turbulent_2022}
Xu, Z., \& Bai, X.-N. 2022, \bibinfo{title}{Turbulent {Dust}-trapping {Rings} as {Efficient} {Sites} for {Planetesimal} {Formation},} The Astrophysical Journal Letters, 937, L4, \dodoi{10.3847/2041-8213/ac8dff}

\bibitem[{C.-C. {Yang} \& A. {Johansen}(2014){Yang} \& {Johansen}}]{YangJohansen14}
{Yang}, C.-C., \& {Johansen}, A. 2014, \bibinfo{title}{{On the Feeding Zone of Planetesimal Formation by the Streaming Instability},} \apj, 792, 86, \dodoi{10.1088/0004-637X/792/2/86}

\bibitem[{C.~C. Yang {et~al.}(2017)Yang, Johansen, \& Carrera}]{Yang2017}
Yang, C.~C., Johansen, A., \& Carrera, D. 2017, \bibinfo{title}{Concentrating small particles in protoplanetary disks through the streaming instability,} Astronomy and Astrophysics, 606, \dodoi{10.1051/0004-6361/201630106}

\bibitem[{C.-C. {Yang} {et~al.}(2009){Yang}, {Mac Low}, \& {Menou}}]{Yang2009}
{Yang}, C.-C., {Mac Low}, M.-M., \& {Menou}, K. 2009, \bibinfo{title}{{Planetesimal and Protoplanet Dynamics in a Turbulent Protoplanetary Disk: Ideal Unstratified Disks},} \apj, 707, 1233, \dodoi{10.1088/0004-637X/707/2/1233}

\bibitem[{C.-C. {Yang} \& Z. {Zhu}(2020){Yang} \& {Zhu}}]{YangZhu20}
{Yang}, C.-C., \& {Zhu}, Z. 2020, \bibinfo{title}{{Morphological signatures induced by dust back reaction in discs with an embedded planet},} \mnras, 491, 4702, \dodoi{10.1093/mnras/stz3232}

\bibitem[{C.-C. {Yang} \& Z. {Zhu}(2021){Yang} \& {Zhu}}]{YangZhu2021}
{Yang}, C.-C., \& {Zhu}, Z. 2021, \bibinfo{title}{{Streaming instability with multiple dust species - II. Turbulence and dust-gas dynamics at non-linear saturation},} \mnras, 508, 5538, \dodoi{10.1093/mnras/stab2959}

\bibitem[{A. Youdin \& A. Johansen(2007)Youdin \& Johansen}]{YJ07}
Youdin, A., \& Johansen, A. 2007, \bibinfo{title}{Protoplanetary Disk Turbulence Driven by the Streaming Instability: Linear Evolution and Numerical Methods,} The Astrophysical Journal, 662, 613, \dodoi{10.1086/516729}

\bibitem[{A.~N. Youdin \& J. Goodman(2005)Youdin \& Goodman}]{YG05}
Youdin, A.~N., \& Goodman, J. 2005, \bibinfo{title}{{STREAMING} {INSTABILITIES} {IN} {PROTOPLANETARY} {DISKS},} The Astrophysical Journal, 620, 459, \dodoi{10.1086/426895}

\bibitem[{A.~N. Youdin \& Y. Lithwick(2007)Youdin \& Lithwick}]{youdin_particle_2007}
Youdin, A.~N., \& Lithwick, Y. 2007, \bibinfo{title}{Particle stirring in turbulent gas disks: {Including} orbital oscillations,} Icarus, 192, 588, \dodoi{10.1016/j.icarus.2007.07.012}

\bibitem[{A. Zsom {et~al.}(2010)Zsom, Ormel, Güttler, Blum, \& Dullemond}]{Zsom10}
Zsom, A., Ormel, C.~W., Güttler, C., Blum, J., \& Dullemond, C.~P. 2010, \bibinfo{title}{The outcome of protoplanetary dust growth: pebbles, boulders, or planetesimals?: {I}. {Mapping} the zoo of laboratory collision experiments,} Astronomy and Astrophysics, 513, \dodoi{10.1051/0004-6361/200912976}

\end{thebibliography}
\bibliographystyle{aasjournalv7}

\end{CJK}
\end{document}